\documentclass[aps,preprint,noshowpacs,nofootinbib,superscriptaddress,tightenlines,groupedaddress, floatfix,superscriptaddress, preprintnumbers]{revtex4-1}  
 \usepackage{slashed}
\usepackage{braket}
\usepackage{amsmath}
\usepackage{bm}
\usepackage{float}
\usepackage{graphicx}
\usepackage{amsfonts}
\usepackage{simpler-wick}
\usepackage{subcaption}
\usepackage{url}

\begin{document}

\title{Nuclei on the Light Front and Electron Scattering}
\author{Dmitriy N. Kim and Gerald A. Miller}
\affiliation{Department of Physics, University of Washington, Seattle, WA 98195, USA}



\begin{abstract}
  This paper 
  develops relativistic nuclear structure, motivated by past, current and future experiments at Jefferson Lab and the Electron-Ion Collider.  Building on existing tools from conventional nuclear
physics, we reformulate them in a relativistic, light-front-quantized framework. Our light-front nuclear structure calculations reproduce nuclear binding energies and shell structure well, and incorporate the physics of nucleon-nucleon short-ranged-correlations (SRCs) through similarity
renormalization group techniques. Our results are  that a purely nucleonic description   is insufficient to capture inclusive electron-nucleus scattering
data. It  does not fully reproduce the plateaus seen at high Bjorken-$x_B$, pointing to the importance of    non-nucleonic dynamics responsible for the EMC  effect.
\end{abstract}
\maketitle





\section {Introduction}\label{chap:introduction}

 Quantum
Chromodynamics (QCD) is accepted as the fundamental theory governing strong interaction
dynamics~\cite{Yang:1954ek,Gross:1973id,Politzer:1973fx,Gross:1973ju}. Understanding how nucleons
and complex nuclei emerge from the underlying quark and gluon interactions has  become an
overarching goal of  nuclear physics. The short-distance behavior of the strong interaction is
perturbatively calculable and well understood.  At the larger distances relevant to hadronic and
nuclear structure  QCD is highly non-perturbative. This non-perturbative regime---the
intersection of low- and high-energy nuclear physics---remains profoundly challenging and   a
primary focus of contemporary research.

The earliest experimental window into this regime was opened in 1983 by the European Muon
Collaboration (EMC). Performing DIS on iron and deuterium targets with
high-energy muons, they observed that the ratio of per-nucleon cross-sections exhibited a distinct
depletion at intermediate momentum fractions~\cite{EuropeanMuon:1983wih}. It had then been  widely
assumed that high-energy partonic observables would be insensitive to low-energy nuclear
dynamics---that, at energies sufficient to resolve partons, the nucleus could be treated as a
collection of free nucleons. The ``EMC effect,'' as this depletion came to be known, sparked intense
theoretical investigation. Early approaches invoking conventional nuclear mechanisms such as Fermi
motion and pionic enhancements failed to fully reproduce the data, and despite more than forty years
of effort and over a thousand publications, the origin of partonic modifications in bound nucleons
remains a major open question with at least two competing ideas; see the review~\cite{Hen:2016kwk}.

Experimental momentum on this question is only growing. Future measurements at the Electron--Ion
Collider (EIC), which will accelerate nuclear targets to energies as high as
100~GeV/nucleon~\cite{Accardi:2012qut,AbdulKhalek:2021gbh}, and current experiments at Jefferson
Lab (JLab) colliding almost 12~GeV electrons against nuclei~\cite{Dudek:2012vr}, will probe quark
substructure in nuclei with unprecedented resolution. This experimental progress demands a parallel
theoretical advance: a description of the initial nuclear state suitable for high-energy processes.
Nuclear structure calculations have historically been performed by solving the Schr\"odinger
equation, but as the energy of the probe increases, relativistic effects cannot   be neglected.
Without a properly relativistic treatment, purely kinematic effects can easily be misinterpreted as
genuine signatures of new physics. Yet the calculation of nuclear wavefunctions within a consistent
relativistic framework remains largely unexplored.

Conventional instant form quantization, using $x^0$ as the time variable, does not adequately meet
this need because it yields frame-dependent descriptions of bound states \cite{Brodsky:2022fqy}. In
this formalism, a nucleus boosted to high momentum appears Lorentz contracted, and reproducing such
a deformed state requires a superposition over all excited states of the system. This contraction is
unphysical in the present context: it is a frame-dependent artifact rather than an intrinsic
property of the nucleus. The resulting calculations are both technically cumbersome---requiring the
explicit inclusion of excited nuclear states---and conceptually opaque, since frame-dependent
effects become entangled with the intrinsic dynamics one aims to study. Such ``fictitious dynamics''
and related pathologies of instant form-based bound-state calculations have historically led to
misleading conclusions, including apparent violations of the Gerasimov--Drell--Hearn sum rule in the
deuteron \cite{Brodsky:1968xc} and more recent efforts to extract time-independent,
three-dimensional densities for systems of relativistically moving constituents
\cite{Miller:2025zte}.

Light-front quantization, using $x^+ = x^0 + x^3$ as the time variable, offers a natural
resolution to these difficulties: its wave functions are invariant under boosts, enabling a
consistent, frame-independent description of nuclear structure. Despite this central advantage,
relativistic light-front descriptions of finite nuclei remain largely undeveloped. Developing such a
framework is essential for interpreting data from the high-energy nuclear experiments underway and
forthcoming at JLab and the EIC, and is the central motivation for the work presented  here.

The remainder of this paper  is organized as follows. \textbf{Section~\ref{chap:lightfront}}
provides an introduction into light-front quantization, with the guiding question: \emph{why should
low-energy nuclear physicists care?} This section  aims to build qualitative
intuition for the importance of the light-front approach and to present the basic theoretical
machinery needed to perform calculations. In particular, the differences between the instant form (IF) and front form (FF) treatments of off-shell effects are explained. Furthermore, an appropriate LF nucleon-nucleon potential is obtained. This potential is applied to obtain the nucleon momentum distribution in the deuteron.,

\textbf{Section~\ref{chap:dft}} begins the calculations using light-front
 theory. The simplest realistic model of nuclear structure, relativistic mean field theory, provides a starting 
pathway to computing inclusive electron--nucleus cross-sections. Here, we highlight a procedure for
obtaining light-front nuclear structure calculations directly from previously published instant form
results.

The primary observable throughout this work is the inclusive electron--nucleus scattering
cross-section, and \textbf{Section~\ref{chap:scattering}} provides a theoretical overview of this
process. The kinematics are defined, the formalism introduced, the treatment of off-shell effects in the electron-nucleon interaction is explained and  applications to electron-deuteron and electron nucleus inclusive scattering are made.

Mean field theory, however, proves inadequate for high-energy nuclear
processes. This shortcoming motivates \textbf{Section~\ref{chap:srg}}, which introduces the
similarity renormalization group (SRG) and its novel application to relativistic descriptions of
nuclear structure on the light-front. We show how incorporating the SRG modifies the cross-section
calculations of the previous section.

\textbf{Section~\ref{chap:fsi}} addresses the role of final-state interactions and finally,
\textbf{Section~\ref{chap:conclusion}} summarizes the results.
Some technical details are presented in~\cite{Kim:2026fen} and in an appendix.
\section{Light-Front Quantization}\label{chap:lightfront}

Dirac's   foundational paper \emph{Forms of Relativistic Dynamics} found
different  distinct Hamiltonian formulations of relativistic quantum mechanics by exploring
different parameterizations of spacetime~\cite{Dirac:1949cp}. His central insight was that the
choice of spacetime foliation, while leaving physical observables unchanged, can offer significant
computational and conceptual advantages depending on the system under investigation---much as
spherical coordinates simplify spherically symmetric problems without altering their physical
content. Dirac identified three classes of foliations that cannot be related by Lorentz
transformations in any non-trivial way, which he termed ``Forms'': the Instant Form (IF), the Front
Form (FF), and the Point Form (PF). The IF, in which states are quantized on surfaces of constant
time $x^0=t$, is the most familiar and most widely adopted, aligning closely with our built-up
intuition about space and time. The FF, in which states are quantized on surfaces of constant time
$x^+ = x^0 + x^3$, departs from this intuition but has proven exceptionally powerful for high-energy
problems and serves as the primary framework of this work. The PF, in which states are
quantized on a hyperboloid, has seen comparatively limited application.

This section aims to convey why formulating nuclear physics on the light front is essential for the
accurate study of high-energy nuclear reactions. This formulation  underpins the   calculations presented
below. It highlights that the FF is the only relativistic Hamiltonian framework in
which  intrinsic bound-state wave functions are, in principle, invariant under relativistic boosts, and  the off-shell
extrapolation of nucleon-level subprocesses is free of contamination from the probe. Both features are
necessary to compute high-energy nuclear cross sections, and only the FF satisfies both. Dirac's FF has been independently rediscovered and rebranded across the
literature as the Infinite Momentum Frame, Null-Plane Quantization, and Light-Cone Quantization, with subtle differences between these formulations.  We
adopt the name Light-Front (LF) Quantization here. The reader
is directed to Ref.~\cite{Brodsky:1997de} and references therein for  a comprehensive  review.  For applications to nuclear
physics, see Refs.~\cite{Frankfurt:1981mk,Frankfurt:1988nt,Miller:2000kv}.

\subsection{Example: Lorentz Boosting the Deuteron}

\begin{figure}
    \centering
 \includegraphics[width=1\linewidth]{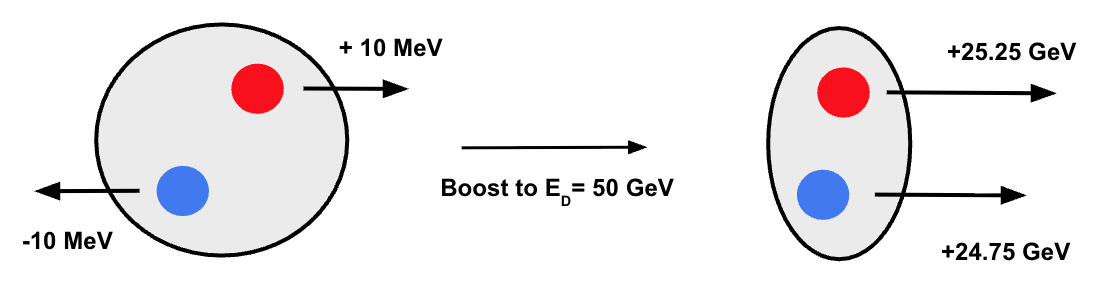}
    \caption{(color online) Cartoon picture of a deuteron configuration where the proton (red) and neutron (blue) have back-to-back momenta of 10 MeV along the $z$-axis. Boosting this configuration to an energy of 50 GeV increases both the absolute and the relative momenta of the constituents.}
    \label{fig:deuteron_cartoon}
\end{figure}

Here we motivate the use of the 
 LF formulation of nuclear dynamics. Consider the interplay between quantum mechanics
and special relativity in a simple system under IF quantization: a deuteron at rest, with its proton
and neutron aligned along the $z$-axis at a relative momentum of 10 MeV
(Fig.~\ref{fig:deuteron_cartoon}). Anticipating the kinematics of the EIC,
where nuclear beams will reach hundreds of GeV per nucleon, apply a longitudinal Lorentz boost of 50
GeV. The individual nucleon momenta can, to a very good  approximation, can be taken as free so that a Lorentz transformation  to approximately 24.75 and 25.25 GeV can be made. Their relative
momentum jumps from 10 MeV to roughly 250 MeV. This is a big change.

This kinematic exercise exposes a fundamental incompatibility between boosts and the traditional
separation of bound-state dynamics into center-of-mass and relative motion. Non-relativistic nuclear
physics---built on the Schr\"odinger equation---relies on this separation, yet the boost has
entangled the two: the relative momentum has changed, despite having no change  to the internal
dynamics of the system. The entanglement carries severe physical consequences. The boost-induced
increase in relative momentum corresponds to a decrease in the relative distance between
constituents; equivalently, the deuteron appears Lorentz-contracted along its direction of motion.
Its spatial density profile in the boosted frame is therefore not the same as in the rest frame, and
no single wave function can describe both.

Within IF quantization, the moving state must instead be represented as a superposition of the
rest-frame Hamiltonian's eigenfunctions. Although the rest-frame and boosted Hamiltonians are
related by a unitary Lorentz transformation, the change in the bound-state wave function means that
boosting effectively induces interactions. These ``fictitious dynamics'' are unphysical,
frame-dependent artifacts rather than intrinsic nuclear properties.

\subsection{General Considerations of Relativistic Hamiltonian Formulations}

The difficulties of the previous sub-section stem from the choice of coordinates used to describe the
bound-state system. In the non-relativistic regime governed by the Schr\"odinger equation, Galilean
invariance enforces the equivalence of inertial frames and permits a clean separation of
center-of-mass and relative variables. As just demonstrated, conventional IF variables are
structurally unsuited for a relativistic description of bound states: they entangle overall motion
with internal structure, introducing fictitious, frame-dependent dynamics. The choice of time
variable lies at the heart of the puzzle. Understanding why requires examining how many-body quantum
mechanics is constructed in general, and revisiting why the non-relativistic case worked in the
first place.

In any canonical quantization framework---relativistic or non-relativistic---one foliates spacetime
into a family of surfaces labeled by a ``time'' coordinate and imposes commutation relations between
fields on each surface. Noether's theorem, applied to the symmetries of the action, then produces a
set of conserved charges that act as generators of unitary transformations on the Hilbert space. For
non-relativistic quantum mechanics, Galilean invariance yields ten such generators: the Hamiltonian
$\hat{H}$, the three momentum components $\hat{P}^i$, the three angular momenta $\hat{J}^i$, and the
three Galilean boost generators $\hat{K}_G^i$. The relativistic case yields ten generators of the
Poincar\'e group: $\hat{H}$, $\hat{P}^i$, $\hat{J}^i$, and the three Lorentz boost generators
$\hat{K}^i$.

The act of foliating exposes a subtlety: a symmetry of the full four-dimensional spacetime need not
act as a symmetry of any single lower-dimensional slice. A cylinder aligned with the $z$-axis is
rotationally symmetric about $z$, yet only certain slices inherit that symmetry---a perpendicular
cut yields a disk, still invariant under rotations about $z$, while a tilted cut yields an ellipse
that is not. This same logic partitions the ten generators into two classes. Those that map a
foliation surface into itself are \emph{kinematic} and contain no interactions; those that move
points off the surface are \emph{dynamical} and carry the interactions of the theory. Different
foliations preserve different subgroups of the full symmetry, and the resulting split between
kinematic and dynamical generators governs which transformations can be performed trivially and
which require solving the dynamics.

In non-relativistic quantum mechanics, the standard choice of $x^0$ as the time variable  
is  optimal: of the ten Galilean generators, only the Hamiltonian is dynamical. The other
nine---spatial translations, rotations, and Galilean boosts---leave equal-time surfaces invariant
and are therefore kinematic. This is   why center-of-mass and relative coordinates separate
cleanly, and why a single non-relativistic wave function describes the bound state in every inertial
frame.

The relativistic case is more delicate. Promoting Galilean boosts to Lorentz boosts changes their
character: a Lorentz boost tilts an equal-time surface rather than mapping it to itself. The three
Lorentz boost generators therefore join the Hamiltonian as dynamical generators, shrinking the
kinematic subgroup to six---spatial translations and rotations alone. The non-intrinsic dynamics
encountered in the boosted deuteron are the direct consequence: boosting a bound state now requires
evaluating an interaction-dependent operator. This is visible explicitly in physical cross sections.
The nuclear tensor, $W^{\mu\nu}$, for a target of momentum {\bf P} that enters in the cross section is given by
\begin{equation}\label{eq:IF_boost_invariance}
\begin{aligned}
W^{\mu\nu}\propto
& \sum_X \braket{A,\bm{P}|\hat{J}^\mu(0)|X}\braket{X|\hat{J}^{\nu}(0)|A, \bm{P}}\delta^4(q+P_A-P_X) \\
= & \sum_X \braket{A,\bm{P}=0|\hat{U}^{\dagger}(\Lambda)\hat{J}^\mu(0)|X}\delta^4(q+P_A-P_X)\braket{X|\hat{J}^{\nu}(0)\hat{U}(\Lambda)|A, \bm{P}=0},
\end{aligned}
\end{equation}
in which spin degrees of freedom are suppressed. 
Restricting the bound state to its rest frame merely transfers the dynamical complexity into
the boost operators $\hat{U}(\Lambda)$ acting on the currents. The nuclear interactions embedded in
those operators are unavoidable so long as boosts are dynamical generators---they cannot be removed,
only relocated. The same argument also affects cases where the final state contains bound nuclear
states. In high-energy nuclear processes with    moving composite objects in the 
initial and final states, boosts are central to the kinematics. Ensuring that bound states
transform kinematically   is essential. 
The FF, introduced next, does precisely this.

\subsection{Light-Front Variables}
\label{sec:relativistic_hamiltonian}

The pathologies identified in the previous section can be traced to the use of the equal-time
hyperplane $x^0 = 0$ as the quantization surface, on which the boost generators are dynamical and
entangle internal motion with the overall frame. The FF, used interchangeably with LF quantization,
instead quantizes on the $x^+ = 0$ hyperplane. The variables
used in FF are called LF variables, they are defined as
   $ x^\mu = (x^+, x^-, x^1, x^2), \,
    x^\pm \equiv x^0 \pm x^3.
    $
The choice of time variable is a matter of convention. We take $x^+$ as the
LF time throughout this work. The corresponding contraction between position and momentum is
$
    x_\mu p^\mu = 
     \tfrac{1}{2} x^+ p^- + \tfrac{1}{2} x^- p^+ - \bm{x}^\perp \cdot \bm{p}^\perp.
$ 
 From this we see that
$x^+$ and $p^-$ are conjugate, and we accordingly identify $p^-$ as the LF energy. For a free
massive particle, the plus component is manifestly positive,
$
    p^+ = p^0 + p^z = \sqrt{m^2 + \bm{p}^2_{\rm IF}} + p^z > 0,
$ and the mass-shell condition $p_\mu p^\mu = m^2$ takes the form
   $ p^- = \frac{m^2 + (\bm{p}^\perp)^2}{p^+} > 0.
$ 
Unlike the equal-time dispersion relation $p^0 = \pm\sqrt{m^2 + \bm{p}^2_{\rm IF}}$, the LF
dispersion relation is rational in the kinematical variables $(p^+, \bm{p}^\perp)$ and admits no
negative-energy branch.

The decisive feature of the FF, however, lies in the structure of its boost generators, which are
kinematical rather than dynamical. To state these definitions, we recall that the Poincar\'e algebra
contains three rotation generators $J^i$ and three boost generators $K^i$, where $J^i$ generates
rotations about the $x^i$-axis and $K^i$ generates ordinary Lorentz boosts along the
$x^i$-direction. The longitudinal LF boost is simply $K^3$, the ordinary Lorentz boost along the
$z$-direction. The transverse LF boosts, however, are the combinations
$
    E^1 = K^1 + J^2, \, E^2 = K^2 - J^1,
$ 
in which an ordinary transverse boost is paired with a rotation in such a way that the quantization
surface $x^+ = 0$ is preserved. Together with the longitudinal boost, the $E^i$ generate the full
set of LF boosts; we refer to these  as longitudinal and transverse LF boosts. Their
action on LF momenta makes their utility manifest.

Defining the rapidity $\omega$ through $e^\omega = \sqrt{(1-v)/(1+v)}$, where $v$ is the velocity of
the boost in natural units, a longitudinal LF boost acts as
\begin{equation}
    \begin{gathered}
        p^+ \rightarrow e^{\omega} p^+, \\
        p^- \rightarrow e^{-\omega} p^-, \\
        \bm{p}^\perp \rightarrow \bm{p}^\perp,
    \end{gathered}
\label{long}\end{equation}
so that the plus and minus components are simply rescaled and the transverse momentum is left
untouched. A transverse LF boost with velocity $\bm{v}^\perp$ instead yields
\begin{equation}
    \begin{gathered}
        p^+ \rightarrow p^+, \\
        \bm{p}^\perp \rightarrow \bm{p}^\perp + p^+ \bm{v}^\perp, \\
        p^- \rightarrow p^- + 2\, \bm{v}^\perp \cdot \bm{p}^\perp + p^+ (\bm{v}^\perp)^2,
    \end{gathered}
\label{perp}\end{equation}
leaving the plus component invariant.

The transformation laws of  Eq.~(\ref{long}) have important consequences for the identifying  intrinsic variables
in nuclear physics. Because longitudinal LF boosts rescale every plus momentum by the same factor
$e^\omega$, the ratio
\begin{equation}\label{eq:LF_x_rel}
    x_i = \frac{p_i^+}{P^+},
\end{equation}
with $P^+ = \sum_i p_i^+$ the total plus momentum of the bound state and the index $i$ running over
its $A$ constituents, is invariant under both longitudinal and transverse LF boosts. Combined with
the positivity of $p^+$, this further implies $0 < x_i < 1$ for each massive
constituent, so that the $x_i$ form a natural set of boost-invariant longitudinal momentum
fractions.

The transverse boost laws  of   Eq.~(\ref{perp})   reveal a remarkable feature of the FF. The
transformations of $p^-$ and $\bm{p}^\perp$ closely mirror the way non-relativistic kinetic energy
and momentum transform under a Galilean boost,
$    \left[ \bm{p} \rightarrow \bm{p} + m\bm{v} \right]_{\rm IF}, \,$ and 
 $   \left[ E_{\text{kin}} \rightarrow E_{\text{kin}} + \bm{v} \cdot \bm{p} + \tfrac{1}{2}m \bm{v}^{2} \right]_{\rm IF},
$
with the plus momentum $p^+$ playing the role of the mass. The subscript IF on the square brackets
is there to emphasize that all variables are IF three-momenta. The   Galilean structure that
broke down under Lorentz transformations re-emerges  in the transverse plane. This
correspondence motivates the definition of the LF relative transverse momenta
\begin{equation}
    \bm{k}_i^\perp = \bm{p}^\perp_i - \frac{p_i^+}{P^+} \bm{P}^\perp = \bm{p}^\perp_i - x_i \bm{P}^\perp.
\end{equation}
which takes  the form of their non-relativistic counterparts
$
    \left[ \bm{k}_i = \bm{p}_i - (m_i/M)\bm{P} \right]_{\rm IF},
$
with $M = \sum_i m_i$. By construction, the $\bm{k}_i^\perp$ are invariant under both longitudinal
and transverse LF boosts, and satisfy the constraint $\sum_i \bm{k}_i^\perp = 0$.

With $(x_i, \bm{k}_i^\perp)$ we have thus identified a set of intrinsic variables that do not mix
with the center-of-mass momentum under LF boosts. The utility of this separation is made manifest by
examining the total LF kinetic energy of an $A$-body system, multiplied by $P^+$,
\begin{equation}\label{eq:LF_kinetic_energy}
    P^+\sum_i p_i^- = P^+\sum_i \frac{m_i^2 + (\bm{p}_i^\perp)^2}{p_i^+} = (\bm{P}^\perp)^2 + \sum_i \frac{m_i^2 + (\bm{k}_i^\perp)^2}{x_i},
\end{equation}
in which the center-of-mass and intrinsic contributions separate cleanly, in direct analogy with the
non-relativistic decomposition $\left[ T = \bm{P}^2/2M + \sum_i \bm{k}_i^2/2m_i\right]_{\rm IF}$.
Since the intrinsic dynamics of the bound state must be independent of its overall center-of-mass
motion, any interaction term added to the Hamiltonian   depends only on the intrinsic variables
$(x_i, \bm{k}_i^\perp)$. Hence, the bound-state wavefunctions obtained by diagonalizing such a
Hamiltonian are then functions of $(x_i, \bm{k}_i^\perp)$ alone, and are accordingly invariant under
both longitudinal and transverse LF boosts.

This is  the structure   absent from  the IF formulation. The fictitious dynamics that
plagued the IF treatment of the boosted deuteron---the recalculation of the wave function in every
frame, the entanglement of internal structure with center-of-mass motion, the migration of dynamical
complexity into the boost operators of Eq.~\eqref{eq:IF_boost_invariance}---are absent here by
construction. The rest-frame and boosted descriptions of the bound state are connected by a
kinematical transformation acting only on $P^+$ and $\bm{P}^\perp$, while the intrinsic wavefunction
$\psi(x_i, \bm{k}_i^\perp)$ is left invariant. The LF variables thus provide the kinematic basis
suited to relativistic bound states.

\subsection{Scattering Amplitudes}
The bound-state wave functions identified in the previous section are one ingredient  that enters in computing a cross
section; the other is a model of the elementary subprocess in which the probe interacts with a
constituent of the bound state. We now turn to this second ingredient and show that LF quantization
is equally indispensable there.

In modern quantum field theory scattering processes are typically evaluated using 
Feynman diagrams that offer the distinct advantage of manifest Lorentz invariance for every
individual graph. When considering scattering by  composite targets, however, this covariant
approach becomes unwieldy: the bound-state input it requires---a fully covariant wave function---is
only available for two-body bound states, where it is given by the Bethe-Salpeter equation
\cite{Salpeter:1951sz}. Even there, its practical use is hampered both by prohibitive computational
costs and by well-known conceptual difficulties such as the treatment of relative time. To make
progress, one reverts to a Hamiltonian framework, in which bound-state wave functions are rigorously
defined and tractable. The cost of this shift is that scattering amplitudes must be computed using
old-fashioned, time-ordered perturbation theory, which increases the number of   graphs to
be summed. The trade-off is favorable: a manageable proliferation of diagrams in exchange for access
to bound-state structure.

\begin{figure}
    \centering
 \includegraphics[width=0.7\linewidth]{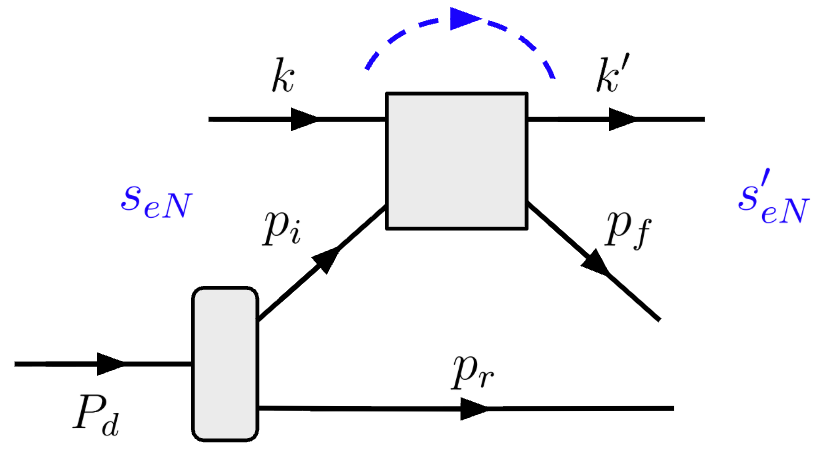}
    \caption{(color online) Feynman diagram for electron-deuteron scattering with an electron-nucleon subprocess. The invariant energy before and after the collision are labeled $s_{eN}$ and $s_{eN}'$ respectively.}
    \label{fig:nuclear_subprocess}
\end{figure}

For high-energy nuclear processes in which the probe resolves the individual nucleons inside the
nucleus, the central theoretical challenge, aside from nuclear structure, is modeling the electron-nucleon
subprocess. The standard practice is an off-shell extrapolation of the on-shell electron-nucleon
amplitude: the bound nucleon, unlike a free one, does not lie on its mass shell, and the
free-nucleon cross section is extended into the off-shell region using a prescription such as that
of de Forest~\cite{DeForest:1983ahx}. Under IF at relativistic kinematics, however, this
extrapolation acquires a contamination unrelated to the nuclear target. The IF and FF, as
Hamiltonian formulations, share a common feature: only three components of momentum are conserved at
each interaction vertex, with intermediate particles placed on their mass shell while the remaining
momentum component absorbs the binding-induced mismatch. They differ in which three components:
spatial three-momentum $\bm{p}_{\rm IF}$ in IF, with energy $p^0$ unconserved; LF three-momentum
$\bm{p}=(p^+, \bm{p}^\perp)$ in FF, with $p^-$ unconserved. This distinction has direct consequences
for the off-shellness of the struck nucleon.

Consider the Feynman diagram in Fig.~\ref{fig:nuclear_subprocess}, with $k$ the incident electron,
$k'$ the scattered electron, $P_d$ the deuteron, $p_r$ the spectator/recoil nucleon, $p_i$ the
initial nucleon, and $p_f$ the scattered nucleon. We quantify the virtuality  by the difference
between the invariant energy of the electron-nucleon subprocess before and after the interaction,
$s_{eN}' - s_{eN}$. Using overall four-momentum conservation, $k' + p_f = k + P_d - p_r$,
$s_{eN}' - s_{eN} = (k' + p_f)^2 - (k + p_i)^2 = (k + P_d - p_r)^2 - (k + p_i)^2\,=\,
                   \left[\, 2k \cdot (P_d - p_r - p_i) \,\right] + \left[\, (P_d - p_r)^2 - p_i^2 \,\right].
$
The square brackets separate two physically distinct contributions. The right-hand bracket, $(P_d -
p_r)^2 - p_i^2$, measures the degree to which  the struck nucleon is off its mass shell: $(P_d - p_r)$ is the
four-momentum that would be assigned to it if four-momentum conservation were enforced at the
deuteron vertex, while $p_i$ is the four-momentum actually assigned by the quantization scheme (on
mass shell, $p_i^2 = m^2$). The two differ because only three momentum components are conserved at
the vertex; the remaining component is fixed instead by the mass-shell condition. The left-hand
bracket depends on the electron's four-momentum and is therefore a property of the probe rather than
the target. A consistent off-shell extrapolation should depend only on the right-hand bracket; the
left-hand bracket is the contamination.

Working in the deuteron rest frame with $(P_d^\mu)_{\rm IF} = (M_d, \bm{0})$ and an incident
electron with $k^\mu_{\rm IF} = (\omega, 0, 0, -\omega)$ in the high-energy limit $\omega > 1$ GeV,
the IF rules at the deuteron vertex give $(\bm{p}_i)_{\rm IF} = -(\bm{p}_r)_{\rm IF}$ and $p_i^0 =
\sqrt{m^2 + (\bm{p}_r)_{\rm IF}^2}$, while $p_r^0 + p_i^0 \neq M_d$---the energy mismatch that
defines the bound state. Evaluating the probe-dependent bracket to leading order in $|\bm{p}_r|_{\rm
IF}/m$ and neglecting the small deuteron binding energy,
\begin{equation}\label{eq:soffshell_IF}
\left[ s_{eN}' - s_{eN} \right]_{\rm IF}\approx -2\omega\,\frac{(\bm{p}_r)_{\rm IF}^2}{m} + \left[\, (P_d - p_r)^2 - p_i^2 \,\right]_{\rm IF},
\end{equation}
which  grows linearly with the incident energy $\omega$ and depends explicitly on the
kinematics of the probe.  At high energies the IF prescription cannot disentangle target structure from probe
kinematics.

LF quantization removes the contamination. Expanding the four-vector dot product in LF components,
\begin{equation}\label{eq:soffshell_LF}
\begin{aligned}
s_{eN}' - s_{eN} &= \left[\, k^+ (P_d - p_r - p_i)^- + k^- (P_d - p_r - p_i)^+ - 2\bm{k}^\perp \cdot (\bm{P}_d - \bm{p}_r - \bm{p}_i)^\perp \,\right] \\
&\quad + \left[\, (P_d - p_r)^2 - p_i^2 \,\right].
\end{aligned}
\end{equation}
With the chosen kinematics, $k^+ = 0$ and $\bm{k}^\perp = 0$, and plus-momentum conservation at the
deuteron vertex enforces $(P_d - p_r - p_i)^+ = 0$. The entire probe-dependent bracket vanishes,
leaving
\begin{equation}\label{eq:soffshell_LF_final}
s_{eN}' - s_{eN} = (P_d - p_r)^2 - p_i^2.
\end{equation}
The effects of virtuality are determined entirely by the nuclear target.  Probe kinematics  do not  enter.


\subsection{Minimal Relativity and the Front Form}
The nucleon-nucleon (NN) interaction plays a central role throughout this
paper. At present, no published work has developed phenomenological
relativistic LF NN potentials capable of describing elastic NN phase shifts to
high accuracy, and constructing such potentials lies outside the scope of this work. We therefore
adopt an alternative prescription: we recast an existing non-relativistic NN
potential---here the Argonne $v_{18}$ (AV18) potential \cite{Wiringa:1994wb}---into relativistic
form, perform a change of variables, and incorporate the effects of LF spinors to obtain an
approximate LF NN potential. See  Refs.
\cite{Frankfurt:1981mk,Frankfurt:1988nt,Strikman:2017koc,Cosyn:2020kwu}.

The starting point is the non-relativistic Lippmann-Schwinger equation, at zero center-of-mass
momenta, for a two-nucleon system in relative-momentum coordinates, neglecting spin and isospin,
\begin{equation}\label{lippmann_schwinger}
    \left [ \frac{k^2}{m_N} \psi(\bm{k}) + \int \frac{d\bm{k}'}{(2 \pi)^3} V(\bm{k},\bm{k}') \psi(\bm{k}') = \epsilon \psi(\bm{k}) \right]_{\rm IF},
\end{equation}
where   the subscript IF on the square brackets is there to emphasize that all variables are IF
three-momenta. To recast this as an effective relativistic equation, we perform the following
sequence of manipulations: multiply throughout by $m_N$, add $m_N^2 \psi(\bm{k}_{\rm IF})$ to both
sides, multiply throughout by $4 \sqrt{E(\bm{k}_{\rm IF})}$, and insert a factor of $E(\bm{k}_{\rm
IF}')/E(\bm{k}_{\rm IF}')$ inside the integral. The result is
\begin{equation}\label{rel_lippmann_schwinger}
    \begin{gathered}
    \left[ 4(k^2+m_N^2) \tilde{\psi}(\bm{k}) + \int \frac{d\bm{k}'}{(2 \pi)^3 E(\bm{k}')} \tilde{V}(\bm{k},\bm{k}') \tilde{\psi}(\bm{k}') = M^2 \tilde{\psi}(\bm{k}) \right]_{\rm IF}, \\
    \left[ \tilde{\psi}(\bm{k}) = \sqrt{E(\bm{k})} \psi(\bm{k}) \right]_{\rm IF}, \\
    \left[ \tilde{V}(\bm{k},\bm{k}') = 4m_N\sqrt{E(\bm{k})}V(\bm{k},\bm{k}')\sqrt{E(\bm{k}')} \right]_{\rm IF},
\end{gathered}
\end{equation}
where $M^2 = 4(m^2_N + m_N \epsilon)$.

This procedure, known as minimal relativity, was first introduced by Brown, Jackson, and Kuo
\cite{Brown:1969tfp}.  Minimal relativity 
supplies the necessary relativistic kinematics without altering the phenomenologically fitted
scattering phase shifts, thereby motivating the high-momentum components that emerge from realistic
NN potentials.
The performance of the AV18 potential offers further support. Although it was
originally fitted to reproduce nucleon-nucleon phase shifts up to a laboratory energy of 350~MeV, it
nonetheless reproduces them well up to 600~MeV in several partial-wave channels as shown in
Ref.~\cite{av18_website}.

The connection to light-front quantization is established by the change of
variables,
\begin{equation}
    \left[ k^2 + m_N^2 \right]_{\rm IF} = \frac{m_N^2 + (\bm{k}^\perp)^2}{\alpha (2-\alpha)},
\end{equation}
where $\alpha=2 p^+_2/(p_1^+ +p_2^+)$ so that  
$
    \int\frac{d\bm{k}'_{\rm IF}}{(2 \pi)^3 E(\bm{k}'_{\rm IF})} = \int\frac{d\alpha'd\bm{k}'^\perp}{(2 \pi)^3 \alpha'(2-\alpha')}.
$
With this substitution, Eq.~\eqref{rel_lippmann_schwinger} takes the form of the two-body Weinberg
equation \cite{Frankfurt:1981mk,Frankfurt:1988nt,Weinberg:1966jm}, where we relabel $ 
\tilde{\psi}_{\rm IF}(\bm{k}_{\rm IF})\equiv \psi(\alpha, {\bf k}_\perp)$ and $V \equiv \tilde{V}_{\rm IF}$ under the new
coordinates,
\begin{equation}\label{weinberg_eq}
    \begin{aligned}
    4\frac{m_N^2 + (\bm{k}^\perp)^2}{\alpha (2-\alpha)}  \psi(\alpha,\bm{k}^\perp) + \int\frac{d\alpha'd\bm{k}'^\perp}{(2 \pi)^3 \alpha'(2-\alpha')} V(\alpha,\bm{k}^\perp;\alpha',\bm{k}'^\perp) & \psi(\alpha',\bm{k}'^\perp)   = M^2 \psi(\alpha,\bm{k}^\perp),
    \end{aligned}
\end{equation}
whose operator form is
\begin{equation}
\begin{gathered}
    \left[ \hat{P}^+\hat{P}^- - \hat{P}^\perp\cdot\hat{P}^\perp \right] \ket{\Psi_{NN}, \bm{P}} = M^2 \ket{\Psi_{NN}, \bm{P}}, 
    ,
\end{gathered}
\end{equation}
with $\hat{P}^- = \hat{P}^-_{\rm kin} + \hat{V}.$

We need to incorporate spin. The  IF
potential is  a matrix element taken between IF Dirac spinors.   What remains is
to transform the IF Dirac spinors into LF Dirac spinors. This  is accomplished through a Melosh
rotation, a unitary relation between the two \cite{Melosh:1974cu,Cosyn:2020kwu},
\begin{equation}
    u(\bm{p},\sigma) = \sum_{\sigma'} u_{\rm IF}(\bm{p}_{\rm IF},
    \sigma') \chi^\dagger(\sigma') \mathcal{U}(\bm{p}_{\rm IF}) \chi(\sigma) 
\end{equation}
with $\mathcal{U}(\bm{p}_{\rm IF}) \equiv \frac{p^+ + m_N + \bm{p}^\perp \cdot \bm{\sigma}^\perp \sigma^3}{\sqrt{2p^+},\sqrt{E(\bm{p}_{\rm IF})+m_N}},\, p^+ \equiv E(\bm{p}_{\rm IF}) + p^3,$ and where $(\bm{\sigma}^\perp,\sigma^3)$ are  Pauli matrices and $\chi(\sigma)$ are   Pauli
spinors.

Including spin, the bound and scattering-state wavefunction solutions of Eq.~\eqref{weinberg_eq} are
therefore
\begin{equation}
    \psi_\beta(\alpha,\bm{k}^\perp,\sigma_1,\tau_1,\sigma_2,\tau_2) = \sum_{\sigma_1' \sigma_2'} \tilde{\psi}^\beta_{\rm IF}(\bm{k}_{\rm IF},\sigma_1',\tau_1,\sigma_2',\tau_2) U^*(\bm{k}_{\rm IF},\sigma_1',\sigma_1)U^*(-\bm{k}_{\rm IF},\sigma_2',\sigma_2),
\end{equation}
where $\beta$ indexes the eigenvalues of the Hamiltonian. The half-off-shell $T$-matrix is
analogously,
\begin{equation}\label{eq:T_matrix_ET_LF}
\begin{gathered}
    T(\alpha_{\rm off}, \bm{k}^\perp_{\rm off}, \sigma_1, \tau_1, \sigma_2, \tau_2;\alpha, \bm{k}^\perp, \sigma_3, \tau_3, \sigma_4, \tau_4) = \\
    \sum_{\sigma_1' \sigma_2'}  \sum_{\sigma_3' \sigma_4'}U^*(\bm{k}_{\rm IF}^{\rm off},\sigma_1',\sigma_1)U^*(-\bm{k}^{\rm off}_{\rm IF},\sigma_2',\sigma_2)  U(\bm{k}_{\rm IF},\sigma_3',\sigma_3)U(-\bm{k}_{\rm IF},\sigma_4',\sigma_4) \\
    \times \tilde{T}_{\rm IF}(\bm{k}^{\rm off}_{\rm IF}, \sigma_1', \tau_1, \sigma_2', \tau_2;\bm{k}_{\rm IF}, \sigma_3', \tau_3, \sigma_4', \tau_4),
\end{gathered}
\end{equation}
where $\tilde{T}_{\rm IF}(\bm{k}_{\rm IF}^{\rm off},\bm{k}_{\rm IF}) = \sqrt{E(\bm{k}_{\rm IF}^{\rm
off})} T_{\rm IF}(\bm{k}_{\rm IF}^{\rm off},\bm{k}_{\rm IF}) \sqrt{E(\bm{k}_{\rm IF})}$.

Taken together, minimal relativity, the transformation to relative LF
coordinates, and the inclusion of LF spinors allow us to employ
well-established non-relativistic potentials within the LF quantization
framework. This prescription has been widely used in high-energy nuclear
physics, for instance in deuteron electrodisintegration \cite{Sargsian:2009hf} and in semi-exclusive
electron-deuteron reactions in both the quasi-elastic and DIS regimes
\cite{Sargsian:2001ax,Strikman:2017koc,Cosyn:2020kwu,Cosyn:2026vpc}. Although
the theoretical uncertainty of the prescription is difficult to quantify, these studies find that it
performs well for relative momenta up to roughly
600~MeV---large enough that we can reliably describe short-range
nucleon-nucleon dynamics within LF quantization. 

\begin{figure}[H]
    \centering
 \includegraphics[width=0.7\linewidth]{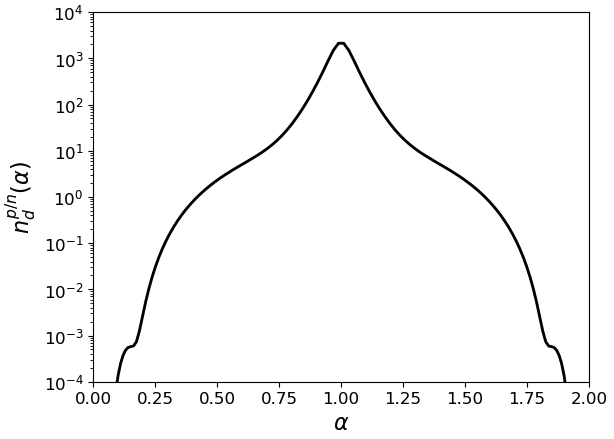}
    \caption{The light-front momentum distribution, integrated over transverse momenta, of deuterium as a function of the Lorentz-invariant relative variable $\alpha=2p^+/P_d^+$.}
    \label{fig:deut_lf_dist}
\end{figure}

The deuteron provides a concrete example of the LF procedure. The
deuteron state is labeled as
$
    \ket{d, M_J; \bm{P}_d}, \, P_d^- = \frac{m_d^2 + (\bm{P}_d^\perp)^2}{P_d^+},
$ 
the overlap with an anti-symmetric two-nucleon state being
  $  \braket{\bm{p}_1,\sigma_1,\tau_1;\bm{p}_2,\sigma_2,\tau_2|d, M_J; \bm{P}_d} = \\(2 \pi)^3 2P_d^+ \delta(p_1^+ + p_2^+ - P_d^+)\delta^{(2)}(\bm{p}_1^\perp + \bm{p}_2^\perp - \bm{P}_d^\perp) \psi_d(\alpha,\bm{k}^\perp,\sigma_1,\tau_1,\sigma_2,\tau_2|M_J),
$ 
with wave function normalization
$
    \frac{1}{3}\sum_{M_J} \frac{1}{2}\sum_{\substack{\sigma_1 \tau_1 \\\sigma_2 \tau_2}}
    \int\frac{d\alpha d\bm{k}^\perp}{(2 \pi)^3\alpha(2-\alpha)} |\psi_d(\alpha,\bm{k}^\perp,\sigma_1,\tau_1,\sigma_2,\tau_2|M_J)|^2 = 1.
$
Fig. \ref{fig:deut_lf_dist}
presents the LF momentum distribution, $n_d^{p/n}$, integrated over transverse momenta, of deuterium. The distribution has broad support caused by the short-ranged component of the nucleon-nucleon interaction. 

\section{ Light Front Mean-Field Theory}\label{chap:dft}

The goal of this section is to provide light front wave functions applicable to high-energy electron scattering calculations.
We start with  Relativistic Mean-Field Theory.  Historically, this 
  began with the Walecka model \cite{WALECKA1974491}.
 The Walecka model is a phenomenological model of
nucleons interacting through the exchange of scalar and vector mesons which mimic nuclear attraction
and repulsion. The extension  to finite nuclear systems was later developed
by Horowitz and Serot  \cite{Horowitz:1981xw}. Using
Dyson's equations for the dressed single-particle propagator, Horowitz and Serot developed a set of
self-consistent equations where nucleon densities act as current sources that generate the meson
fields, which in turn influence the individual nucleon dynamics. Here,  we extend this
framework  to the LF, with a further generalization to
density-dependent couplings.

The starting point is a phenomenological field-theoretic Lagrangian, $\mathcal{L}$, describing
nucleons interacting through the exchange of isoscalar-scalar $\sigma$-mesons, isoscalar-vector
$\omega$-mesons, isovector-vector $\rho$-mesons, and photons.
\begin{equation}
\mathcal{L} = \mathcal{L}_N + \mathcal{L}_m + \mathcal{L}_{int},
\label{eq:lagrangian_total}
\end{equation}
$\mathcal{L}_N$ is the Dirac Lagrangian describing free nucleons
$
\mathcal{L}_N = \bar{\psi}(i\gamma_\mu \partial^\mu - m_N)\psi,
$ 
$\mathcal{L}_m$ corresponds to the free meson and photon fields,
\begin{equation}
\begin{split}
\mathcal{L}_m =&\ \frac{1}{2}\partial_\mu \sigma \partial^\mu \sigma - \frac{1}{2}m_\sigma^2 \sigma^2 - \frac{1}{4}\Omega_{\mu\nu}\Omega^{\mu\nu} + \frac{1}{2}m_\omega^2 \omega_\mu \omega^\mu \\
&- \frac{1}{4}\vec{R}_{\mu\nu} \cdot \vec{R}^{\,\mu\nu} + \frac{1}{2}m_\rho^2 \vec{b}_\mu \cdot \vec{b}^{\,\mu} - \frac{1}{4}F_{\mu\nu}F^{\mu\nu},
\end{split}
\label{eq:lagrangian_meson}
\end{equation}
where the field tensors are given by
$
\Omega_{\mu\nu} = \partial_\mu \omega_\nu - \partial_\nu \omega_\mu,
$
$
\vec{R}_{\mu\nu} = \partial_\mu \vec{b}_\nu - \partial_\nu \vec{b}_\mu,
$
$
F_{\mu\nu} = \partial_\mu A_\nu - \partial_\nu A_\mu.
$
The interaction terms between the meson/electromagnetic fields are encapsulated in
$\mathcal{L}_{int}$
\begin{equation}
\mathcal{L}_{int} = -g_\sigma \bar{\psi}\psi\sigma - g_\omega \bar{\psi}\gamma^\mu \psi \omega_\mu - \frac{1}{2}g_\rho \bar{\psi}\gamma^\mu \vec{\tau}\psi \cdot \vec{b}_\mu - e\bar{\psi}\gamma^\mu \frac{1+\tau_3}{2}\psi A_\mu,
\label{eq:lagrangian_int}
\end{equation}
In the above expressions, $m_N$ is the nucleon mass, while $m_m$ and $g_m$ ($m = \sigma, \omega,
\rho$) denote the meson masses and meson--nucleon coupling constants. The electromagnetic field is
denoted $A^\mu$, with coupling $e$, and the $\rho$-meson field is denoted $\vec{b}^{\, \mu}$.

RMF treats the meson mediators as classical fields. Because the meson fields are not quantized,
quantum-mechanical exchange and correlation interactions are neglected so that  RMF is
intrinsically a Hartree-level theory. Furthermore, by restricting our focus to static,
spherically-symmetric nuclei,  pionic fields do not contribute to the ground state due to parity
conservation. These meson fields phenomenologically capture distinct aspects of the nucleon-nucleon
interaction: the $\sigma$-meson simulates intermediate-range attraction, the $\omega$-meson provides
short-range repulsion, and the $\rho$-meson accounts for the isospin dependence. It is well known
that pions play a critical role in the nuclear force, providing the long-range attraction and the
short-range tensor force. Although the underlying assumptions of the RMF framework prevent the pion
fields from contributing explicitly to the ground state, the specific combination of the remaining
effective meson fields is designed to implicitly absorb and replicate their overall physical
effects. For example, the $\sigma$ meson roughly accounts for the effects of two-pion exchange. 

From Noether's Theorem, the conserved four-momenta follows from the energy momentum tensor
$\mathcal{T}^{\mu\nu} = \sum_n \frac{\partial \mathcal{L}}{\partial(\partial_\mu\phi_n)}
\partial^\nu\phi_n - g^{\mu\nu}\mathcal{L}$, with $\phi_n$ running over all fields:
\begin{equation}\label{EMT}
    \begin{gathered}
        \mathcal{T}^{\mu\nu} = \bar{\psi} i \gamma^\mu \partial^\nu \psi - \Omega^{\mu \alpha} \partial^\nu \omega_\alpha - F^{\mu \alpha} \partial^\nu A_\alpha \\ -  \vec{R}^{\mu \alpha} \cdot \partial^\nu \vec{b}_\alpha + \partial^\mu \sigma \partial^\nu \sigma- g^{\mu \nu} \mathcal{L},
    \end{gathered}
\end{equation}
where the equations of motion have been used to simplify the expression. The
conserved LF momenta on surfaces of constant $x^+$ are
\begin{equation}\label{noether_charge}
    P^\mu = \frac{1}{2} \int dx^- d\bm{x}^\perp \mathcal{T^{+ \mu}},
\end{equation}
with $\mu=(-,+,\perp)$.

In developing LF RMF one must   select a viable energy density functional. The natural
choice of the light-front Hamiltonian, $P^-$, is fundamentally flawed in this context; given the
light-front dispersion relation $P_A^- = \frac{m_A^2 + (\bm{P}_A^\perp)^2}{P^+_A}$, the system can
trivially minimize its energy by driving the longitudinal momentum $P_A^+ \rightarrow \infty$. The
linear combination $\frac{1}{2}(P^+ + P^-)$  avoids the unphysical $P^+_A
\rightarrow \infty$ minimization limit, and connected to the rest mass condition on the LF $P^+_A =
P^-_A = m_A$ \cite{Miller:1999mi,Blunden:1999gq}. We use  this combination.

At this stage, we   introduce several critical simplifications.  First, because we focus on static, ground-state nuclei, the
meson and photon fields exhibit no dependence on LF time, $x^+$. Second, by restricting our analysis
to spherically symmetric, even-even nuclei, the perpendicular components of the meson and photon
fields vanish due to the time-reversal invariance of the nuclear ground state, and the plus and
minus components of the vector fields are rendered equal. Hence, we remove the Lorentz indicis from
the vector meson and photon fields, labeling them with subscript 0. Finally, the charged $\rho$
meson fields do not contribute; this arises because we operate within the Hartree
approximation, neglecting exchange-correlation effects and because   the nucleus has a definite
total charge.

We quantize the nucleon fields into field operators on the $x^+=0$ hypersurfacr by expanding them
in terms of single-particle orbitals indexed by $j$,
$
    \hat{\psi}(x) = \sum_j \left[ \psi_j(\bm{x}) \,\hat{a}_j + \phi_j(\bm{x})  \, \hat{b}_j^\dagger \right],
$ 
where $\hat{a}_j$ ($\hat{b}_j^\dagger$) is the annihilation (creation) operator for a nucleon
(anti-nucleon) in state $j$, and $\psi_j(\bm{x})$ ($\phi_j(\bm{x})$) is the corresponding
single-particle wavefunction. The creation and annihilation operators obey fermionic anti-commutation relations
$
    \{ \hat{a}_i, \hat{a}_j^{\dagger}\} = \delta_{ij} \;\;\; \mbox{and} \;\;\; \{ \hat{a}_i, \hat{a}_j\} = \{ \hat{a}_i^{\dagger}, \hat{a}_j^{\dagger}\} = 0,
$ 
and the nucleon and anti-nucleons anti-commute. The normalized ground state of the nucleus with baryon number
$A$ is constructed by a product of the $A$-lowest lying nucleon single particle states, i.e. the
Slater determinant
\begin{equation}\label{gs_wf}
    \ket{\Psi_A^{MF}} = \prod_{i=1}^A \hat{a}_i^{\dagger} \ket{\Omega}  
\end{equation}
Note that our nuclear model contains no anti-nucleons and that the nuclear state is built on top of
the non-interacting vacuum $\ket{\Omega}$. When evaluating matrix elements in the nuclear ground state,
one must normal-order with respect to the trivial LF vacuum in order to remove the zero-point
energy. Doing so yields two terms: one corresponding to the positive-energy nucleons and the other
to contributions from the Dirac sea, the latter of which we discard \cite{Reinhard:1989zi}. Building
the nucleus exclusively from the positive-energy solutions together
with this prescription for evaluating matrix elements constitutes the no-sea scheme, which
explicitly omits the dynamical effects of anti-nucleons. This is 
understood to be a foundational modeling choice \cite{doi:10.1142/9872}.

Applying the simplifications above together with the quantization just introduced, the Hamiltonian
density evaluated with respect to the nuclear ground state is
\begin{equation}\label{LF_hamiltonian_density}
    \mathcal{H}_{\rm LF}^{RMF} =\langle \Psi_{\rm A}^{\rm MF}| \frac{1}{2}\mathcal{T}^{+-} + \frac{1}{2}\mathcal{T}^{++} |\Psi_{\rm A}^{\rm MF}\rangle. 
\end{equation}
Full details are presented in Ref.~\cite{Kim:2026fen}.
 The LF energy
density functional is obtained by integrating the Hamiltonian density over the LF spatial volume,
\begin{equation}\label{eq:energy_functional}
    E^{RMF}_{\rm LF}[\psi,\bar{\psi},\sigma,\omega,\rho_3,A] = \frac{1}{2}\int dx^-\, d\bm{x}^\perp\, \mathcal{H}_{\rm LF}^{RMF}(x^-,\bm{x}^\perp),
\end{equation}

As discussed in \cite{Lepage:1980fj,Brodsky:1997de}
only the  $+$ projection of the nucleon field,
$\psi_i^+ \equiv \Lambda^+\psi_i\,(\Lambda^+=1/2 \gamma^0\gamma^+), $ is dynamical on the LF; the $-$ projection is a constraint solved
for in terms of $\psi_i^+$. We therefore impose the normalization on the $+$ components only,
\begin{equation}\label{eq:normalization}
    \sum_{i=1}^A \frac{1}{2}\int dx^-\, d\bm{x}^\perp\, |\psi_i^+(\bm{x})|^2 = A.
\end{equation}
Minimizing $E^{RMF}_{\rm LF}$ with respect to $\psi_i^\dagger$ subject to
Eq.~\eqref{eq:normalization} introduces a set of Lagrange multipliers $p_i^-$ that play the role of
single-particle LF energies. The stationarity condition reads
\begin{equation}\label{eq:variation}
    \frac{\delta \mathcal{H}^{RMF}_{\rm LF}}{\delta \psi_i^\dagger(\bm{x})} - p_i^-\,\Lambda^+ \psi_i(\bm{x}) = 0.
\end{equation}
Performing the functional derivatives on $\mathcal{H}_{\rm LF}^{RMF}$ yields the LF Dirac equation
\begin{equation}\label{eq:sp_dirac_RMF}
    \left[\,i\bm{\alpha}^{\perp} \!\cdot \bm{\partial}^{\perp} + i\alpha^3\partial^+ + \gamma^0m_N + \Sigma_H(\bm{x})\,\right]\psi_i(\bm{x}) = p_i^- \Lambda^+ \psi_i(\bm{x}),
\end{equation}
with the Hartree self-energy
\begin{equation}\label{eq:hartree_selfenergy}
    \Sigma_H(\bm{x}) = g_\sigma\,\gamma^0\sigma(\bm{x}) + g_\omega\,\omega_0(\bm{x}) + \tfrac{1}{2}g_\rho\tau_3\,b_0(\bm{x}) + \tfrac{1}{2}e(1+\tau_3)A_0(\bm{x}),
\end{equation}
where $\alpha^\mu = \gamma^0 \gamma^\mu$. The meson and photon fields obey constrained Klein-Gordon
equations~\cite{Kim:2026fen}
that are solved numerically and
self-consistently. A notable computational difficulty in the LF approach is that the Dirac
single-particle Hamiltonian must be solved in cylindrical coordinates, with states labeled by their
energy and the total angular momentum projection quantum number $m_j$. Ref.~\cite{Blunden:1999gq}
carried out this calculation and verified that rotational invariance is restored by checking that
the $m_j$ states exhibit the correct degeneracy. In the same paper, along with
Ref.~\cite{Smith:2002ci}, the authors further develop a procedure for recasting the LF Dirac
single-particle Hamiltonian into a spherically symmetric form. We now discuss this procedure and show that
this is not an approximation.

\subsection{Connecting IF and LF Formulations of MFT}
We perform two successive steps 
to recast equation~\eqref{eq:sp_dirac_RMF} into a form that makes the underlying spherical symmetry
of the nucleus manifest, 

{Step 1, Replace $x^-$ with the spatial $z$ coordinate.} Since we solve for the states of
the Hamiltonian at $x^+=0$, $x^- = -2z$. In other words, the LF longitudinal direction can be mapped
onto the ordinary spatial coordinate $z$. The LF plus-derivative becomes
\begin{equation}\label{eq:deriv_change}
    \partial^+ \equiv 2\,\frac{\partial}{\partial x^-} = -\frac{\partial}{\partial z} = \partial^3,
\end{equation}
where in the last equality we used $\partial^3 = -\partial/\partial z$ in the mostly-minus metric.
We further relabel the spatial argument $\bm{x}\to\bm{x}_{\rm IF}$.
All Instant Form variables will be denoted with ``IF'' subscript or superscript. Boldface with the
subscript ``IF'' denotes spatial three-vectors, $\bm{p}_{\rm IF} = (p^1, p^2, p^3) = (\bm{p}^\perp,
p^3)$, while boldface symbols without subscript denote light-front three-vectors, $\bm{p} = (p^+,
\bm{p}^\perp)$. The energy component is conventionally listed first in both forms.

 {Step 2, Phase redefinition.} Writing
\begin{equation}\label{eq:phase}
    \psi_i(\bm{x}_{\rm IF}) = \Psi_i(\bm{x}_{\rm IF})\,e^{ip_i^- z/2},
\end{equation}
the action of $i\alpha^3\partial^3$ on $\psi_i$ produces two terms,
\begin{equation}\label{eq:phase_action}
    i\alpha^3\partial^3\psi_i(\bm{x}_{\rm IF}) = i \alpha^3 \left[\, \partial^3\Psi_i(\bm{x}_{\rm IF}) - i\tfrac{p_i^-}{2} \Psi_i(\bm{x}_{\rm IF})\,\right] e^{ip_i^- z/2}.
\end{equation}
Moving the second term in Eq. (\ref{eq:phase_action}) to the right-hand side of
Eq.~\eqref{eq:sp_dirac_RMF} collapses it to $\tfrac{p_i^-}{2}\Psi_i(\bm{x}_{\rm IF})$.

  Putting the two steps together, the set of LF RMF equations
takes the form
\begin{equation}\label{eq:dirac_IF}
    \left[\,i\bm{\alpha}^{\perp}\!\cdot\bm{\partial}^{\perp} + i\alpha^3\partial^3 + \gamma^0 m_N + \Sigma_H(\bm{x}_{\rm IF})\,\right]\Psi_i(\bm{x}_{\rm IF}) = \frac{p_i^-}{2}\,\Psi_i(\bm{x}_{\rm IF}),
\end{equation}
\begin{equation}\label{eq:meson_k_IF}
    \begin{gathered}
        \left[ \bm{\partial}^{\perp} \cdot \bm{\partial}^{\perp}  + \partial^3 \partial^3  - m_\sigma^2 \,\right]\sigma(\bm{x}_{\rm IF}) = g_\sigma \rho_s(\bm{x}_{\rm IF}), \\
        \left[ \bm{\partial}^{\perp} \cdot \bm{\partial}^{\perp} + \partial^3 \partial^3 - m_\omega^2 \, \right]\omega_0(\bm{x}_{\rm IF}) = -g_\omega \rho_B(\bm{x}_{\rm IF}), \\
        \left[ \bm{\partial}^{\perp} \cdot \bm{\partial}^{\perp} + \partial^3 \partial^3 - m_\rho^2 \, \right]b_0(\bm{x}_{\rm IF}) = -g_\rho \rho_3(\bm{x}_{\rm IF}), \\
        \left[ \bm{\partial}^{\perp} \cdot \bm{\partial}^{\perp} + \partial^3 \partial^3 \, \right]A_0(\bm{x}_{\rm IF}) = -e \rho_p(\bm{x}_{\rm IF}). \\
    \end{gathered}
\end{equation}
in which spherical symmetry can now be exploited directly. In fact, Eqs.~\eqref{eq:dirac_IF} and
~\eqref{eq:meson_k_IF} are the exactly same equations that are solved in the original IF MFT
formulation, with $p_i^-/2$ as the single-particle energy of the nucleon \cite{Horowitz:1981xw}.
Hence, IF MFT calculations can be repurposed for LF MFT.

The worry  of Ref.~\cite{Blunden:1999gq} was   that the
Fourier transform of the single-particle LF wavefunction \textemdash obtained from the IF wavefunction \textemdash has
non-zero support at $p^+=0$, which cannot occur for massive particles. On this basis, the authors
argued that the equivalence between IF and LF MFT is only approximate, attributing the discrepancy
to a breaking of the LF spectral condition. However, the non-zero support at $p^+=0$ does not signal
physics that the IF formulation fails to capture; it is instead a consequence of the
independent-particle nature of MFT models. The relation $p^+>0$ for massive particles holds only for
momentum eigenstates, and single-particle orbitals are not momentum eigenstates. The support at
$p^+=0$ is therefore to be expected: it would persist even if the LF--IF relation were never invoked
and the entire calculation were carried out directly on the LF. More precisely, single-particle
momentum distributions of the nucleus acquire support at $p^+=0$ because the nuclear state is not an
eigenstate of LF momentum---MFT calculations of nuclear structure pin the nucleus to the origin.
This interpretation is reinforced by studies of infinite nuclear matter on the light front, where
translational invariance is preserved, the nucleons are momentum eigenstates, and the support at
$p^+=0$ vanishes~\cite{Miller:1999mi, Miller:2000kv}. One can correspondingly expect that properly
projecting the nuclear state onto the zero-momentum eigenstate would remove the support at $p^+=0$.
 Methods for carrying out this projection, and for incorporating
it into the variational calculation, have been outlined in
Refs.~\cite{Peierls:1957er,PEIERLS1962154}; their application    lies
outside the scope of the present efforts.

Furthermore, Eq.~\eqref{eq:phase} admits a natural geometric interpretation. In IF quantization,
states are defined on surfaces MFT the Dirac fields are quantized at $t=0$, hence the wavefunctions
solved for on that surface. LF quantization proceeds analogously, but on surfaces of constant $x^+$,
with the Dirac fields of LF MFT quantized at $x^+=0$. Equation~\eqref{eq:phase} bridges the two. It
takes the IF wave function defined at $t=0$ and applies a $z$-dependent time-evolution that
transports it onto the $x^+=0$ hypersurface, thereby yielding the corresponding LF wavefunction. The relation between LF and IF wavefunctions has been examined previously
in Ref.~\cite{Miller:2009fc}. Focusing on two-body bound states, the authors showed that the IF and
LF wavefunctions can each be recovered from the Bethe--Salpeter wavefunction by integrating over
$k^0$ and $k^-$, respectively, but that neither can be obtained directly from the other. That a
direct map exists in our case is a consequence of the single-particle nature of the problem.

From here, one can obtain the single-particle Dirac equation and the Klein-Gordon equations for the
meson fields that characterize the original RMF formulation in Ref. \cite{Horowitz:1981xw}. However,
despite its successes, the original RMF formulation with constant meson-nucleon couplings failed to
reproduce several key properties of nuclear matter and finite nuclei. Most notably, it predicted a
nuclear-matter incompressibility of roughly 500--600 MeV, far in excess of the empirical value of $K
\approx 210$--$240$ MeV \cite{Garg:2018uam}, and yielded poor agreement with nuclear surface
properties and the equation of state away from saturation. Furthermore, the original results
underbind finite nuclei by roughly 2 MeV per nucleon. These deficiencies were traced to the absence
of medium dependence in the effective interaction. Dirac--Brueckner--Hartree--Fock (DBHF)
calculations, which resum ladder diagrams built from a realistic bare nucleon-nucleon interaction in
the nuclear medium \cite{Brockmann:1990cs}, demonstrated that the scalar and vector nucleon
self-energies acquire a pronounced density dependence through Pauli blocking, short-range
correlations, and implicit three-body effects---features that cannot be captured by fixed couplings.
This microscopic insight motivated the introduction of density-dependent meson-nucleon vertices
\cite{Fuchs:1995as,Typel:1999yq}, which phenomenologically encode the physics missing from a Hartree
description.

\begin{figure}[H]
    \centering
 \includegraphics[width=0.7\linewidth]{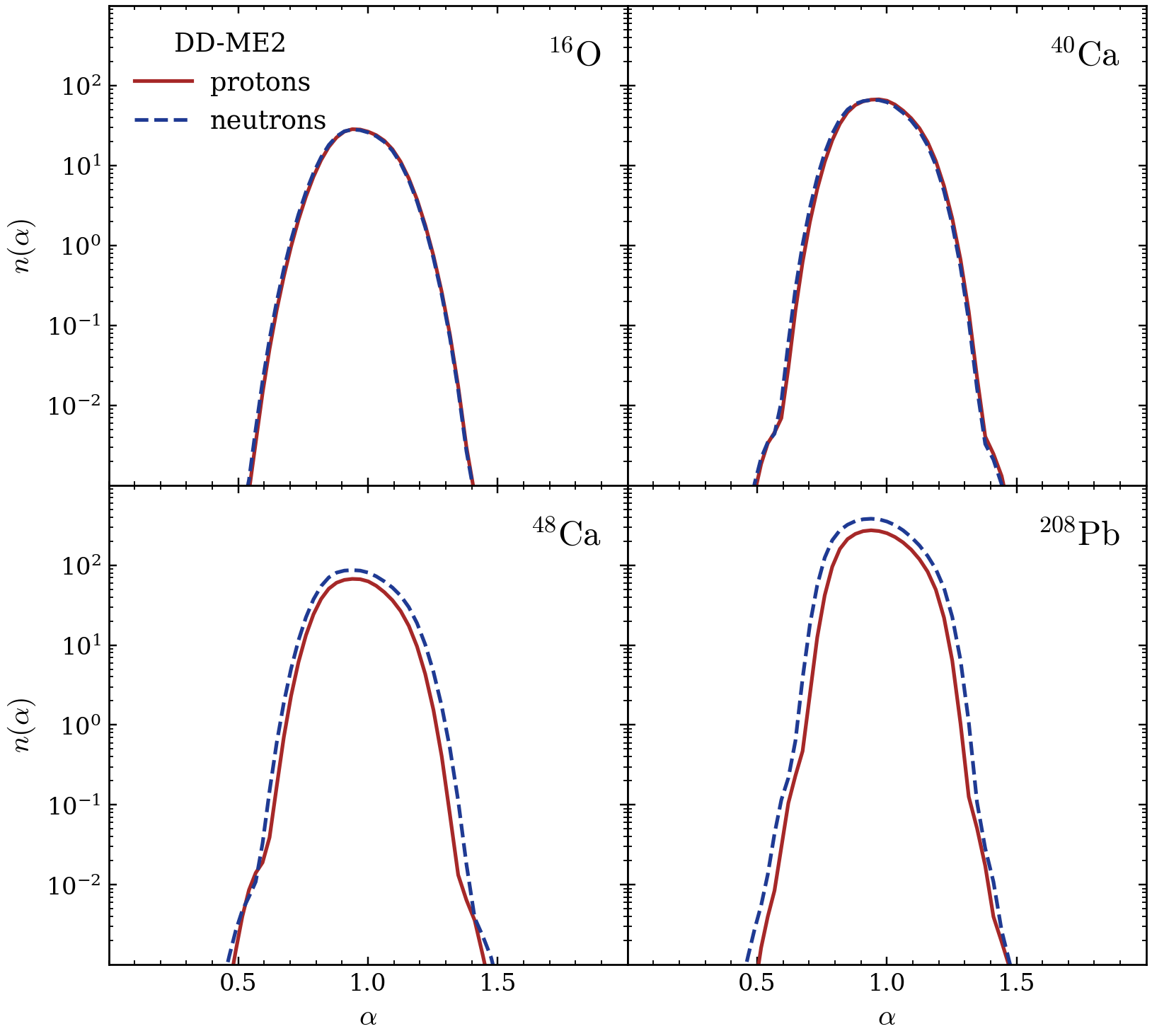}
\caption{(color online) Transverse-momentum-integrated light-front momentum distributions for $^{16}$O, $^{40}$Ca, $^{48}$Ca, and $^{208}$Pb versus $\alpha = A p^+/P_A^+$. $A$ is the mass number and $P^+_A$ is the nuclear plus momentum ($m_A$ in the rest frame). Solid red lines denote protons; dashed blue lines denote neutrons.}    \label{fig:LF_DFT_k_dist}
\end{figure}

Incorporating this density dependence into the meson coupling constants does not affect the
equivalence between the IF and LF formulations.   This equivalence allows existing IF codes to be employed directly. We
use the DIRHB package \cite{Niksic:2014dra}, a set of Fortran codes that compute the ground-state
properties of even-even nuclei within relativistic self-consistent mean-field theory with
density-dependent couplings. For the effective interaction we adopt the DD-ME2 functional of Ref.
\cite{Lalazissis:2005de}, which reproduces total nuclear binding energies for approximately 200
nuclei to within 5 MeV. Figure~\ref{fig:LF_DFT_k_dist} presents the LF momentum distributions,
integrated over longitudinal momenta, for the four doubly magic nuclei considered in this paper. To
interpret these distributions qualitatively, consider the LF momentum distribution of a free proton,
which would be a delta function at $\alpha=1$. The spread in $\alpha$ in the nuclear case is
therefore due to Fermi motion inside the nucleus, while the offset of the peak from $\alpha=1$
reflects the effect of nuclear binding.

\section{Inclusive Electron-Nucleus Scattering}\label{chap:scattering}

Inclusive electron-nucleus scattering $A(e,e')$, the main subject of this work, provides two
independent kinematic  dials: the energy transfer, $\nu$, and the three-momentum transfer, $\vec q$, that can
be tuned to isolate specific physical regimes.  
To develop nuclear models applicable to future high-energy nuclear physics experiments at
the EIC and JLab, one must first understand QE scattering---a process that probes the nucleonic
structure of the nucleus---and then build on that baseline to understand how nucleonic
configurations in the nucleus interplay with their partonic dynamics. The experimental data
considered in this section  are from JLab E08-018, an inclusive electron-nucleus scattering
measurement with an electron beam energy of $3.356$~GeV and a scattered-electron angle of $25^\circ$
\cite{Zhang:2025nst}. Restricting the kinematics  to $x_B = Q^2/(2 m_N \nu) > 1$ yields $Q^2 > 1$~GeV$^2$,
places the calculations  in a kinematic window with momentum transfer high enough to resolve short-distance
nuclear structure, but energy transfer low enough that we need not consider the transition to quark
degrees of freedom in DIS.

\subsection{Collinear Frames}
\begin{figure}
    \centering
  \includegraphics[width=0.65\linewidth]{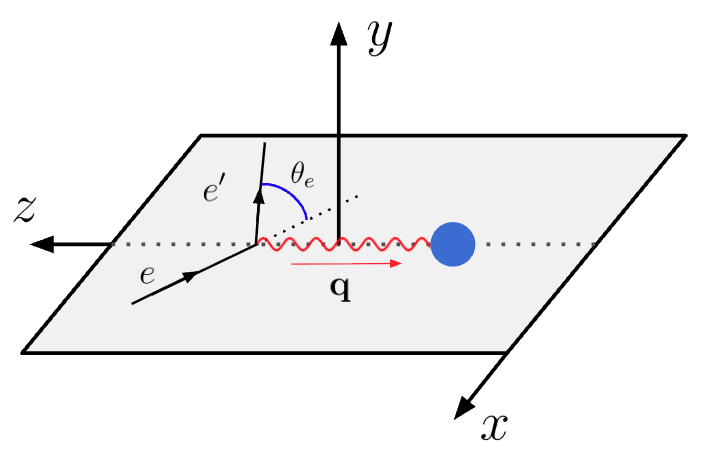}
    \caption{(color online) Nuclear rest frame, with the virtual photon momentum, $\bm{q}_{\rm IF}$, 
             defining the $z$-axis and the $x-z$ plane defined by the electron and scattered electron 
             momentum vectors. The nucleus is represented by the filled blue circle.}
    \label{fig:collinear_frame}
\end{figure}
Throughout this paper, all calculations are performed using the nucleus
rest frame, with the virtual photon's IF three-momentum defining the negative
$z$-axis (see Fig.~\ref{fig:collinear_frame}). The nucleus rest frame belongs to a broader
class of frames---known as \emph{collinear frames}---that are related to one another by boosts along
the $z$-axis. In any such frame, the LF momentum components of the nucleus, $P_A^\mu$,  and the virtual photon,$q^\mu$,
take the form $P_A^\mu=(m_A^2/P_A^+,P_A^+,{\bf 0}_\perp),\,q^\mu= (Q^2/(\xi_A\,P_A^+),-\xi_AP_A^+,{\bf 0})$,
with $x_A=Q^2/(2P_A\cdot q),$
and where the parameter $\xi_A$ is fixed by the kinematic invariant
\begin{equation}
    2P_A \cdot q = P_A^+ q^- + P_A^- q^+ = \frac{Q^2}{x_A}.
\end{equation}
Solving this constraint for $\xi_A$ in terms of $x_A$ yields
\begin{equation}
    \xi_A = \frac{2x_A}{1 + \sqrt{1+4m_A^2 x_A^2/Q^2}},
\end{equation}
 known as the Nachtmann variable \cite{Nachtmann:1973mr}. 
In the scaling limit $Q^2 \gg m_A^2$ this
reduces to
   $\xi_A = x_A + \mathcal{O}(m_A^2/Q^2).
$ 
In the nucleus rest frame, $P_A^+ = P_A^- = m_A$, and $x_A$ is related to
the Bjorken variable $x_B = Q^2/(2 m_N \nu)$ by
\begin{equation}
    x_B = \frac{m_A}{m_N}\, x_A.
\end{equation}

\subsection{Unpolarized Inclusive Electron-Nucleus Cross-Section}

We start with the general parameterization of the cross-section for unpolarized inclusive
electron-nucleus scattering where the final states can be anything, $X$. Because the electromagnetic
interaction between the electron and nucleus is governed by the fine structure constant $\alpha \approx 1/137$, one  models the electron-nucleus interaction through the
exchange of a single virtual photon:
\begin{equation}\label{cs_parameterization}
    \frac{d\sigma}{dE'd\Omega'} = \frac{E'}{E} \frac{\alpha_e^2}{Q^4} L_{\mu \nu} W^{\mu \nu}_A,
\end{equation}

\noindent where $q^\mu$ is the virtual photon's four momentum and $Q^2 = -q^{\mu}q_{\mu}$, $E$ and
$E'$ are the energies of the incident and scattered electron respectively, and $\alpha_e$ is the
fine structure constant. Ignoring the mass of the electron, the leptonic tensor, $L_{\mu \nu}$,
is given by
$
    L_{\mu \nu} =    2 (k'_{\mu} k_{\nu} + k'_{\nu} k_{\mu} - k' \cdot k \, g_{\mu \nu}),
$
where $k^{\mu}$ and $k'^{\mu}$ are the four momenta of the incident and scattered
electrons respectively, $\gamma^{\mu}$ are the Dirac gamma matrices, and $g^{\mu \nu}$ is the
Minkowski metric tensor. The unpolarized nuclear tensor, $W_A^{\mu \nu}$, is given by
\begin{equation}\label{nuclear_tensor}
    \begin{split}
    W_A^{\mu \nu} = \frac{1}{4 \pi m_A} \frac{1}{2J + 1}\sum_{M_J} \sum_{X} \int d\Pi_{X} \bra{\Psi_A, \bm{P}_A} \hat{J}^{\mu}_A (0) \ket{X} \bra{X} \hat{J}_A^{\nu}(0) \ket{\Psi_A, \bm{P}_A} \\ \times \, (2 \pi)^4 \delta^4(q^\mu + P_A^\mu - P_X^\mu),
\end{split}
\end{equation}
where $\Psi_A$ denotes the nuclear state which implicitly contains the mass $m_A$, 
total angular momentum $J$, and total angular momentum projection $M_J$ labels. $d\Pi_X$ is the Lorentz 
invariant phase space measure for the final states and $P_A^{\mu}$ and $P_X^\mu$ are the four momenta 
of the nucleus and final states.

It is worthwhile to re-write Eq. (\ref{cs_parameterization}) in terms of response functions that
describe the scattering of the target by a virtual photon with a given polarization. This is done
by including explicit factors of the Minkowski metric in the contraction between the leptonic and
nuclear tensors, $L_{\mu \nu} W_A^{\mu \nu} = L^{\rho \sigma} g_{\rho \mu} g_{\sigma \nu} W_A^{\mu
\nu}$, and $\sum_{\lambda} \epsilon_{\mu}^\lambda(q) \epsilon_{\nu}^{*\lambda}(q) = -g_{\mu
\nu} + \frac{q_\mu q_\nu}{q^2}$. Using  the symmetric nature of the unpolarized nuclear   and 
the Ward identity, $q_\mu W_A^{\mu \nu} = 0$, we obtain the following decomposition of Eq.
(\ref{cs_parameterization}):

\begin{equation}\label{cs_response_func}
     \frac{d\sigma}{dE'd\Omega'} = \left( \frac{d\sigma}{d\Omega'} \right)_{Mott} \left\{  \frac{Q^4}{|\bm{q}_{\rm IF}|^4} W_A^L(x_B, Q^2) + \left( \frac{1}{2}\frac{Q^2}{|\bm{q}_{\rm IF}|^2} + \mbox{tan}^2\frac{\theta_e}{2}\right) W_A^T(x_B,Q^2)   \right\},
\end{equation}
with 
$
     \left( \frac{d\sigma}{d\Omega'} \right)_{Mott} = \frac{\alpha_e \mbox{cos}^2\frac{\theta_e}{2}}{4 E^2 \mbox{sin}^4 \frac{\theta_e}{2}},
$
$
    W_A^T = W_A^{11} + W_A^{22},
$
$
    W_A^L = \frac{|\bm{q}_{\rm IF}|^2}{4Q^2} \left( \frac{Q^2}{(q^{+})^2}W_A^{++} + 2 W_A^{+-}+ \frac{(q^{+})^2}{Q^2}W_A^{--} \right),
$ 
where $W_A^T$ and $W_A^L$ are the transverse and longitudinal nuclear response functions, and $\theta_e$ is the angle between the incident and scattered electron. With this, the 
 calculation  is reduced  to calculating $W_A^T$ and $W_A^L$.

\subsection{Electromagnetic Current}
The following discussion closely follows Refs.~\cite{Vera:2018orr,Vera:2021rnw}
and is presented  for purposes of completeness.
The present effort focuses on the QE regime, in which the
momentum transfer is large enough to resolve individual nucleons inside the nucleus. In this
kinematic domain the electromagnetic probe couples primarily to a single nucleon, which justifies
the replacement $\hat{J}_A^\mu(0) \to \hat{J}_N^\mu(0)$ in Eq.~(\ref{nuclear_tensor}). The operator
$\hat{J}_N^\mu(0)$ is not, however, the free-nucleon electromagnetic (EM) current, because the
nucleons inside the target are bound and therefore off-shell. Modeling this \emph{off-shell} EM
current operator is the central theoretical challenge of this section.

Historically, the off-shell problem rose to prominence in the early 1980s
with the advent of intermediate-energy $A(e,e'N)$ experiments at
SACLAY~\cite{Alde:1990im,Turck-Chieze:1984ido} and
NIKHEF~\cite{VanDenBrand:1988pv}. The most widely used prescriptions to this
day are those of de Forest~\cite{DeForest:1983ahx}, which extrapolate the
on-shell EM current to off-shell kinematics while retaining on-shell spinors
for the struck nucleon. This construction is 
not compatible with the covariant Feynman-diagram evaluations in which it is
typically embedded, because  a bound nucleon in such a calculation is described
by a propagator rather than by on-shell spinors.

\begin{figure}
    \centering
\includegraphics[width=0.4\linewidth]{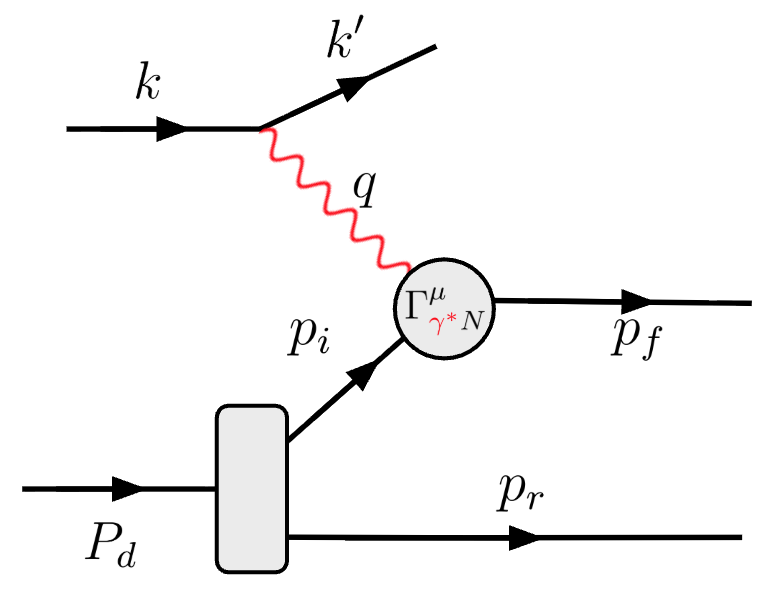}
    \caption{Covariant Feynman diagram describing electron--nucleon scattering inside a deuterium target.}
    \label{fig:e_d_scattering}
\end{figure}

To understand the problem, consider the covariant Feynman diagram
for electron--deuteron scattering shown in Fig.~\ref{fig:e_d_scattering}.
Evaluating the diagram with the standard Feynman rules gives the matrix element that connects the nucleons denoted by $p_i$ and $p_f$ as
\begin{equation}
\begin{aligned}
    \bar{u}_{\rm IF}(\bm{p}_f,\sigma_f,\tau_f)\, \Gamma_{\textcolor{red}{\gamma^*}N}^\mu\, \frac{\slashed{p}_i + m_N}{p_i^2 - m_N^2 + i\epsilon}\, \bar{u}_{\rm IF}(\bm{p}_r,\sigma_r,\tau_r)\, \Gamma^\nu_{dNN}\, \chi_\nu(M_J),
\end{aligned}
\end{equation}
where $\Gamma_{\textcolor{red}{\gamma^*}N}^\mu$ is the effective electromagnetic vertex between the
virtual photon and bound nucleon, $\Gamma_{{dN}N}^\mu$ represents the vertex of the $d \rightarrow
NN$ transition, and $\chi_\nu(M_J)$ is the polarization vector of the deuteron. Note that  the EM vertex
$\Gamma^\mu_{\textcolor{red}{\gamma^*}N}$ is not
sandwiched between two on-shell spinors: the intermediate nucleon enters through a propagator rather
than as an external on-shell state.  

The advent of high-energy eA experiments, in which electrons transfer large momenta to bound
nucleons and yield final-state nucleons with momenta above a few GeV
\cite{Fomin:2017ydn,Boeglin:2015cha,Arrington:2011xs},  motivated renewed interest in formulating
off-shell EM currents directly within the LF framework. References~\cite{Vera:2018orr,Vera:2021rnw}
carry out   this
construction. Using effective LF diagrammatic rules to sum all time-orderings of
Fig.~\ref{fig:e_d_scattering}---a step required to correctly account for the LF off-shellness of the
intermediate nucleon---they obtain an off-shell EM current operator of the form
\begin{equation}
    \hat{J}_{N}(0) = \sum_{\sigma \sigma' \tau} \int [dp][dp']\, \sqrt{2p'^+}\sqrt{2p^+}\, J_N(\bm{p}',\sigma',\tau;\,\bm{p},\sigma,\tau)\, \hat{a}^\dagger(\bm{p}',\sigma',\tau)\, \hat{a}(\bm{p},\sigma,\tau),
\end{equation}
with single-particle matrix element
\begin{equation}
    J_N(\bm{p}',\sigma',\tau;\,\bm{p},\sigma,\tau) = \bar{u}(\bm{p}',\sigma',\tau) \Gamma^\mu_{\textcolor{red}{\gamma^*}N}\!\left(1 + \frac{\slashed{\Delta}_{p}}{2m_N}\right)\!u(\bm{p},\sigma,\tau),
\end{equation}
where $u(\bm{p},\sigma,\tau)$ is a LF spinor and the EM vertex is parametrized as
\begin{equation}
    \Gamma_{\textcolor{red}{\gamma^*}N}^\mu = \gamma^\mu F_1 + i\sigma^{\mu\nu}q_\nu\, \frac{\kappa F_2}{2m_N} . \end{equation}
Here $\sigma^{\mu\nu} \equiv \frac{i}{2}[\gamma^\mu,\gamma^\nu]$ is the Pauli tensor, $\kappa$
is the anomalous magnetic moment of the nucleon, $F_1$, $F_2$  are the free elastic form
factors, taken to be functions of $Q^2$ alone  and $q^\mu \neq p'^\mu-p^\mu$. 
The off-shell information is carried by the operator $\slashed{\Delta}_p$ appearing
alongside $\Gamma^\mu_{\textcolor{red}{\gamma^*}N}$:
\begin{equation}\label{EM_offshell_factors}
    \slashed{\Delta}_p = \frac{\gamma^+}{2}\bigl(p^-_{\rm c} - p^- \bigr),
    \qquad
    p^- = \frac{m_N^2 + (\bm{p}^\perp)^2}{p^+},
\end{equation}
where $p^-$ is the on-shell LF energy and $p^-_{\rm c}$ (subscript ${\rm c}$ for
``conserved'') is the LF energy that the nucleon would carry if energy were conserved at each vertex. The difference $p^-_c - p^-$ is the LF
virtuality  of the bound nucleon.

Refs. \cite{Vera:2018orr,Vera:2021rnw} compared their LF predictions against the widely used de
Forest prescriptions. In essentially all kinematic cases examined, the LF approach predicts smaller
off-shell effects than de Forest at $Q^2 >1~\mathrm{GeV}^2$. More importantly, the LF approach
predicts a pronounced suppression of these effects with increasing $Q^2$, which is understood
intuitively as a decreasing sensitivity of the hard scattering process to the virtuality of the
target nucleon.

\subsection{Inclusive Quasi-Elastic Electron--Deuteron Scattering}
We  start  by 
computing the inclusive electron--deuteron scattering cross section using the plane-wave impulse
approximation (PWIA).  This process must be understood first as the deuteron cross section is the denominator appearing the ratios that display the plateau. The deuteron nuclear tensor reads
\begin{equation}\label{deuteron_tensor}
\begin{aligned}
    W_d^{\mu\nu} = \frac{1}{4\pi m_d}\cdot\frac{1}{3}\sum_{M_J}\,\frac{1}{2}\sum_{\substack{\sigma_1 \sigma_2 \\ \tau_1 \tau_2}} \int [dp_1][dp_2]\; &\bra{\Psi_d}\hat{J}^{\mu}_N(0)\ket{p_1,p_2}\bra{p_1,p_2}\hat{J}^{\nu}_N(0)\ket{\Psi_d} \\
    &\times (2\pi)^4\,\delta^4(q^\mu + P_d^\mu - p_1^\mu - p_2^\mu),
\end{aligned}
\end{equation}
where the $\tfrac{1}{3}\sum_{M_J}$ averages over the three deuteron polarizations, the factor of
$\tfrac{1}{2}$ accounts for the indistinguishability of the two-nucleon final state, and we drop
the momentum label for nuclear states at rest. 

The current matrix element receives two contributions, corresponding to the
virtual photon striking nucleon~1 or nucleon~2 of the deuteron:
\begin{equation}
\begin{aligned}
    \braket{p_1,p_2|\hat{J}_N^\mu(0)|\Psi_d} = \sum_{\sigma_i}\biggl[\,
    &\frac{m_d}{p_1^+ - q^+}\, J_N\!\bigl(\bm{p}_1,\sigma_1,\tau_1;\,\bm{p}_1-\bm{q},\sigma_i,\tau_1\bigr)\\[-2pt]
    &\quad\times \psi_d\!\left(\tfrac{2(p_1^+ - q^+)}{P_d^+},\bm{p}_1^\perp,\sigma_i,\sigma_2,\tau_1,\tau_2\,\bigg|\,M_J\right) \\
    -\;&\frac{m_d}{p_2^+ - q^+}\, J_N\!\bigl(\bm{p}_2,\sigma_2,\tau_2;\,\bm{p}_2-\bm{q},\sigma_i,\tau_2\bigr)\\[-2pt]
    &\quad\times \psi_d\!\left(\tfrac{2(p_2^+ - q^+)}{P_d^+},\bm{p}_2^\perp,\sigma_i,\sigma_1,\tau_2,\tau_1\,\bigg|\,M_J\right)
    \biggr].
\end{aligned}
\end{equation}
Squaring this matrix element produces interference terms between the two contributions, which are
routinely neglected in calculations of this type, { but are kept here}. These terms are negiigible.

\begin{figure}
    \centering
\includegraphics[width=1\linewidth]{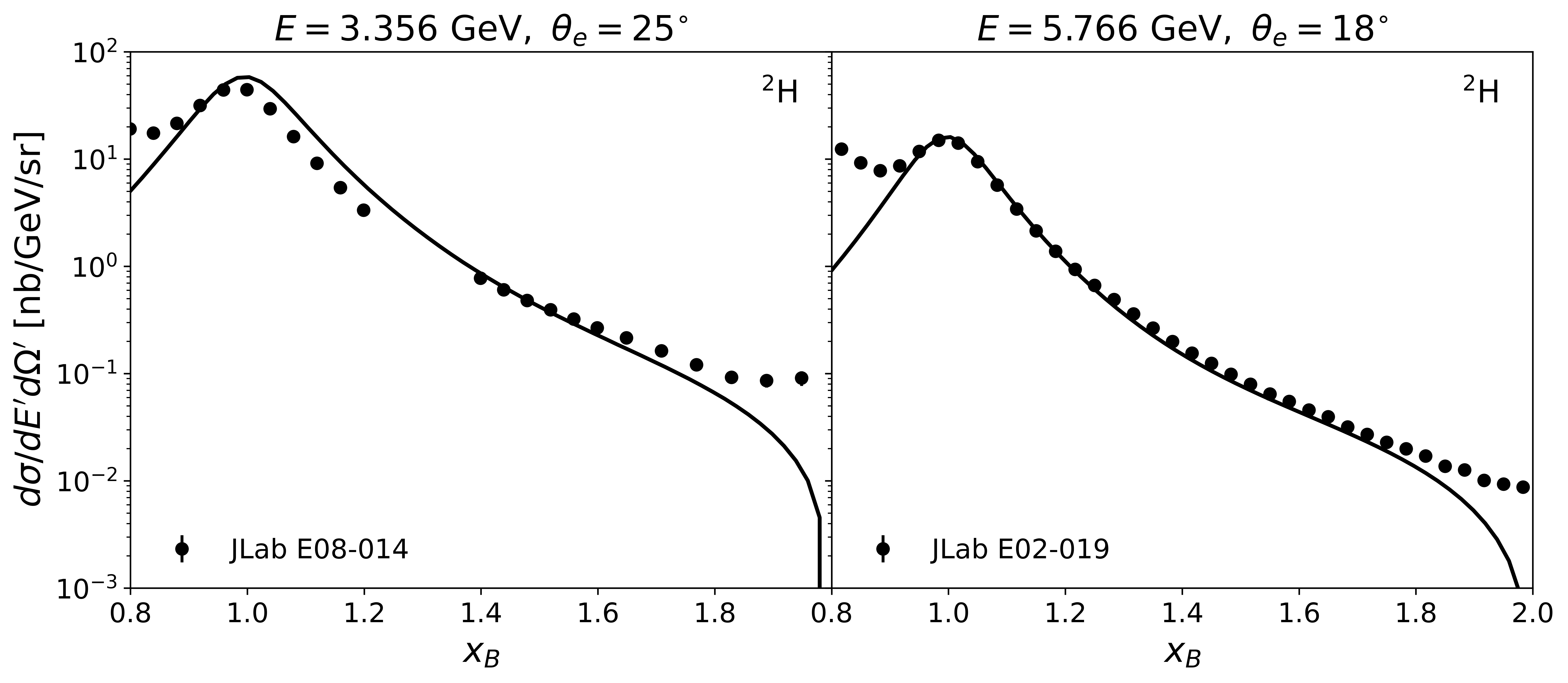}
    \caption{Inclusive $e-D$ cross-section
     as a function of the Bjorken variable $x_B$}
        \label{fig:H2_cs_pw}
\end{figure}

Figure~\ref{fig:H2_cs_pw} presents the inclusive electron–deuteron cross section as a function of
the Bjorken variable $x_B$, compared to JLab data at two kinematic settings: $E = 3.356$ GeV,
$\theta_e = 25^\circ$ from JLab E08-014~\cite{Zhang:2025nst} (left panel) and $E = 5.766$ GeV,
$\theta_e = 18^\circ$ from JLab E02-019~\cite{Fomin:2011ng} (right panel). The approximate
LF deuteron wave function obtained from the procedure outlined in
section~\ref{chap:lightfront} reproduces the E02-019 data well, owing to its larger $Q^2$, which
provides higher spatial resolution of the high-momentum tail of the deuteron than the more modest
$Q^2$ of E08-014. The residual discrepancy near the QE peak in the E08-014 panel reflects
physics absent both from the plane-wave impulse approximation and from the AV18 interaction itself —
principally two-body meson-exchange currents and the low-energy tail of the $\Delta(1232)$
excitation, neither of which is generated by a one-body current acting on a purely nucleonic NN
potential without explicit meson-exchange or $\Delta$ degrees of freedom. Both contributions fall in
importance at the higher $Q^2$ of E02-019. Overall, both datasets are well described using the AV18
NN interaction, supporting the use of the minimal-relativity AV18 potential as an approximate
LF NN potential,  one capable of capturing the high-momentum two-nucleon dynamics necessary
for calculations developed later in this dissertation.

\subsection{Inclusive Electron-Nucleus Cross-Sections}
\begin{figure}
    \centering
  \includegraphics[width=0.8\linewidth]{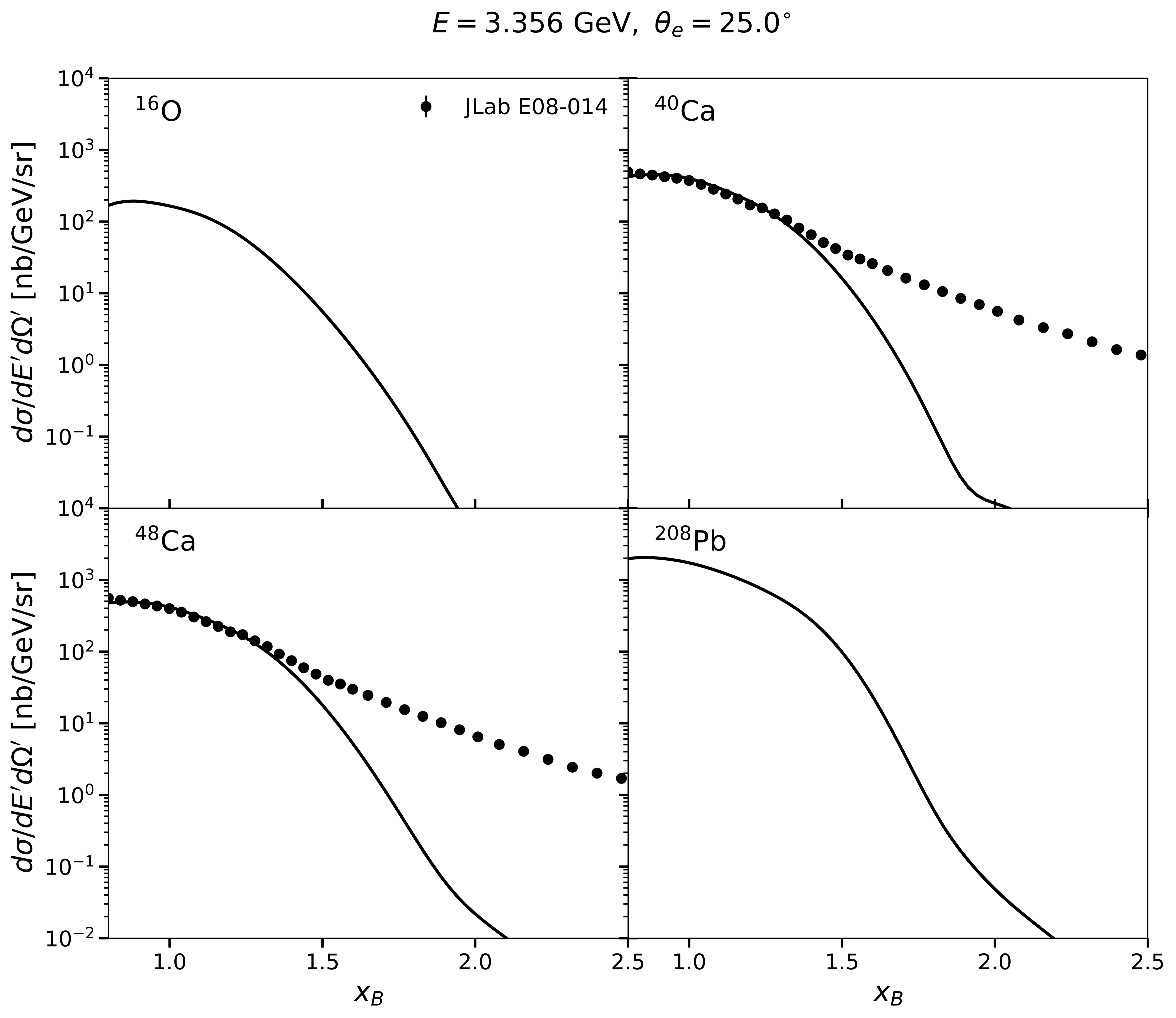}
    \caption{Inclusive electron-nucleus cross section at $E = 3.356$~GeV and $\theta_e = 25^\circ$ as a function of the 
    Bjorken variable $x_B$, for $^{16}\mathrm{O}$ (top left), $^{40}\mathrm{Ca}$ (top right), $^{48}\mathrm{Ca}$ (bottom left),
    and $^{208}\mathrm{Pb}$ (bottom right). The solid black curves are the mean-field contribution,
    evaluated with relativistic mean-field single-particle wave functions. Black points are the JLab
    E08-014 data, available for $^{40}\mathrm{Ca}$ and $^{48}\mathrm{Ca}$ at these kinematics.}
    \label{fig:MF_cs}
\end{figure}

The appendix  provides a detailed derivation of the inclusive electron-nucleus cross section 
calculated within the FF, utilizing LF MFT states to evaluate the intrinsic nuclear matrix elements. The resulting cross 
sections are displayed in Fig.~\ref{fig:MF_cs}. We find that the LF MFT framework successfully reproduces the data in the 
region $0.8 < x_B < 1.3$, capturing the quasi-elastic peak. However, a significant discrepancy 
emerges at $x_B > 1.3$, where the theoretical calculations systematically underestimate the cross section. This shortfall 
is directly attributable to the absence of short-distance nuclear structure---specifically, the high-momentum short-range 
correlations that are inherently missing from independent-particle models.

\section{Similarity Renormalization Group}\label{chap:srg}

Renormalization Group (RG) techniques originated during the development of Quantum Electrodynamics 
in the 1940s \cite{Bethe:1947id, Dyson:1949bp,Gell-Mann:1954yli}.  Wilson generalized this framework by investigating the scale dependence of Hamiltonians 
\cite{PhysRevB.4.3174, RevModPhys.47.773} and  made mathematically precise
the idea that scale dictates our perception of the physical world.

In the 1990s, Glazek and Wilson sought to understand the phenomenological  success of constituent quark models, by studying the RG evolution of the QCD Hamiltonian under light-front
quantization \cite{Glazek:1993rc, Glazek:1994qc}.  Glazek and Wilson, and independently Wegner for condensed matter
applications \cite{Wegner:1994fdg}, developed continuous unitary RG techniques, known as 
the Similarity Renormalization Group (SRG). Instead of discarding high-energy modes, these RG flow equations systematically drive the bare
ultraviolet Hamiltonian toward an energy-diagonal form. Because these transformations are
strictly unitary, the original physical spectrum is exactly preserved.  
In the nuclear domain, induced many-body operators are found to be suppressed; as the rank
of the interaction ($N$) increases, the corresponding $N$-body forces become increasingly
negligible. Because this allows the interaction space to be safely truncated, SRG techniques became widely used in nuclear physics.

Recently, and of particular importance to this work, the SRG has been utilized to reconcile
short-range correlation (SRC) phenomenology with independent-particle nuclear models.  
 The methods and applications of the SRG have been extensively documented. For an
excellent pedagogical introduction, complete with explicit numerical implementations, see Ref.
\cite{Hergert:2016iju}; for broader foundational reading, see Refs. \cite{Binder:2013xaa,
Roth:2013fqa, Furnstahl:2013oba, Bogner:2009bt}.  This section focuses on describing the minimal set of theoretical fundamentals 
required to support our calculations. 

\subsection{Theory Fundamentals}

The basic idea of the SRG method is   to
construct a continuous unitary transformation, $\hat{U}(s)$ that  simplifies the Hamiltonian by
making it more band-diagonal when represented as a matrix \cite{Wegner:1994fdg,Hergert:2016iju}.

\begin{equation}\label{srg_hamiltonian}
    \hat{H}(s) = \hat{U}(s) \,
    \hat{H}(s = 0) \, \hat{U}^{\dagger}(s).
\end{equation}
\noindent By convention $\hat{H}(s = 0)$ is the starting, bare Hamiltonian and $s$ is called the
flow parameter, which parameterizes $\hat{U}$. To elucidate,  if we split up the
Hamiltonian into diagonal and off-diagonal parts,
\begin{equation}\label{hamiltonian_split}
    \hat{H}(s) = \hat{H}_{d}(s) + \hat{H}_{od}(s),
\end{equation}
  we want to construct $\hat{U}(s)$ such that,
\begin{equation}\label{SRG_hamiltonian_asymptotics}
\hat{H}(s) \underset{s \rightarrow \infty}{\longrightarrow} \hat{H}_{d}(s), \,\, \hat{H}_{od}(s) \underset{s \rightarrow \infty}{\longrightarrow} 0.
\end{equation}

The necessary procedure is implemented by 
defining the anti-Hermitian operator
\begin{equation}\label{SRG_generator}
    \hat{\eta}(s) \equiv \frac{d\hat{U}(s)}{ds} \hat{U}^{\dagger}(s) = - \hat{\eta}^{\dagger}(s).
\end{equation}
Then 
 the SRG-flow equation is:
\begin{equation}\label{SRG_hamiltonian_flow_equation}
    \frac{d\hat{H}(s)}{ds} = [\hat{\eta}(s),\hat{H}(s)].
\end{equation}
The operator  $\hat{\eta}(s)$ is the dynamical generator of the SRG-transformation. Eq.
(\ref{SRG_hamiltonian_flow_equation}) is a first-order linear differential equation. Hence, by calculating $\hat{H}(s)$ one can build $\hat{U}(s)$ directly by
using the eigenvectors of the evolved and initial Hamiltonians. There is no set procedure to
determine $\hat{\eta}(s)$, which generates an $s$-dependent Hamiltonian that obeys Eq.
(\ref{SRG_hamiltonian_asymptotics}).
For momentum-space SRG, Wegner showed that the off-diagonal part of the Hamiltonian flows to zero,
as $s \rightarrow \infty$, if   the generator  is chosen to be
\begin{equation}\label{momentum_space_SRG_generator}
\hat{\eta}(s) = [\hat{T}(s=0), \hat{H}(s)],
\end{equation}
where $\hat{T}(s=0)$ is the bare kinetic energy operator \cite{Wegner:1994fdg}. It is
convenient to perform a change of variables, defining a new flow parameter $\lambda = s^{-1/4}$ that  has units of [$\mbox{fm}^{-1}$]. The flow parameter $\lambda$ provides a measure of the spread
of the band-diagonal in the SRG-evolved Hamiltonian \cite{Hergert:2016iju}. Under this change of
variables, the SRG flow equation is written as

\begin{equation}\label{SRG_hamiltonian_flow_equation_lambda}
\frac{d \hat{H}(\lambda)}{d\lambda} = -\frac{4}{\lambda^5} \, [\hat{\eta}(\lambda),\hat{H}(\lambda)].
\end{equation}

\subsection{SRG Transformation of Operators: Connecting Low- and High- Resolution Pictures}

Now we discuss how the SRG can be used to bridge low- and high-resolution pictures of the nucleus.
This can be immediately seen by examining  the matrix element for the momentum distribution of a
single nucleon inside a nucleus,
\begin{equation}\label{lf_momentum_distribution}
        n_A^{\tau}(\bm{p})   
        = \sum_{\sigma} \bra{\Psi_A} \hat{n}^{\tau}(\bm{p},\sigma) \ket{\Psi_A},
\end{equation}
with $\hat{n}^{\tau}(\bm{p},\sigma)=a^{\dagger}(\bm{p},\sigma,\tau) \, a(\bm{p},\sigma,\tau),$ 
 where $\sigma$ is the spin projection, $\tau$ is the isospin projection, and
$\ket{\Psi_A}$ is the nuclear many-body state. Applying $\hat{U}^{\dagger}(\lambda) \hat{U}(\lambda)
= \mathbb{I}$ before and after the operator we find,

\begin{equation}\label{srg_momentum_distribution}
        n_A^{\tau}(\bm{p})  
         = \sum_{\sigma} \bra{\Psi_A(\lambda)} \hat{n}^{\tau}_{\lambda}(\bm{p},\sigma) \ket{\Psi_A(\lambda)},
\end{equation}
with $\hat{n}^{\tau}_{\lambda}(\bm{p},\sigma) =U^\dagger(\lambda)\hat n^\tau({\bf p},\sigma)U(\lambda)$ and 
  $\hat{U}(\lambda)$ is the unitary SRG operator at flow parameter $\lambda$,
$\ket{\Psi_A(\lambda)}$ is the SRG-evolved nuclear ground state, and
$\hat{n}^{\tau}_{\lambda}(\bm{p},\sigma)$ is the SRG-evolved momentum distribution operator for
nucleon $\tau$. The SRG transformation
decouples low- and high-momentum states. Hence, as we SRG-evolve the Hamiltonian, it's
wave functions begin to reflect a dominantly mean-field, or
"low-resolution" description of nuclei. However, since SRG transformations are unitary, the
high-resolution physics is shifted from the wave function to the operators. Hence, within the SRG
framework, low- and high-resolution pictures of nuclei are equivalent. Both can output the same
results as long as consistently uses  the SRG-scale of the reaction (operators) and structure
(states).

A crucial theoretical consideration is determining, a priori, whether the chosen probing operator is
truly the bare operator. Within the context of nuclear physics, the SRG transformation is driven
entirely by the strong nuclear interactions governing the initial state of the target. Therefore, if
the external probe—such as an electromagnetic or weak current—does not inherently contain nuclear
force dynamics, it constitutes the bare, high-resolution operator. For theoretical consistency, this
unevolved operator must strictly be evaluated using exact high-resolution many-body eigenstates.
However, as demonstrated by Eq. (\ref{srg_momentum_distribution}), when we apply the SRG
transformation to evolve the high-resolution nuclear state to a low-resolution scale, the unitary
transformation is formally shifted onto the operator. Consequently, the originally bare operator
absorbs the short-range nuclear dynamics, naturally inducing two-body and higher-rank many-body
currents which originate from nuclear forces. This has important implications for high-energy
approximations in nuclear scattering, as we will shortly see.   

\subsection{Application to Light-Front Physics}

The formal connection between the low- and high-resolution pictures of nuclei provided by the SRG
led Tropiano et al. to approximate high-resolution observables using shell-model wavefunctions
\cite{Tropiano:2024bmu}. Specifically, they substituted the exact evolved state
$\ket{\Psi_A(\lambda)}$ in Eq. (\ref{srg_momentum_distribution}) with an uncorrelated mean-field
nuclear wavefunction, $\ket{\Psi_A^{MF}}$. In principle, $\ket{\Psi_A(\lambda)}$ should be obtained
by solving the many-body Schrödinger equation associated with the evolved Hamiltonian; however, as
discussed previously, this low-resolution state reflects a dominantly independent-particle
description of the nucleus. Introducing this mean-field approximation breaks the formal SRG
invariance of the matrix element, inherently introducing a dependence on the resolution scale
$\lambda$. This scale dependence is the theoretical tradeoff for extracting correlated observables
from uncorrelated wavefunctions. Consequently, $\lambda$ becomes a parameter that must be
constrained. By focusing on contributions from the two-body nucleon-nucleon potential and neglecting
higher-rank SRG-induced interactions, Tropiano et al. demonstrated that setting $\lambda = 1.5
\text{ fm}^{-1}$ yielded high-resolution momentum distributions that showed excellent agreement with
exact high-resolution Quantum Monte Carlo calculations.

This operator-evolution approach to extracting high-resolution observables significantly simplifies
computational efforts by circumventing the need to explicitly solve the correlated nuclear many-body
problem. Within the domain of light-front nuclear physics, this methodology proves particularly
advantageous. Given the historical scarcity of full many-body calculations performed under
light-front quantization, relying on evolved operators paired with independent-particle
wave functions provides a highly practical and effective path forward.

This is    accomplished this by mirroring 
the procedure of Ref. \cite{Tropiano:2024bmu}. The  unitary transformation is constructed from 
the light-front, two-body nucleon-nucleon interaction of \ref{chap:lightfront}. The light-front mean-field  theory model developed in section
\ref{chap:dft} is used input for a low-resolution wave function.

The two-body unitary light-front SRG transformation is expressed  as the inner product between
eigenstates of the un-evolved and SRG-evolved Hamiltonians,
\begin{equation}\label{two_body_LF_SRG}
    \hat{U}(\lambda) = \sum_{\beta} \int [dP] \, \ket{\beta \, , \lambda\, ;\bm{P}} \bra{\beta \, ;\bm{P}},
\end{equation}
where $\beta$ indexes the eigenvalues of the two-body light-front Hamiltonian,
$\bm{P}=(P^+, \bm{P}^\perp)$ are the plus and transverse momenta of the center of mass, and
$\lambda$ is the SRG-scale. To be explicit,

\begin{equation}\label{beta_sum}
    \begin{gathered}
        \sum_{\beta} = \sum_{\text{energy}} \sum_{J M_J} \sum_{T M_T} \sum_{S} \, \, , \\
        \sum_{\text{energy}} \xrightarrow{\text{bound states}} \sum_{M^2} \, \, ,\\
        \sum_{\text{energy}} \xrightarrow{\text{scattering states}} \frac{1}{2} \sum_{\substack{\sigma_1\tau_1 \\ \sigma_2 \tau_2}} \int  \frac{d\alpha d\bm{k}^\perp}{(2 \pi)^2 \alpha (2-\alpha)},
    \end{gathered}
\end{equation}
where $(\alpha, \bm{k}^\perp)$ are the intrinsic LF relative variables. The expansion of the two-body state in terms of anti-symmetrized plane wave nucleonic
states is defined by,

\begin{equation}\label{LF_2body_expansion}
\begin{aligned}
     \ket{\beta \, , \lambda\, ; \bm{P}}  =  \frac{1}{2}\sum_{\substack{\sigma_1\tau_1 \\ \sigma_2 \tau_2}} \int \frac{d\alpha d\bm{k}^\perp}{(2 \pi)^2 \alpha (2-\alpha)}
     \psi_{\beta}^{\lambda}(\alpha, \bm{k}^\perp,\sigma_1,\tau_1,  \sigma_2, \tau_2) \ket{p_1, p_2}.
\end{aligned}
\end{equation}
In Eq. (\ref{LF_2body_expansion}), $\psi_{\beta}^{\lambda}$ is the two-body light-front
wavefunction for eigenstate $\beta$, SRG-evolved to $\lambda$. In the above expression, $(\alpha,\bm{k}^\perp)$ are defined by
\begin{equation}\label{relative_lf_coordinates}
\begin{aligned}[c]
    p_1^+  &= \frac{\alpha}{2} P^+ , \\
    \bm{p}_1^\perp &= \frac{\alpha}{2} \bm{P}^\perp + \bm{k}^\perp,
\end{aligned}
\qquad
\begin{aligned}[c]
    p_2^+  &= \frac{2-\alpha}{2} P^+ ,\\
    \bm{p}_2^\perp &= \frac{2-\alpha}{2} \bm{P}^\perp - \bm{k}^\perp,
\end{aligned}
\end{equation}
One important observation can explain why SRG techniques are most suitable for relativistic physics
formulated using LF quantization. Notice that Eq. (\ref{two_body_LF_SRG}) includes an integral
over the center-of-mass momenta. If we instead attempted to use SRG techniques formulated under
relativistic IF Quantization, we will obtain  a computationally intractable problem. The
IF wave functions are not boost invariant, and depend on center-of-mass momenta. Hence, one
must re-calculate the relativistic two-body problem for every frame, making the construction of
$\hat{U}(\lambda)$ extremely difficult. On the LF, however, we don't encounter this problem
as our two-nucleon wave functions are boost invariant.


\subsection{High-Resolution Light-Front Momentum Distributions}

It is instructive to re-write $\hat{U}(\lambda) = \hat{\mathbb{I}} + \delta \hat{U}(\lambda)$.
Replacing $\ket{\Psi_A(\lambda)}$ with $\ket{\Psi_A^{MF}}$, Eq. (\ref{srg_momentum_distribution})
is rewritten as a sum of four terms.
with detailed derivations of each term  presented in the Appendix.  
Figure \ref{fig:O16_SRG_momentum_distribution} presents the approximated high-resolution
light-front momentum distribution of $^{16}$O. The second and third SRG-interference terms  linear in $\delta \hat U$from  are represented by the green dashed line; the left and right panels are
included to illustrate the isolated behavior of these contributions. Qualitatively, these
interference terms deplete probability from the low-momentum mean-field region, while the $\delta U
\delta U^{\dagger}$ term subsequently redistributes it to higher momenta. This generates the
high-momentum tails characteristic of SRCs, consistent with both phenomenological understanding and
exact non-relativistic high-resolution calculations. To our knowledge, this represents the first
  calculation of light-front momentum distributions exhibiting these high-momentum tails.
Figure \ref{fig:lf_dist_4nuclei} presents the approximated high-resolution light-front momentum
distributions for $^{16}$O, $^{40}$Ca, $^{48}$Ca, and $^{208}$Pb.

\begin{figure}
    \centering
  \includegraphics[width=1\linewidth]{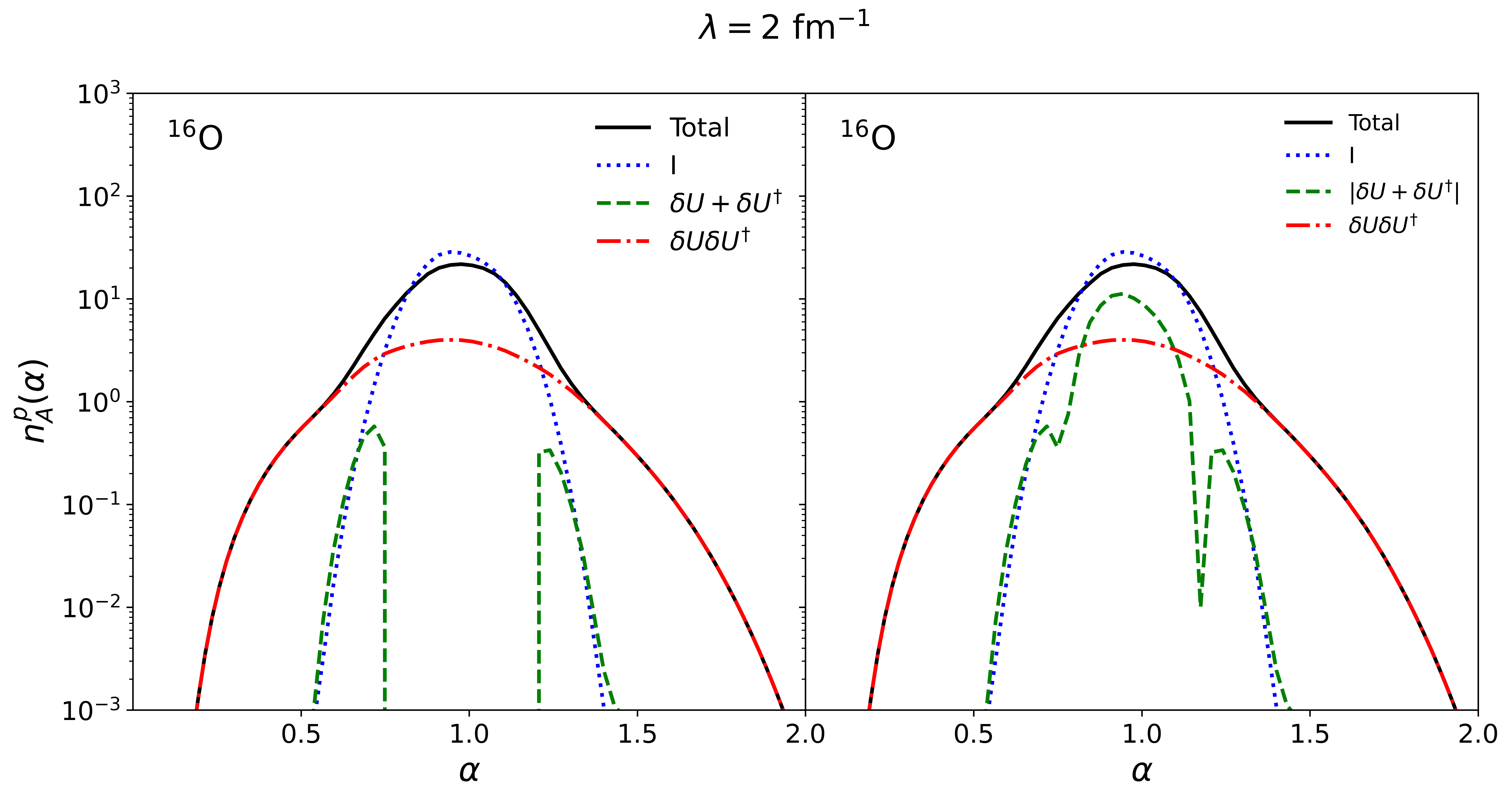}
    \caption{(color online) The approximated high-resolution light-front momentum distribution of Oxygen-16 using similarity renormalization group techniques. The left and right panels differ in their presentation of the green dashed line. The right figure presents the absolute value of the $\delta U + \delta U^\dagger$ term, while the left figure does not.}
    \label{fig:O16_SRG_momentum_distribution}
\end{figure}

\begin{figure}
    \centering
  \includegraphics[width=1\linewidth]{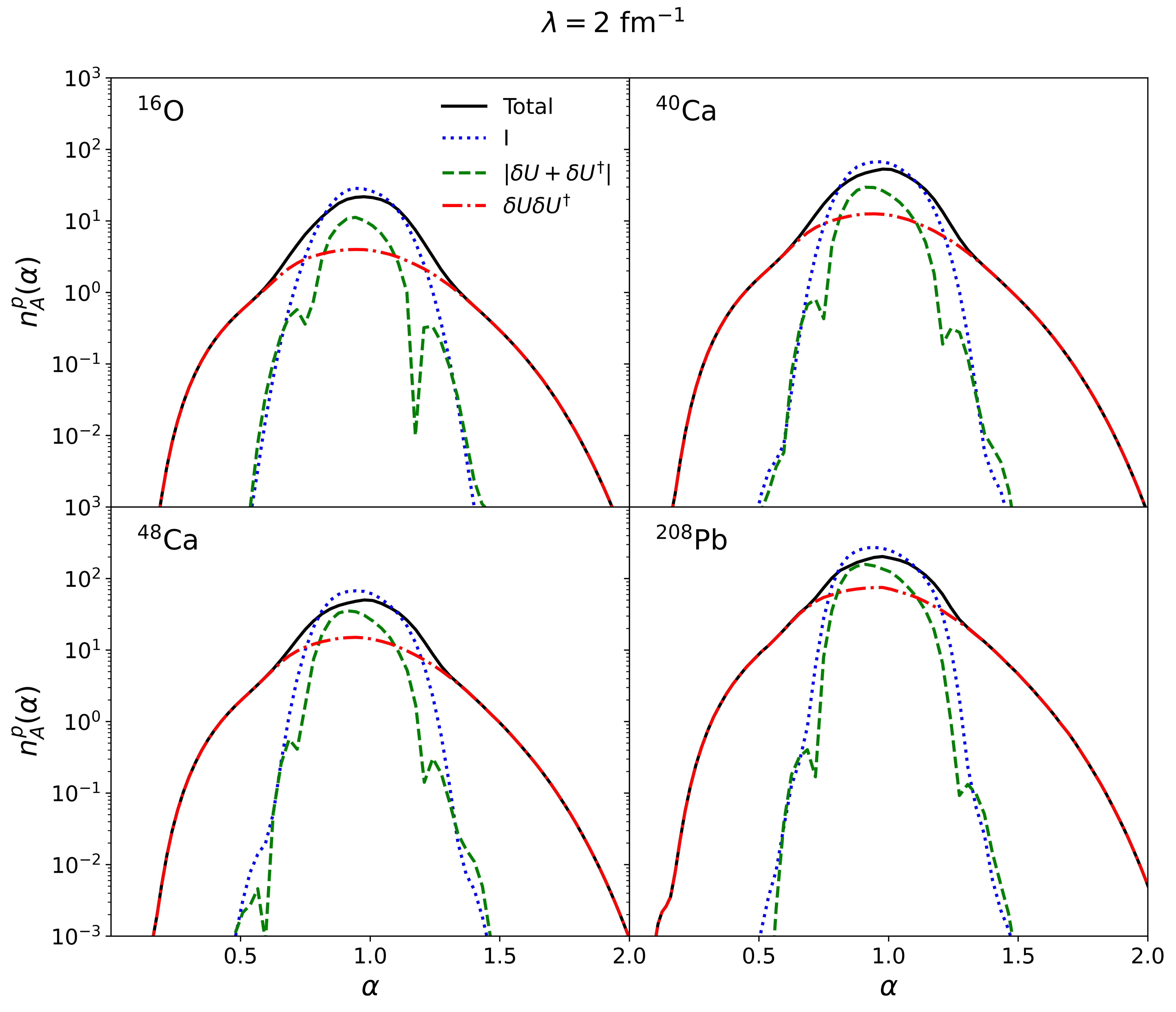}
    \caption{Light-front proton momentum distribution $n^p_A(\alpha)$ as a function
    of the longitudinal momentum fraction $\alpha = A\,p^+/M_A$, for
    $^{16}\mathrm{O}$ (top left), $^{40}\mathrm{Ca}$ (top right),
    $^{48}\mathrm{Ca}$ (bottom left), and $^{208}\mathrm{Pb}$ (bottom right). The
    solid black curve is the total distribution
    $n^p = I + \delta U + \delta U^{\dagger} + \delta U\,\delta U^{\dagger}$; the
    mean-field contribution $I$ (blue dotted), the two-body cross-term
    $|\delta U + \delta U^{\dagger}|$ (green dashed), and the two-body
    contribution $\delta U\,\delta U^{\dagger}$ (red dash-dot) are shown
    separately. SRG evolution scale $\lambda = 2~\mathrm{fm}^{-1}$.}
    \label{fig:lf_dist_4nuclei}
\end{figure}

\subsection{Inclusive Electron-Nucleus Scattering}
The SRG framework carries crucial implications for our previous cross-section
calculations utilizing mean-field nuclear states. Our prior analysis relied on the plane-wave
impulse approximation—a high-energy approximation which assumes that the incident electron interacts
with only a single nucleon inside the nucleus. In formal terms, this restricts the interaction to a
bare, one-body current. Viewed through the perspective of the SRG, we now understand that employing
such a simplified, high-energy approximation for the probe implicitly requires the target to be
described by a high-resolution state. Consequently, it is unsurprising that uncorrelated mean-field
results fail to reproduce experimental results; there is an inherent mismatch between the reaction
mechanism and nuclear structure. This discrepancy can be systematically resolved by formally
applying the SRG transformation to evolve the probing operators, generating higher-rank currents
that incorporate the nuclear dynamics missing from the low-resolution state.

Applying the SRG transformation systematically refines our previous mean-field calculations by 
depleting probability from the independent single-particle orbitals and redistributing it into 
short-range correlated (SRC) configurations. We implement this refinement in two steps. First, we 
update the one-body results of the previous section. Second, we incorporate 2N plus 
(A-2)-nuclear final-state configurations to explicitly capture the observable contributions from SRCs. 
Crucially, in evaluating these matrix elements, we only apply the SRG operator terms that correspond 
to the initial nuclear state. By intentionally neglecting the SRG evolution of the final scattering 
states, we isolate the specific impact of initial-state short-range correlations on the observable 
cross section.

\subsection{Updating LF MFT Results}

To incorporate the initial-state short-range correlations discussed above, we first update the one-body 
results derived in the Appendix. This is achieved by formally applying the 
SRG transformation operator, $\hat{U}(\lambda)$, to the intrinsic nuclear matrix element $M_i^A(\bm{p},\sigma,\tau)$. 
The transition from the bare matrix element to the SRG-evolved one is given by
\begin{equation}
    M_i^A(\bm{p},\sigma,\tau) =  \Braket{\Psi_i^{MF}|\hat{a}(\bm{p},\sigma,\tau)|\Psi_A^{MF}} \longrightarrow \Braket{\Psi_i^{MF}|\hat{a}(\bm{p},\sigma,\tau) \, \hat{U}(\lambda)|\Psi_A^{MF}}.
\end{equation}
By decomposing the unitary operator as $\hat{U}(\lambda) = \mathbb{I} + \delta\hat{U}(\lambda)$, we can 
separate the pure mean-field contribution from the SRG-induced corrections.

\begin{figure}
    \centering
  \includegraphics[width=1\linewidth]{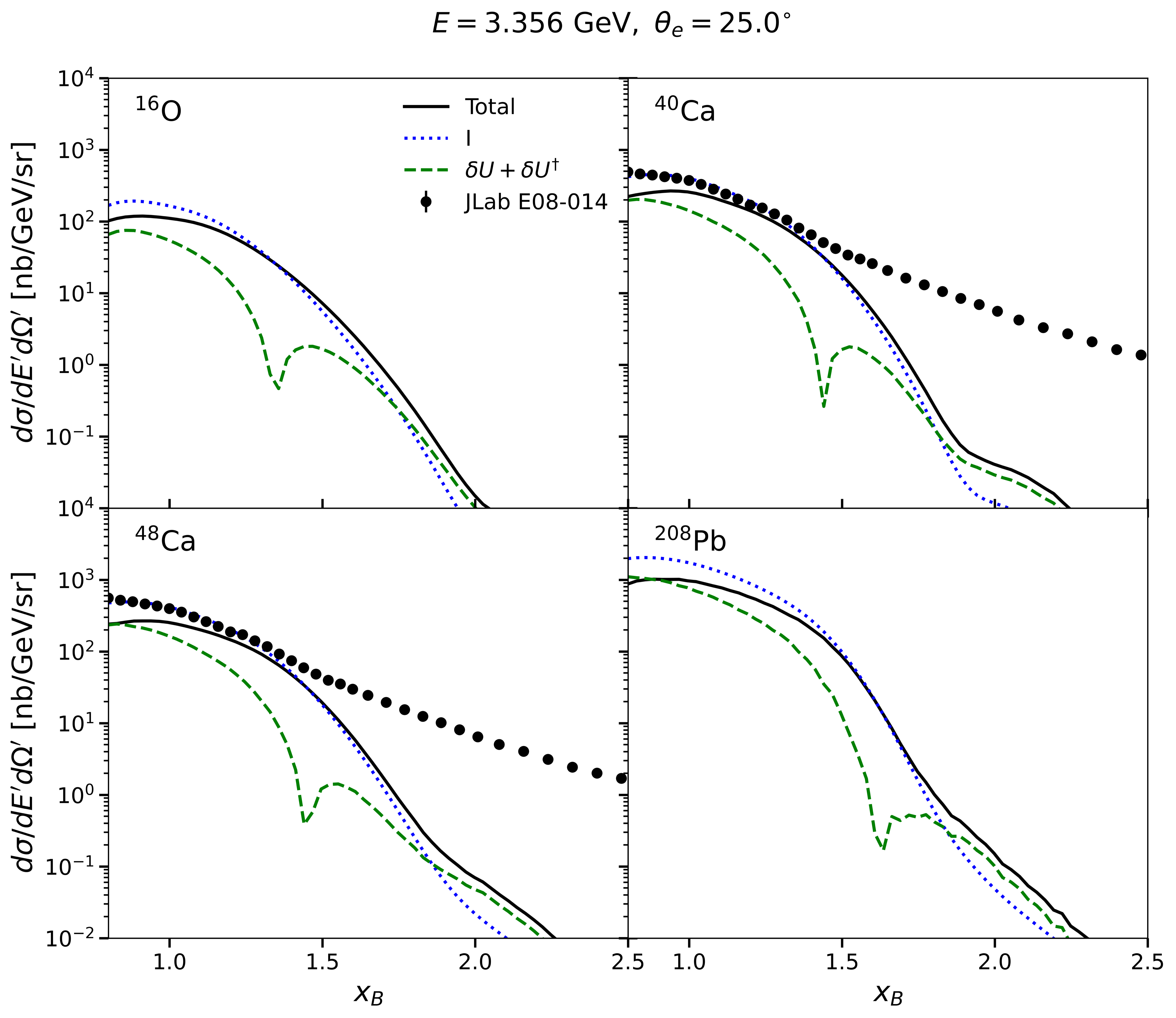}

    \caption{Same as Fig.~\ref{fig:MF_cs}, with the SRG operator
            corrections added. The solid black curve is the total
            $\mathbb{I} + \delta \hat{U} + \delta \hat{U}^{\dagger}$ contribution evaluated at an SRG scale of
            $\lambda = 2~\mathrm{fm}^{-1}$. The pure mean-field contribution (blue dotted) and the absolute 
            value of the one-body cross-term $|\delta \hat{U} + \delta \hat{U}^{\dagger}|$ (green dashed), which
            depletes the cross section near the quasi-elastic peak.}
    \label{fig:cs_4nuclei_A1}
\end{figure}

Figure~\ref{fig:cs_4nuclei_A1} displays the updated inclusive cross sections evaluated at an SRG scale 
of $\lambda = 2~\mathrm{fm}^{-1}$. The total updated one-body cross section (solid black curve) is composed 
of the bare mean-field baseline ($\mathbb{I}$, blue dotted curve) and the linear interference terms 
($\delta \hat{U} + \delta \hat{U}^{\dagger}$, green dashed curve). Consistent with the behavior of SRG 
evolution, these linear interference terms are negative in the quasi-elastic peak; they act to deplete probability 
from the low-momentum mean-field region. This depleted probability is precisely what gets redistributed to higher 
momenta via the two-nucleon final-state configurations, which we address in the following subsection.

\subsection{Inclusion of 2N + (A-2) Nuclear Final States}
SRG techniques allow us to extend our previous calculations. Because the SRG-evolved current
operator now contains two-body terms, the current can couple to two bound nucleons and promote them
into scattering states — that is, it introduces 2N plus (A-2)-nuclear final-state configurations
into our calculations. For such configurations, the terms in Eq. \eqref{nuclear_tensor} become

\begin{equation}
    \sum_X \rightarrow \frac{1}{2}\sum_{\substack{\sigma_1 \tau_1 \\ \sigma_2 \tau_2}} \sum_f \int [dP_{A-2}] [dP_{cm}'] \frac{d\alpha'd\bm{k}'^\perp}{(2 \pi)^3 \alpha' (2-\alpha')},
\end{equation}

\begin{equation}
\begin{gathered}
    \ket{X}\bra{X} \rightarrow \ket{\Psi_{A-2}^f,\bm{P}_{A-2}}\bra{\Psi_{A-2}^f,\bm{P}_{A-2}} \, \otimes \, \ket{ p_1, p_2 }\bra{ p_1, p_2 }, \\
    \ket{p_1, p_2} = \ket{\bm{p}_1, \sigma_1, \tau_1; \bm{p}_2, \sigma_2, \tau_2},
\end{gathered}
\end{equation}
with $ 
    \bm{p}_1 = \left(\tfrac{\alpha'}{2} P_{cm}'^{+}, \, \tfrac{\alpha'}{2} \mathbf{P}_{cm}'^{\perp} + \bm{k}'^\perp\right), \,
    \bm{p}_2 = \left(\tfrac{2-\alpha'}{2} P_{cm}'^{+}, \, \tfrac{2-\alpha'}{2} \mathbf{P}_{cm}'^{\perp} - \bm{k}'^\perp\right)$
and where $f$ denotes all possible (A-2)-nuclear final states. Utilizing the delta function in
light-front momenta, we integrate over $P_{A-2}$ and use $\bm{P}_{cm}' = \bm{P}_{cm} + \bm{q}$ to get

\begin{equation}
\begin{aligned}
    W_A^{\mu\nu} = & \, \frac{1}{2 m_A} \frac{1}{2P_{A-2}^+} \frac{1}{2} \sum_{\substack{\sigma_1 \tau_2 \\ \sigma_2, \tau_2}} \sum_f \int \frac{dP_{cm} d\bm{P}_{cm}^{\perp}}{(2 \pi)^3} \frac{d\alpha'd\bm{k}'^\perp}{(2 \pi)^3 \alpha' (2-\alpha')} \\
    & \times \braket{\Psi_A|J^\mu_A(0)|\Psi_{A-2}^f,\bm{P}_{A-2};p_1,p_2} \braket{\Psi_{A-2}^f,\bm{P}_{A-2};p_1,p_2|J^\nu_A(0)|\Psi_A} \\
    & \times \left[ \frac{1}{P^+_{cm} + q^+} \delta(q^- + P_A^- - (P_{A-2}^{f})^- - P_{cm}'^{-}) \right], 
\end{aligned}
\end{equation}
where
   $ (P_{A-2}^f)^- = \frac{M_f^2 + (\bm{P}_{cm}^{\perp})^2}{P_A^+ - P_{cm}^+}, \qquad P_{cm}'^- = \frac{4\frac{m^2 + (\bm{k}'^\perp)^{2}}{\alpha' (2 - \alpha')} + (\bm{P}_{cm}^\perp)^2}{P_{cm}^+ + q^+} .$

The nuclear current matrix element is,
\begin{equation}
\begin{aligned}
    \braket{\Psi_{A-2}^f,\bm{P}_{A-2};p_1,p_2|J^\nu_A(0)|\Psi_A} & = \frac{1}{2}\sum_{\substack{s_1 t_1 \\ s_2 t_2}} \sum_{\sigma \sigma' \tau} \int [dp][dp'][dq_1][dq_2] \\
    & \times  \sqrt{2p^+} \sqrt{2p'^+} \sqrt{2q_1^+} \sqrt{2q_2'^+} \\
    & \times J_N^\nu(\bm{p}',\sigma',\tau;\bm{p},\sigma,\tau) \braket{p_1,p_2|\hat{a}^\dagger(\bm{p}',\sigma',\tau) \hat{a}(\bm{p},\sigma,\tau)|q_1, q_2} \\
    & \times \braket{\Psi_{A-2}^f,\bm{P}_{A-2}|\hat{a}(\bm{q}_2,s_2,t_2) \hat{a}(\bm{q}_1,s_1,t_1)|\Psi_A}.
\end{aligned}
\end{equation}

\noindent Following the same procedure as the (A-1) calculation in Appendix \ref{appendix:A-1
implementation}, we get

\begin{equation}
\begin{aligned}
    & \braket{\Psi_{A-2}^f,\bm{P}_{A-2};p_1,p_2|J^\nu_A(0)|\Psi_A} \approx \sqrt{2P_A^+} \sqrt{2P_{A-2}^+}\\
    & \times \sum_\sigma \Bigg[ \sqrt{\frac{p_2^+}{p_1^+ - q^+}} J_N^\nu(\bm{p}_1,\sigma_1,
    \tau_1;\bm{p}_1-\bm{q},\sigma,\tau_1) M_{ij}^A(\bm{p}_2,\sigma_2,\tau_2;\bm{p}_1-\bm{q},\sigma,\tau_1|\lambda)  \\
    & \qquad-\sqrt{\frac{p_1^+}{p_2^+ - q^+}} J_N^\nu(\bm{p}_2,\sigma_2,
    \tau_2;\bm{p}_2-\bm{q},\sigma,\tau_2)  M_{ij}^A(\bm{p}_1,\sigma_1,\tau_1;\bm{p}_2-\bm{q},\sigma,\tau_2|\lambda)  \Bigg],
\end{aligned}
\end{equation}

\begin{equation}
     M_{ij}^A(\bm{p}_2,\sigma_2,\tau_2;\bm{p}_1-\bm{q},\sigma,\tau_1|\lambda) =\braket{\Psi^{MF}_{ij}|\hat{a}(\bm{p}_2,\sigma_2,\tau_2) \hat{a}(\bm{p}_1-\bm{q},\sigma,\tau_1) \delta \hat{U}^{\dagger}(\lambda)|\Psi^{MF}_A},
\end{equation}

\noindent where we substitute the high-resolution (A-2)-nuclear states with their low-resolution
mean-field counterparts, approximating them as two-hole states relative to the target A-nucleus
state, i.e. $\ket{\Psi^{MF}_{ij}} = \hat{a}_j \hat{a}_i\ket{\Psi^{MF}_A}$. In doing so, we replace
$\sum_f \rightarrow \frac{1}{2} \sum_{ij}$. Furthermore, note that we use $\delta \hat{U}^\dagger(
\lambda)$ instead of $\hat{U}^\dagger(\lambda)$ in the above expression. This is because the $\mathbb{I}$
term  cannot promote two bound state nucleons into free plane-waves.
\begin{figure}
    \centering
\includegraphics[width=1\linewidth]{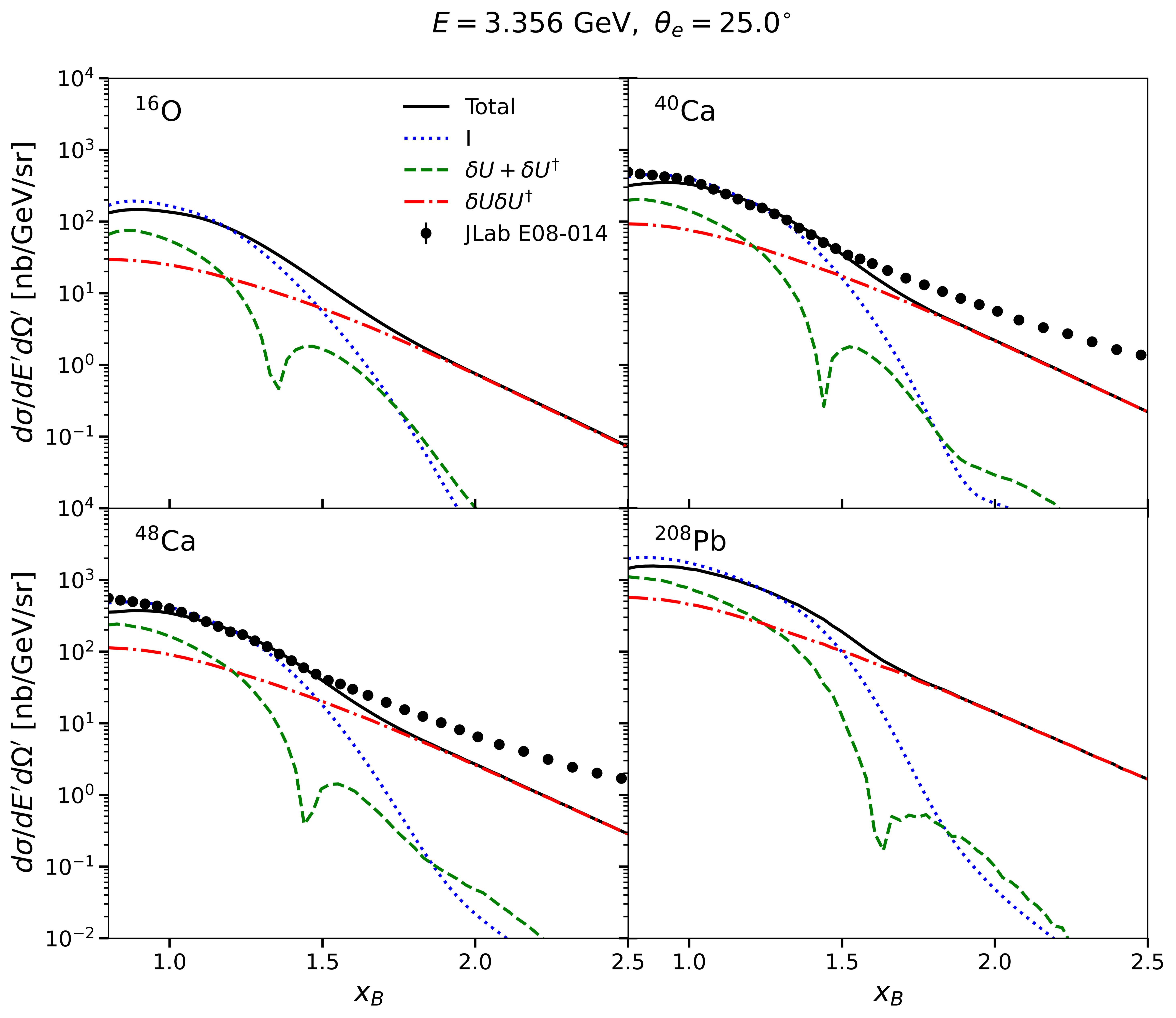}
    \caption{Same as Fig.~\ref{fig:cs_4nuclei_A1}, with the two-body SRG operator
    contribution $\delta U\,\delta U^{\dagger}$ (red dash-dot) added. The
    solid black curve now contains all SRG operator contributions through
    second order in the unitary transformation, $I + \delta U +
    \delta U^{\dagger} + \delta U\,\delta U^{\dagger}$, again at
    $\lambda = 2~\mathrm{fm}^{-1}$.}
    \label{fig:cs_4nuclei_A1_A2}
\end{figure}

Squaring this nuclear current matrix element to compute the nuclear tensor   generates the 
quadratic $\delta\hat{U}\,\delta\hat{U}^{\dagger}$ contribution to the cross section. Physically, 
this term corresponds to the virtual photon striking a nucleon that belongs to a short-range correlated pair, 
resulting in the emission of both the struck nucleon and its correlated partner, while leaving the residual 
nucleus in a two-hole (A-2) spectator state.

Figure~\ref{fig:cs_4nuclei_A1_A2} presents the total inclusive cross section incorporating these 2N + (A-2) 
final states. The isolated contribution from the quadratic $\delta U \,\delta U^{\dagger}$ term is 
represented by the red dash-dot curve. This term acts to explicitly redistribute the probability that was 
depleted from the quasi-elastic peak by the linear interference terms (green dashed curve). Consequently, 
this missing strength is shifted to higher values of missing momentum, manifesting as an enhancement in 
the cross section at $x_B > 1.3$. The total inclusive cross section---evaluated at $\lambda = 2~\mathrm{fm}^{-1}$ and incorporating all SRG 
operator contributions through second order in the unitary transformation ($\mathbb{I} + \delta U + 
\delta U^{\dagger} + \delta U \,\delta U^{\dagger}$)---is shown as the solid black curve. It 
should be noted that the specific scale parameter $\lambda = 2~\mathrm{fm}^{-1}$ was not determined through 
a rigorous fitting procedure, but was instead chosen phenomenologically to best reproduce the cross section 
in the intermediate $1.3 < x_B < 1.5$ region. While this systematic inclusion of initial-state SRCs successfully 
generates the high-momentum tails missing from the pure mean-field results, the theoretical calculation still 
underestimates the experimental data in the far tail ($x_B > 1.5$). Attempting to compensate for this deficit 
by further decreasing $\lambda$ to generate more high-momentum configurations proves unviable, as doing so 
leads to an overprediction of the data in the $1.3 < x_B < 1.5$ window. This residual shortfall indicates 
that initial-state correlations alone cannot fully account for the high-momentum reaction dynamics, confirming 
that the plane-wave impulse approximation and using only nucleon degrees of remains insufficient in this regime.

\section{Final State Interactions}\label{chap:fsi}

The inclusive cross section is the probability of detecting a scattered electron at given
kinematics, characterized by $x_B$ and $Q^2$, or equivalently by the scattered electron's energy and
angle. The electron interacts with the initial target and emerges into the detector,
while the rest of the system is left to evolve and re-interact unobserved. The yield of detected
electrons is therefore insensitive to the dynamics that occur after the initial electron-nucleus
scattering: whatever rescattering takes place inside the target system cannot retroactively alter
the electron that has already been measured. This translates into the statement that
the inclusive cross section can be expressed without any reference to the final state,
\begin{equation}\label{cs_without_final_state}
\begin{aligned}
    \frac{d\sigma}{dE'd\Omega'} \propto & \sum_X \braket{A|\hat{J}^\mu(0)|X} \braket{X|\hat{J}^\nu(0)|A} \delta(P_A^- + q^- - p_X^-) \\
    = & \braket{A|\hat{J}^\mu(0)\,\delta(P_A^- + q^- - \hat{P}^-)\,\hat{J}^\nu(0)|A}.
\end{aligned}
\end{equation}
\noindent The sum over final states in the first line is a book-keeping device for tabulating the
contributions of individual channels.  The second line shows that the inclusive observable is   a ground-state matrix
element of an operator built from the full Hamiltonian, with no final-state basis appearing
anywhere. Why, then, do the results of the previous section suggest missing physics from final-state
interactions (FSI)?

The answer is that the calculations performed so far have captured only a portion of the inclusive
cross section within each final-state channel. The resolution lies in recognizing what is required
to pass from the second line of Eq.~(\ref{cs_without_final_state}) back to the first. The collapse
of the sum over $|X\rangle$ into the operator $\delta(P_A^- + q^- - \hat{P}^-)$ is a spectral
decomposition, which is valid only when the states $|X\rangle$ are eigenstates of the  
Hamiltonian operator $\hat{P}^-$ appearing inside the delta function. The energy-conserving delta function
contains an operator that contains interactions in the final state.  Therefore, for inclusive scattering, FSI effects are   required to properly evaluate the delta function.

\subsection{Elastic Two-Nucleon Re-scattering}

In the $x_B > 1.5$ regime the dominant initial-state configurations are short-range correlated (SRC)
pairs. Such pairs are characterized by a large relative momentum and a small center-of-mass
momentum, so that the two correlated nucleons are close to one another while the pair as a whole is
spatially separated from the residual A-2 system. Consequently, when the virtual photon strikes one
member of the pair, the struck nucleon is far more likely to re-scatter from its correlated partner
than from the remaining A-2 nucleons as it leaves the nucleus. We model this final-state interaction
FSI by replacing the two-nucleon plane-wave final state with the scattering state generated by the LF
minimal-relativity AV18 potential,
\begin{equation}
\begin{gathered}
\ket{p_1,p_2}\bra{p_1,p_2}
\rightarrow \ket{\Psi_{NN}^{(-)},\bm{P}}\bra{\Psi_{NN}^{(-)},\bm{P}}, \\
\ket{\Psi_{NN}^{(-)},\bm{P}} = \ket{p_1,p_2}
+ \frac{1}{\hat{P}^+}\frac{1}{p_1^- + p_2^- - \hat{P}^- - i\epsilon}\,
\hat{P}^+\hat{T}^{(-)}\ket{p_1,p_2},
\end{gathered}
\end{equation}
where the $(-)$ superscript denotes the physical final-state scattering solution, $\hat{T}^{(-)}$ is
the half-off-shell $T$-matrix, and the limit $\epsilon \to 0^+$ is understood. The factor
$\hat{P}^+/\hat{P}^+$ is inserted because the $T$-matrix obtained from Eq.~\eqref{weinberg_eq} is
proportional to the matrix element of $\hat{P}^+\hat{T}^{(-)}$.

Rather than solving for $\hat{T}^{(-)}$ directly, we first compute the half-off-shell $K$-matrix.
The $K$-matrix generates standing-wave (principal-value) scattering states \cite{Taylor:1972pty}.
It is real-valued  and satisfies an integral equation that can be rendered non-singular, it is
thus numerically more convenient than using the imaginary $T$-matrix. Partial-wave projection of
Eq.~\eqref{rel_lippmann_schwinger}, together with the standing-wave boundary condition, leads to the
coupled integral equations for the K-matrix,
\begin{equation}\label{K_matrix_eq}
    \left[\, \tilde{K}_{LL'}^{JTS}(k_{\rm off},p)
    = \tilde{V}_{LL'}^{JTS}(k_{\rm off},p)
    + \sum_{L''}\mathcal{P}\!\int \frac{dk'\,k'^2}{(2\pi)^3 E(k')}\,
    \frac{\tilde{V}_{LL''}^{JTS}(k_{\rm off},k')\,
    \tilde{K}_{L''L'}^{JTS}(k',p)}{4(p^2 - k'^2)} \,\right]_{\rm IF},
\end{equation}
where the labels $(J,T,S,L,L')$ specify the partial-wave channel and $\mathcal{P}$ denotes the
principal value.

We solve Eq.~\eqref{K_matrix_eq}   by discretizing the momentum integral on a
Gauss--Legendre grid $\{k_i, w_i\}$, which converts the integral equation into a matrix equation that is solved by matrix inversion.
The half-off-shell $T$-matrix
is recovered from $\tilde{K}$ through the (relativistic) Heitler relation,
\begin{equation}
\begin{gathered}
    \left[ \left( \tilde{T}_{LL'}^{JTS}(k_{\rm off},p) \right)^{(-)}  = \frac{\tilde{K}_{LL'}^{JTS}(k_{\rm off},p)}{\hat{\mathbb{I}}-i\rho_{\rm ps}(p)\tilde{K}_{LL'}^{JTS}(p,p)}\right]_{\rm IF} \\
    \left[ \rho_{\rm ps}(p) = \frac{p}{64\pi^2E(p)} \right]_{\rm IF}
\end{gathered}
\end{equation}
where $\rho_{\rm ps}$ is the on-shell phase-space factor fixed by the phase-space factors in
Eq.~\eqref{rel_lippmann_schwinger}. Finally, the LF analogue is obtained from Eq.
\eqref{eq:T_matrix_ET_LF}.

Incorporating FSI requires the struck nucleon to remain off-shell even
after interacting with the virtual photon. Consequently, the single-nucleon electromagnetic current
matrix element introduced in Sect.~\ref{chap:lightfront} must be generalized to account for the effect of virtuality in the 
final-state:
\begin{equation}
    J_N(\bm{p}',\sigma',\tau;\,\bm{p},\sigma,\tau) = \bar{u}(\bm{p}',\sigma',\tau)\!\left(1 + \frac{\slashed{\Delta}_{p'}}{2m_N}\right)\! \Gamma^\mu_{\textcolor{red}{\gamma^*}N}\!\left(1 + \frac{\slashed{\Delta}_{p}}{2m_N}\right)\!u(\bm{p},\sigma,\tau),
\end{equation}
where the final-state off-shell operator, $\slashed{\Delta}_{p'}$, follows the   same structure
as Eq.~\eqref{EM_offshell_factors}. We note that the recoil spectator nucleon involved in the
elastic rescattering is technically also off-shell. However, we neglect its corresponding off-shell
corrections here because they are kinematically suppressed. Specifically, these factors only become
significant in the integration region where $\alpha_{r} \equiv 2p_r^+/P^+ \to 0$, with $p_r$
the momentum of the SRC recoil nucleon. Physically, this limit describes a strongly virtual
spectator nucleon emerging from the deuteron vertex and instantaneously interacting with the struck
nucleon---a configuration whose contribution to the overall integral is negligible \cite{Vera:2021rnw}.

\subsection{Results and Discussion for Deuterium and Calcium-40}\label{sec:fsi_deuterium_CA40}

We first apply our FSI formalism  to the deuteron in which the  final-state rescattering is the elastic
re-interaction between the two nucleons modeled here. We solve the K-matrix
equation~\eqref{K_matrix_eq} for the LF minimal-relativity AV18 potential in all $S$, $P$, and $D$ partial-wave
channels, retaining the coupled $^3S_1$--$^3D_1$ and $^3D_3$--$^3G_3$ channels, and construct
the LF half-off-shell $T$-matrix via Eq.~\eqref{eq:T_matrix_ET_LF}.

\begin{figure}
    \centering
 \includegraphics[width=1.0\linewidth]{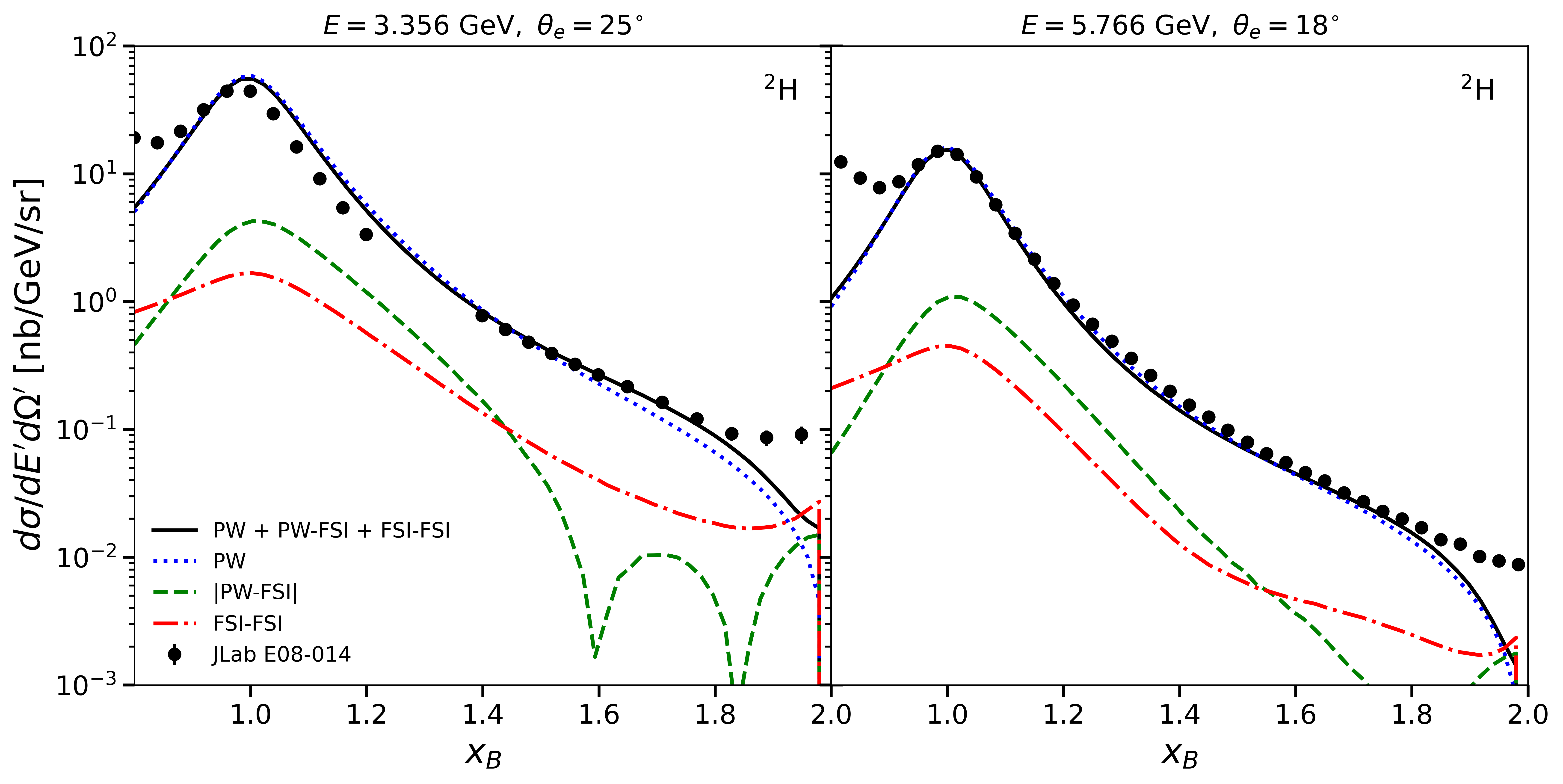}
    \caption{Inclusive electron-deuteron cross section as a function of the Bjorken
             variable $x_B$ at the JLab E08-014 kinematics ($E = 3.356$~GeV,
             $\theta_e = 25^\circ$, left) and the JLab E02-019 kinematics
             ($E = 5.766$~GeV, $\theta_e = 18^\circ$, right), decomposed into plane-wave
             and FSI contributions. The solid black curve is the full theory result,
             $\sigma_{\rm PW} + \sigma_{\rm PW-FSI} + \sigma_{\rm FSI-FSI}$; the blue dotted curve is the
             plane-wave piece; the green dashed curve is the PW–FSI interference $|\sigma_{\rm PW-FSI}|$; the red dash-dot curve
             is the FSI quadratic contribution $\sigma_{\rm FSI-FSI}$. Black points are the experimental
             data.}
    \label{fig:H2_FSI}
\end{figure}

Figure~\ref{fig:H2_FSI} shows the resulting inclusive cross sections compared to JLab data at 
two kinematic settings: $E = 3.356$ GeV, $\theta_e = 25^\circ$ from JLab E08-014~\cite{Zhang:2025nst} (left panel) 
and $E = 5.766$ GeV, $\theta_e = 18^\circ$ from JLab E02-019~\cite{Fomin:2011ng} (right panel), together 
with the individual FSI contributions. The interference term includes both cross terms from the squared matrix element, 
PW-FSI + FSI-PW; we label it simply as PW-FSI. Since it removes probability from the plane-wave channel by unitarity,
it is plotted as its absolute value. The quadratic term is positive and redistributes probability.

The central result is that the full cross section is nearly indistinguishable from the
plane-wave result alone across the entire $x_B$ range shown for both kinematics. The interference
correction lies one to two orders of magnitude below the cross section at most kinematics, as does the
quadratic correction.

We now extend the same calculation to $^{40}$Ca. We only apply FSI to the SRG-induced 2N + (A-2) final state term.
The FSI is again described by the two-nucleon $T$-matrix from the minimal-relativity AV18 potential. Results
are compared to JLab E08-014 data for $^{40}$Ca \cite{Zhang:2025nst}. The results, shown in Fig.~(16) for $^{40}$Ca are identical in 
character to those for the deuteron: the FSI corrections are negligible and the full cross section is again nearly 
indistinguishable from the plane-wave result alone.

\begin{figure}
    \centering
  \includegraphics[width=0.8\linewidth]{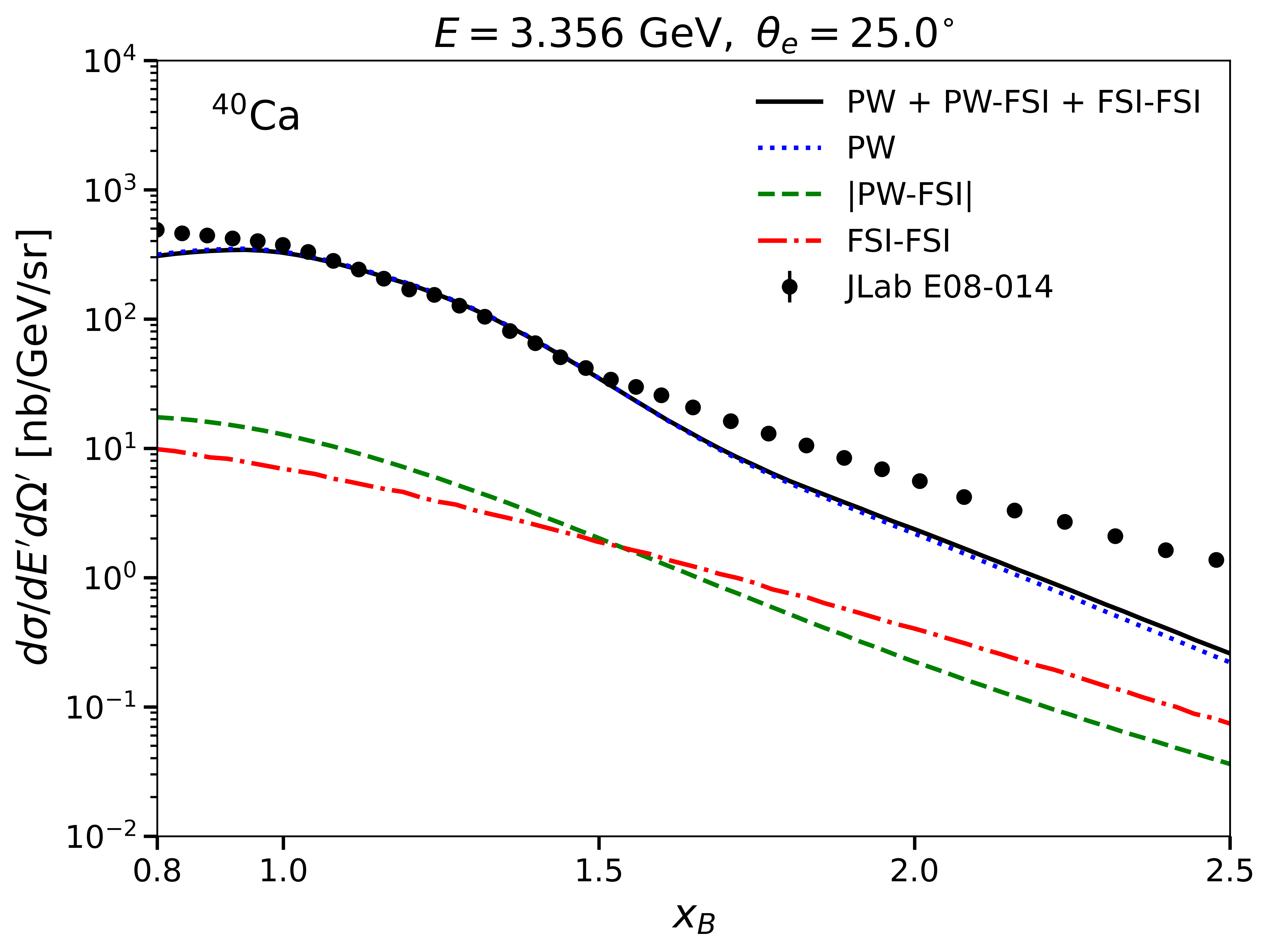}
    \caption{Inclusive electron-$^{40}$Ca cross section as a function of $x_B$ for JLab
    E08-014 kinematics. Curves as in Fig.~\ref{fig:H2_FSI}. Data (black points) from
    JLab E08-014. The `PW' term is the same as the `Total' term labeled in Fig. \ref{fig:cs_4nuclei_A1_A2}.}
    \label{fig:Ca40_FSI}
\end{figure}

Our results differ quantitatively from the similar elastic rescattering study of
Ref.~\cite{CiofidegliAtti:1994ys}. Both approaches employ elastic two-nucleon rescattering.  The
difference is mainly due  to the use of an approximate calculation of  the effects of an optical potential-the interaction between the outgoing proton  and the residual nucleus. 
Our view is that the use of an optical potential is inappropriate  for calculating an inclusive cross section because the optical potential restricts the final state to be the $A-1$ ground-state.

More broadly, the AV18 $T$-matrix captures only elastic
$NN$ degrees of freedom and is therefore missing non-nucleonic effects. For example, there could be the production of $\Delta$-isobars in the final state. There could be other effects such as the those explicitly involving quark degrees of freedom. 


These missing inelastic effects  leave a clear empirical signature. While our
purely nucleonic model --- two-nucleon SRC configurations with elastic re-scattering --- reproduces
the deuterium data well for $x_B>1$ shown in Fig.~\ref{fig:H2_FSI}, the same
model falls short of the $^{40}$Ca data at large $x_B$ (Fig.~\ref{fig:Ca40_FSI}), where the
cross section is dominated by the SRC contribution. The growing discrepancy with nuclear mass
reflects the non-nucleonic interactions  in the denser nuclear medium of heavier nuclei, which our
nucleonic framework does not capture.

This has a direct consequence for the interpretation of the SRC plateau in the experimental
ratio $\sigma(A)/\sigma(D)|_{x_B > 1.5}$. Because our computed  cross section for heavy nuclei 
falls below the data while the deuterium cross section does not, the experimental plateau exceeds the purely nucleonic ratio
$\sigma_{\rm theory}(A)/\sigma_{\rm theory}(D)$. The standard identification of the plateau
value with nucleonic SRC abundance $a_2(A)$ \cite{Fomin:2017ydn} therefore overestimates the
purely nucleonic content: the experimental ratio is inflated by non-nucleonic channels that
our theory is missing, and which do not cancel between the heavy nucleus and deuterium. By the
same reasoning, theoretical and phenomenological extractions of the average number of SRC pairs
from momentum distribution ratios \cite{Weiss:2020bkp} inherit the same sensitivity, and the
pair counts extracted from experimental data should be understood as an upper bound on the
purely nucleonic SRC content.

\begin{figure}
    \centering
     \includegraphics[width=0.8\linewidth]{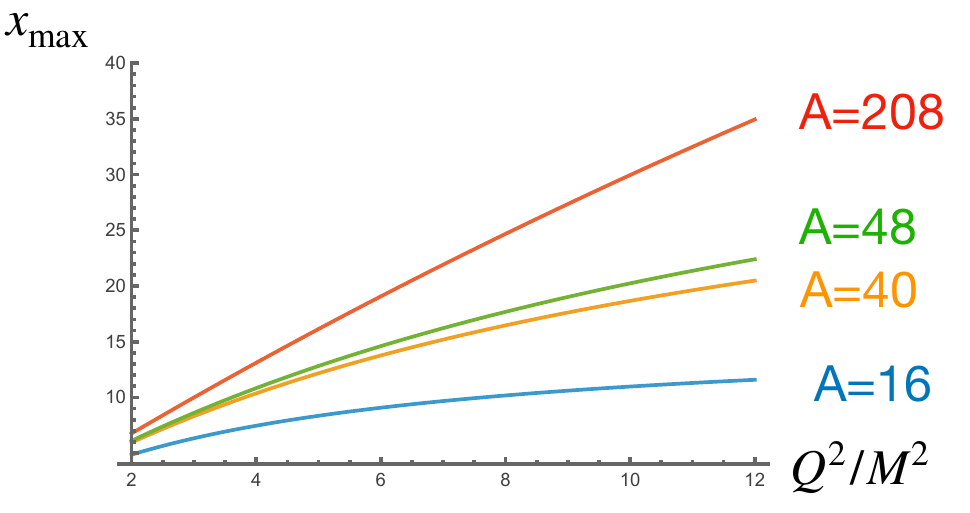}
    \caption{ Maximum value of $x_B$ that can produce a pion as a function of $A$ and $Q^2$}
    \label{xmax}
\end{figure}

The importance of pion production and the production of other inelastic states at large values of $x_B$ for nuclear targets
is somewhat of a surprise because inelasticities are associated with values of $x_B<1.$  That inelastic production is possible at larger values of $x_B$ is a simple consequence of the  large target mass. This is explained here by using conservation of 4-momentum for the example of pion production. We compute the threshold value of  $s_A=W_A^2$
as
\begin{eqnarray}
&W_A^2=(\nu +MA)^2-\vec q^2=(M A +m_\pi)^2\\
&{Q^2\over M^2}\left({A\over x_{\rm max}}-1\right)=2A {m_\pi\over M},
\label{xm}
\end{eqnarray}
where $M$ is the nucleon mass and $x_{\rm max}$ is the largest value of $x_B$ that can produce a pion.  The factor $A$ replaces replaces the unity that would appear for a proton target. Eq.~(\ref{xm}) is solved for $x_{\rm max}$ as a function of $A,Q^2$ and the results are shown in Fig.(\ref{xmax}).
The figure shows that a pion can be produced at very large values of $x_B$. The production is not limited to pion production. For inelastic baryon production one would replace $m_\pi$ appearing in Eq.(\ref{xm}) by the mass excess over a nucleon and the qualitative nature of Fig.~(\ref{xmax}) would not change.

\section{Conclusion}\label{chap:conclusion}

The advent of high-energy experiments at Jefferson Lab and the planned Electron-Ion Collider, (EIC) both
of which aim to probe non-perturbative aspects of QCD using nuclear targets, places a new demand on
low-energy nuclear physics: the structure models that feed these programs must account for the relativistic motion of the nuclei and the particles within.
Light-front quantization is the natural framework for this regime, but the path to relativistic
many-body calculations is not obvious. Correlated \textit{ab initio} approaches based on the
many-body Dirac equation are out of reach---the requisite two- and three-body relativistic
interactions have not been developed, and even the numerical machinery for solving the Dirac
many-body problem is in its infancy. Furthermore, the EIC's heavy-nuclei program compounds the
difficulty, pushing well beyond what even non-relativistic \textit{ab initio} methods can currently
address.

The strategy adopted here is to be as conservative as possible with the existing,
successful machinery of IF nuclear structure, and to reformulate it  to the LF. The two key
developments along this line and the implications are presented next. 

\subsection*{Light-front mean-field  theory}

The first development, presented in Sect.~\ref{chap:dft}, is the formulation of LF mean-field theory (MFT) and, more importantly, a demonstration of its formal equivalence to IF MFT. This
equivalence is exact: no approximation is required to map one onto the other. The practical
consequence is that existing and future developments in IF MFT transfers entirely to the LF. Applied
to inclusive electron-nucleus scattering at high momentum transfer and low energy transfer, LF MFT
reproduces the quasi-elastic peak well. It fails, however, at high Bjorken-$x_B$.  In line with what
is already established from many previous electron scattering measurements: mean-field theory does not describe the
high-momentum tail of the nuclear wave function.

\subsection*{Beyond mean field: SRG and short-range correlations}

Short-range correlation phenomenology identifies the missing strength at high $x_B$ with
two-nucleon configurations of high relative momentum and low center-of-mass momentum. The second
development of this program incorporates these correlations into the LF framework by using the
Similarity Renormalization Group. SRCs do increase the inclusive cross section in the expected
direction. They are, however, insufficient: even after including elastic nucleon-nucleon
rescattering as a final-state interaction, the predicted cross section  undershoots the data.
This suggests beyond purely nucleonic, elastic dynamics---toward non-nucleonic channels as necessary 
ingredients in the high-$x_B$ regime, currently missing in phenomenology.

\subsection*{Implications for the SRC plateau}

The cross-section ratio $\sigma(A)/\sigma(D)\big|_{x_B > 1.5}$, which exhibits a characteristic
plateau, is typically  identified with the per-nucleon SRC probability $a_2(A)$~\cite{Fomin:2017ydn}.
This identification rests on the implicit assumption that the dominant non-nucleonic contributions
cancel between numerator and denominator. The present analysis shows this cancellation is
incomplete: the experimental cross section for nuclei contain inelastic 
contributions, whereas the deuterium cross section is well described by the
purely nucleonic model. The experimental ratio therefore exceeds the purely nucleonic ratio
$\sigma_{\rm theory}(A)/\sigma_{\rm theory}(D)$, and the conventional $a_2(A)$ overestimates the
purely nucleonic SRC content. The nucleonic theory developed here provides a lower bound on the
ratio; the gap between the experimental plateau and the theoretical prediction quantifies the inelastic
excess that inflates it. By the same reasoning, phenomenological extractions of SRC pair counts from
ratios of nuclear momentum distributions~\cite{Weiss:2020bkp} inherit the same sensitivity, and
those pair counts should be understood as upper bounds on the purely nucleonic content.

The failure to reproduce the plateau provides support the conclusion~\cite{Hen:2016kwk} that conventional nuclear dynamics are not sufficient to explain the nuclear dependence of quark structure functions known as the EMC effect.
\section*{Acknowledgments} This work is supported in part by the U.S. Department of Energy, Office of Science, Office of Nuclear Physics, under contract numbers   DE-SC0026252  and Award No. DE- FG02-97ER41014.
\subsection*{Code availability}

The codes developed for this paper are publicly available at
\url{https://github.com/dima-kim/Light-Front-Quasi-Elastic-Scattering}.

\appendix
\newpage
\appendix{\bf Appendix}

\subsection{ Inclusive Electron-Nucleus Cross Section for (A-1)-Nuclear Final States}\label{appendix:A-1 implementation}
\label{appendix:A_1}
Restricting ourselves to doubly-magic spherically symmetric ($J=0$) nuclear initial states, with
single nucleon plus (A-1)-nuclear final states, terms in Eq. (\ref{nuclear_tensor}) become

\begin{equation}
    \sum_X\int d\Pi_X \rightarrow \sum_{\sigma_1 \tau_1} \sum_f \int [dP_{A-1}][dp_1],
\end{equation}

\begin{equation}
    \ket{X}\bra{X} \rightarrow \ket{\Psi_{A-1}^f,\bm{P}_{A-1}}\bra{\Psi_{A-1}^f,\bm{P}_{A-1}} \otimes \ket{\bm{p}_1,\sigma_1,\tau_1}\bra{\bm{p}_1,\sigma_1,\tau_1},
\end{equation}

\noindent where $f$ denotes all the possible (A-1)-nuclear states. The nuclear tensor becomes

\begin{equation}
\begin{aligned}
    W_A^{\mu \nu} = & \, \frac{1}{4 \pi m_A} \sum_{\sigma_1 \tau_1} \sum_f \int [dP_{A-1}][dp_1] (2 \pi)^4 \delta^4(q^\mu + P_A^\mu - P_{A-1}^\mu - p_{1}^\mu)\\
    & \qquad\qquad\qquad\qquad    \times \bra{\Psi_A} \hat{J}^{\mu}_A (0) \ket{\Psi_{A-1}^f,\bm{P}_{A-1};\bm{p}_1,\sigma_1,\tau_1} \\
    & \qquad\qquad\qquad\qquad    \times \bra{\Psi_{A-1}^f,\bm{P}_{A-1};\bm{p}_1,\sigma_1,\tau_1} \hat{J}_A^{\nu}(0) \ket{\Psi_A} \\
    = & \, \frac{1}{2 m_A} \frac{1}{2 P_{A-1}^+}\sum_{\sigma_f \tau_f} \sum_f \int \frac{dp_i^+ d\bm{p}_{i}^{\perp}}{(2 \pi)^3} \left[ \frac{1}{p_i^+ + q^+}\delta(q^- + P_A^- - (P_{A-1}^f)^- - p_{1}^-) \right] \\
    & \qquad\qquad\qquad\qquad\qquad \times \bra{\Psi_A} \hat{J}^{\mu}_A (0) \ket{\Psi_{A-1}^f,\bm{P}_{A-1};\bm{p}_i+\bm{q},\sigma_f,\tau_f} \\
    & \qquad\qquad\qquad\qquad\qquad \times \bra{\Psi_{A-1}^f,\bm{P}_{A-1};\bm{p}_i+\bm{q},\sigma_f,\tau_f} \hat{J}_A^{\nu}(0) \ket{\Psi_A}, \\
\end{aligned}
\end{equation}

\noindent where in the second equality we integrated over $\bm{P}_{A-1}$, changed variables to
$\bm{p}_i = \bm{p}_1-\bm{q}$, and re-labeled $(\sigma_1, \tau_1) \rightarrow (\sigma_f, \tau_f)$.
Furthermore, we drop the $J=0$ label for the initial nuclear state. Since we work in the collinear
frame, where the virtual photon defines the negative $z$-axis,

\begin{equation}
    (P_{A-1}^f)^- = \frac{m_f^2 + (\bm{p_{i}^{\perp}})^2}{P_A^+ - p_i^+}, \qquad p_1^- = \frac{m^2 + (\bm{p_{i}^{\perp}})^2}{p_i^+ + q^+}.
\end{equation}

\noindent The nuclear current matrix element is

\begin{equation}
\begin{aligned}
    & \braket{\Psi_{A-1}^f,\bm{P}_{A-1};\bm{p}_i+\bm{q},\sigma_i,\tau_i| \hat{J}_A^{\nu}(0)| \Psi_A} \\
    & \qquad\qquad\qquad\qquad = \sum_{s_1 t_1} \sum_{\sigma \sigma' \tau} \int [dp][dp'][dq_1] \sqrt{2p^+} \sqrt{2p'^+} J_N^\nu(\bm{p}',\sigma',\tau;\bm{p},\sigma,\tau) \\
    & \qquad\qquad\qquad\qquad\quad \times \braket{\bm{p}_i+\bm{q},\sigma_f,\tau_f|\hat{a}^\dagger(\bm{p}',\sigma',
    \tau)\hat{a}(\bm{p},\sigma,
    \tau)|\bm{q}_1,s_1,t_1} \\
    & \qquad\qquad\qquad\qquad\quad \times \braket{\Psi_{A-1}^f, \bm{P}_{A-1};\bm{q}_1,s_1,t_1|\Psi_A} \\
    & \qquad\qquad\qquad\qquad = \sum_{s_1 t_1} \sum_{\sigma \sigma' \tau} \int [dp][dp'][dq_1] \sqrt{2p^+} \sqrt{2p'^+} \sqrt{2(p_i^+ + q^+)}(2q_1^+) \\
    & \qquad\qquad\qquad\qquad\quad \times \braket{0|\hat{a}(\bm{p}_i+\bm{q},\sigma_f,\tau_f)\hat{a}^\dagger(\bm{p}',\sigma',
    \tau)\hat{a}(\bm{p},\sigma,
    \tau)\hat{a}^{\dagger}(\bm{q}_1,s_1,t_1)|0} \\
    & \qquad\qquad\qquad\qquad\quad \times \, J_N^\nu(\bm{p}',\sigma',\tau;\bm{p},\sigma,\tau)  \braket{\Psi_{A-1}^f, \bm{P}_{A-1}|\hat{a}(\bm{q}_1,s_1,t_1)|\Psi_A}.
\end{aligned}
\end{equation}

\noindent Here $J_N^\nu(\bm{p}',\sigma',\tau;\bm{p},\sigma,\tau)$ is the single-nucleon
electromagnetic current matrix element defined in Sect.~\ref{chap:scattering}.

Using Wick's Theorem to evaluate $\braket{0|\hat{a}\hat{a}^\dagger\hat{a}\hat{a}^\dagger|0}$ 
yields two delta functions which we use to integrate over $q_1$ and $p'$. This gives

\begin{equation}
\begin{aligned}
    \braket{\Psi_{A-1}^f,\bm{P}_{A-1};\bm{p}_i+\bm{q},\sigma_i,\tau_i| \hat{J}_A^{\nu}(0)| \Psi_A}
    & = \sum_{\sigma} \int [dp]\sqrt{2p^+} J_N^\nu(\bm{p}_i+\bm{q},\sigma_f,\tau_f;\bm{p},\sigma,\tau_f)\\
    & \qquad\quad \times  \braket{\Psi_{A-1}^f, \bm{P}_{A-1}|\hat{a}(\bm{p},\sigma,\tau_f)|\Psi_A}.
\end{aligned}
\end{equation}

\noindent Now we focus on the nuclear matrix element term.

\begin{equation}\label{A-1_NME_decomp}
\begin{gathered}
    \braket{\Psi_{A-1}^f, \bm{P}_{A-1}|\hat{a}(\bm{p},\sigma,\tau_f)|\Psi_A} = \\\sqrt{2P_A^+}\sqrt{2P_{A-1}^+} (2 \pi)^3\delta(P_{A-1}^+ + p^+-P_A^+)\delta^{(2)}(\bm{P}_{A-1}^{\perp} + \bm{p}^{\perp}) M^A_f(\bm{p},\sigma,\tau_f),
\end{gathered}
\end{equation}

\noindent where $M^A_f$ is a quantity constructed from the overlap of intrinsic A- and (A-1)-nuclear
light-front wavefunctions. $M^A_f$ can be determined by looking at the matrix element of the
momentum distribution operator

\begin{equation}
\begin{gathered}
    \Braket{\Psi_A, \bm{P}_A'|\hat{a}^\dagger(\bm{p},\sigma,\tau)\hat{a}(\bm{p},\sigma,\tau)|\Psi_A,\bm{P}_A} = \\ 
    \qquad (2 \pi)^3 2P_A^+ \delta(P^+ - P'^+) \delta^{(2)}(\bm{P}^{\perp} - \bm{P}'^{\perp}) \, n(\bm{p},\sigma,\tau).
\end{gathered}
\end{equation}

\noindent By inserting a complete set of (A-1) states in between the creation and annihilation
operator and using Eq. (\ref{A-1_NME_decomp}) we find the following relation,

\begin{equation} \label{M_F^A true relation}
    n(\bm{p},\sigma,\tau) = \sum_f M_f^A(\bm{p},\sigma,\tau)^* M_f^A(\bm{p},\sigma,\tau).
\end{equation}

\noindent Up to this point, everything has remained exact and completely independent of any specific
nuclear model. The inherent model dependence is encapsulated within $M_f^A$, which we evaluate by
employing mean-field theory. Within the mean-field framework, we have the following relation:

\begin{equation}\label{M_F^A MFT relationn}
\begin{aligned}
    n^{MF}(\bm{p},\sigma,\tau) & =\Braket{\Psi_A^{MF}|\hat{a}^\dagger(\bm{p},\sigma,\tau)\hat{a}(\bm{p},\sigma,\tau)|\Psi_A^{MF}} \\
    &= \sum_i \Braket{\Psi_A^{MF}|\hat{a}^\dagger(\bm{p},\sigma,\tau) \Ket{\Psi_i^{MF}} \Bra{\Psi_i^{MF}}\hat{a}(\bm{p},\sigma,\tau)|\Psi_A^{MF}},
\end{aligned}
\end{equation}

\noindent comparing Eq. (\ref{M_F^A true relation}) and Eq. (\ref{M_F^A MFT relationn}) we find

\begin{equation}
\begin{gathered}
    \sum_f \rightarrow \sum_i \\
    M_f^A \rightarrow M_i^A \\
    M_i^A(\bm{p},\sigma,\tau) =  \Braket{\Psi_i^{MF}|\hat{a}(\bm{p},\sigma,\tau)|\Psi_A^{MF}}
\end{gathered}
\end{equation}

\noindent where we approximate the (A-1)-nuclear states as one-hole states relative to the target
A-nucleus state, i.e. $\ket{\Psi^{MF}_i} = \hat{a}_i\ket{\Psi_A^{MF}}$. The final result of the
nuclear current is

\begin{equation}\label{A1_current}
\begin{gathered}
    \braket{\Psi_{A-1}^f,\bm{P}_{A-1};\bm{p}_i+\bm{q},\sigma_f,\tau_f| \hat{J}_A^{\nu}(0)| \Psi_A} \approx \\
    \sum_{\sigma} \frac{1}{\sqrt{2p_i^+} }J_N^\nu(\bm{p}_i+\bm{q},\sigma_f,\tau_f;\bm{p},\sigma,\tau_f) M_i^A(\bm{p}_i,\sigma,\tau_f)
\end{gathered}
\end{equation}

  \subsection{Derivation of Similarity Renormalization Group Terms}\label{SRG Implementation Chapter}

In this section, we provide explicit formulas for each of the four terms need to compute the FF momentum distribution.   To avoid redundancy, we present a detailed step-by-step derivation
solely for the $\delta \hat{U}(\lambda) \,\hat{n}^{\tau}(p, \sigma) \, \mathbb{\hat{I}}$
interference term, as the underlying procedure straightforwardly generalizes to the remaining
components, which are subsequently outlined. The corresponding non-relativistic results are
documented in the appendix of Ref. \cite{Tropiano:2024bmu}, and the light-front adaptation presented
here closely follows that study.

Firstly, the SRG unitary transformation at flow parameter $\lambda$ in the plane-wave basis is given
by

\begin{equation}
    \begin{aligned}
        \hat{U}(\lambda) = \mathbb{\hat{I}} + \frac{1}{4}\sum_{\substack{\sigma_1 \sigma_2 \\ \sigma_3 \sigma_4}} \sum_{\substack{\tau_1  \tau_2 \\ \tau_3 \tau_4}} & \int [dP] \frac{d\alpha d\bm{k}^\perp}{(2\pi)^3 \sqrt{\alpha (2 - \alpha)}} \frac{d\alpha'd\bm{k}'^\perp}{(2\pi)^3 \sqrt{\alpha' (2 - \alpha')}}  (P^+)^2 \\
        & \times \braket{\alpha, \bm{k}^\perp, \sigma_1, \tau_1, \sigma_2, \tau_2|\delta U (\lambda)|\alpha', \bm{k}'^\perp, \sigma_3, \tau_3, \sigma_4, \tau_4}\\
        & \times \, \hat{a}^{\dagger}(p_1) \hat{a}^{\dagger}(p_2) \hat{a} (p_4) \hat{a}(p_3) \, + \, ...
    \end{aligned}
\end{equation}

\noindent where $(\alpha, \bm{k}^\perp)$ and $(\alpha', \bm{k}'^\perp)$ are the relative momenta
and $(P^+, \bm{P}^\perp)$ is the center of mass momenta. $\bm{p}_i = (p^+_i, \bm{p}_i^\perp)$ for
$i=1,2,3,4$ are defined as follows:

\begin{equation}
    \begin{gathered}
    \bm{p}_1 = \left(\frac{\alpha}{2} P^+, \frac{\alpha}{2} \bm{P}^\perp + \bm{k}^\perp \right), \qquad \bm{p}_2 = \left(\frac{2-\alpha}{2} P^+, \frac{2-\alpha}{2} \bm{P}^\perp - \bm{k}^\perp \right), \\
    \bm{p}_3 = \left(\frac{\alpha'}{2} P^+, \frac{\alpha'}{2} \bm{P}^\perp + \bm{k}'^\perp \right), \qquad \bm{p}_4 = \left(\frac{2-\alpha'}{2} P^+, \frac{2-\alpha'}{2} \bm{P}^\perp - \bm{k}'^\perp \right).
    \end{gathered}
\end{equation}

\noindent Also, the matrix element of $\delta \hat{U}$ is antisymmetrized, i.e.

\begin{equation}
\begin{aligned}
    & \braket{\alpha, \bm{k}^\perp, \sigma_1, \tau_1, \sigma_2, \tau_2|\delta U (\lambda)|\alpha', \bm{k}'^\perp, \sigma_3, \tau_3, \sigma_4, \tau_4}  \\
    & \qquad\qquad\qquad\qquad = -\braket{\alpha, \bm{k}^\perp, \sigma_1, \tau_1, \sigma_2, \tau_2|\delta U (\lambda)|2-\alpha', -\bm{k}'^\perp,\sigma_4, \tau_4, \sigma_3, \tau_3} \\
    & \qquad\qquad\qquad\qquad = -\braket{2-\alpha, -\bm{k}^\perp, \sigma_2, \tau_2, \sigma_1, \tau_1|\delta U (\lambda)|\alpha', \bm{k}'^\perp, \sigma_3, \tau_3, \sigma_4, \tau_4}.
\end{aligned}
\end{equation}

\noindent In this thesis, we focus on incorporating SRG-evolution at the two-body level only. We
will enforce this by using Wick's theorem to truncate all operators of rank $N>2$. For the momentum
distribution operator, we have four terms to consider:

\begin{enumerate}
    \item $\mathbb{\hat{I}} \,\hat{n}^{\tau}(\bm{p}, \sigma) \, \mathbb{\hat{I}}$
    \item $\delta \hat{U}(\lambda) \,\hat{n}^{\tau}(\bm{p}, \sigma) \, \mathbb{\hat{I}}$
    \item $\mathbb{\hat{I}} \,\hat{n}^{\tau}(\bm{p}, \sigma) \, \delta \hat{U^{\dagger}}(\lambda)$
    \item $\delta \hat{U}(\lambda) \,\hat{n}^{\tau}(\bm{p}, \sigma) \, \delta
\hat{U}^{\dagger}(\lambda)$
\end{enumerate}

\noindent The first term is just the bare momentum distribution operator. To begin evaluating the
second term, let us examine the string of creation and annihilation operators it involves:

\begin{equation}\label{SRG_interference_term_1}
    \hat{a}^{\dagger}(p_1) \, \hat{a}^{\dagger}(p_2) \, \hat{a} (p_4) \, \hat{a}(p_3) \,\hat{a}^{\dagger}(\bm{p},\sigma,\tau) \, \hat{a}(\bm{p},\sigma,\tau).
\end{equation}

\noindent By applying Wick's theorem, we can isolate operators of rank $N<2$,

\begin{equation}\label{interference_wicks_thm}
    \begin{aligned}
        \wick{a^{\dagger} a^{\dagger} a \c1 a \c1 a^{\dagger}  a} + & \wick{a^{\dagger} a^{\dagger} \c1 a a \c1 a^{\dagger}  a} \\
        &  = (2 \pi)^3 \delta(p_3^+ - p^+) \delta(\bm{p}_3^\perp - \bm{p}^\perp) \delta_{\sigma_3 \sigma} \delta_{\tau_3 \tau} \\
        & \qquad - (2 \pi)^3 \delta(p_4^+ - p^+) \delta(\bm{p}_4^\perp - \bm{p}^\perp)   \delta_{\sigma_4 \sigma} \delta_{\tau_4 \tau} \\
        & = \frac{2}{P^+} \Bigg[  (2 \pi)^3 \delta\left(\alpha' - \frac{2p^+}{P^+}\right) \delta\left(\frac{\alpha'}{2}\bm{P}^\perp + \bm{k}'^\perp -\bm{p}^\perp\right) \delta_{\sigma_3 \sigma} \delta_{\tau_3 \tau} \\
        & \qquad - (2 \pi)^3 \delta\left(2 - \alpha' - \frac{2p^+}{P^+}\right) \delta\left(\frac{2-\alpha'}{2}\bm{P}^\perp - \bm{k}'^\perp - \bm{p}^\perp\right) \delta_{\sigma_4 \sigma} \delta_{\tau_4 \tau} \Bigg].
    \end{aligned}
\end{equation}

\noindent First, we focus on the first delta function term in Eq (\ref{interference_wicks_thm}).

\begin{align}\label{srg_inter_first_term}
    \left[\delta \hat{U}(\lambda) \,\hat{n}^{\tau}(\bm{p}, \sigma) \, \mathbb{\hat{I}}\right]_1 = & \,\frac{1}{4}\sum_{\substack{\sigma_1 \sigma_2 \\ \sigma_3 \sigma_4}} \sum_{\substack{\tau_1  \tau_2 \\ \tau_3 \tau_4}} \int \frac{dP^+ d\bm{P}^\perp}{(2 \pi)^3} \frac{d\alpha d\bm{k}^\perp}{(2\pi)^3 \sqrt{\alpha (2 - \alpha)}} \frac{d\alpha'd\bm{k}'^\perp}{ \sqrt{\alpha' (2 - \alpha')}} \nonumber \\
    & \times \braket{\alpha, \bm{k}^\perp, \sigma_1, \tau_1, \sigma_2, \tau_2|\delta U (\lambda)|\alpha', \bm{k}'^\perp, \sigma_3, \tau_3, \sigma_4, \tau_4} \nonumber \\
    & \times \delta\left(\alpha' - \frac{2p^+}{P^+}\right) \delta\left(\frac{\alpha'}{2}\bm{P}^\perp + \bm{k}'^\perp -\bm{p}^\perp\right) \delta_{\sigma_3 \sigma} \delta_{\tau_3 \tau} \nonumber \\
    & \times\hat{a}^{\dagger}(p_1) \hat{a}^{\dagger}(p_2) \hat{a} (p_4) \hat{a}(p,\sigma,\tau) \nonumber \displaybreak \\
    = & \,\frac{1}{4}\sum_{\substack{\sigma_1 \sigma_2 \\ \sigma'}} \sum_{\substack{\tau_1  \tau_2 \\ \tau'}} \int \frac{dP^+ d\bm{P}^\perp}{(2 \pi)^3} \frac{d\alpha d\bm{k}^\perp}{(2\pi)^3 \sqrt{\alpha (2 - \alpha)}} \frac{1}{ \sqrt{p^+ (P^+ - p^+)}} \frac{P^+}{2} \nonumber \\
    & \times \Braket{\alpha, \bm{k}^\perp, \sigma_1, \tau_1, \sigma_2, \tau_2|\delta U (\lambda)|\frac{2p^+}{P^+}, \bm{p}^\perp - \frac{p^+}{P^+}\bm{P}^\perp, \sigma, \tau, \sigma', \tau'} \nonumber \\
    & \times\hat{a}^{\dagger}(p_1,\sigma_1,\tau_1) \hat{a}^{\dagger}(p_2,\sigma_2,\tau_2) \hat{a} (p_4,\sigma',\tau') \hat{a}(p,\sigma,\tau). 
\end{align}

\noindent Where in the second equality, we used the delta function to integrate over
$(\alpha',\bm{k}'^\perp)$ and re-labeled $(\sigma_4,\tau_4)\rightarrow(\sigma',\tau')$. In
evaluating the second delta function term in Eq. (\ref{interference_wicks_thm}), using the asymmetry
of $\delta \hat{U}(\lambda)$ and re-labeling $(\sigma_3,\tau_3)\rightarrow(\sigma',\tau')$, we find
that it is identical to Eq. (\ref{srg_inter_first_term}). Hence the final result is,

\begin{equation}\label{srg_inter_U}
\begin{aligned}
    \delta \hat{U}(\lambda) \,\hat{n}^{\tau}(\bm{p}, \sigma) \, \mathbb{\hat{I}} = & \,\frac{1}{2}\sum_{\substack{\sigma_1 \sigma_2 \\ \sigma'}} \sum_{\substack{\tau_1  \tau_2 \\ \tau'}} \int \frac{dP^+ d\bm{P}^\perp}{(2 \pi)^3} \frac{d\alpha d\bm{k}^\perp}{(2\pi)^3 \sqrt{\alpha (2 - \alpha)}} \frac{1}{ \sqrt{p^+ (P^+ - p^+)}} \frac{P^+}{2}\\
    & \times \Braket{\alpha, \bm{k}^\perp, \sigma_1, \tau_1, \sigma_2, \tau_2|\delta U (\lambda)|\frac{2p^+}{P^+}, \bm{p}^\perp - \frac{p^+}{P^+}\bm{P}^\perp, \sigma, \tau, \sigma', \tau'} \\
    & \times\hat{a}^{\dagger}(p_1) \hat{a}^{\dagger}(p_2) \hat{a} (p_4,\sigma',\tau') \hat{a}(p,\sigma,\tau). \\
\end{aligned}
\end{equation}

\noindent The third term is similar

\begin{equation}\label{srg_inter_Udag}
\begin{aligned}
    \mathbb{\hat{I}} \,\hat{n}^{\tau}(\bm{p}, \sigma) \, \delta \hat{U}^{\dagger}(\lambda) = & \,\frac{1}{2}\sum_{\substack{\sigma_3 \sigma_4 \\ \sigma'}} \sum_{\substack{\tau_3  \tau_4 \\ \tau'}} \int \frac{dP^+ d\bm{P}^\perp}{(2 \pi)^3} \frac{d\alpha' d\bm{k}'^\perp}{(2\pi)^3 \sqrt{\alpha' (2 - \alpha')}} \frac{1}{ \sqrt{p^+ (P^+ - p^+)}} \frac{P^+}{2}\\
    & \times \Braket{\frac{2p^+}{P^+}, \bm{p}^\perp - \frac{p^+}{P^+}\bm{P}^\perp, \sigma, \tau, \sigma', \tau' | \delta U^{\dagger}(\lambda) |\alpha', \bm{k}'^\perp, \sigma_3, \tau_3, \sigma_4, \tau_4} \\
    & \times  \hat{a}^{\dagger}(p,\sigma,\tau) \hat{a}^{\dagger}(p_2,\sigma',\tau') \hat{a} (p_4) \hat{a}(p_3). \\
\end{aligned}
\end{equation}

\noindent For the fourth term, looking at the string of creation and annihilation operators it
involves, we anticipate four terms at the two-body level:

\begin{itemize}
    \item $\wick{a^{\dagger} a^{\dagger} \c2 a \c1 a \,\, \c1 a^{\dagger} \c1 a \,\, \c1 a^{\dagger}
\c2 a^{\dagger} a a}$
    \item $\wick{a^{\dagger} a^{\dagger} \c2 a \c1 a \,\, \c1 a^{\dagger} \c1 a \,\, \c2 a^{\dagger}
\c1 a^{\dagger} a a}$
    \item $\wick{a^{\dagger} a^{\dagger} \c1 a \c2 a \,\, \c1 a^{\dagger} \c1 a \,\, \c1 a^{\dagger}
\c2 a^{\dagger} a a}$
    \item $\wick{a^{\dagger} a^{\dagger} \c1 a \c2 a \,\, \c1 a^{\dagger} \c1 a \,\, \c2 a^{\dagger}
\c1 a^{\dagger} a a}$
\end{itemize}

\noindent Again, utilizing the asymmetry of the $\delta U$ matrix element and re-labeling spin and
isospin projections we find that we only need to evaluate the first contraction above and multiply
the result by four. This gives

\begin{equation}\label{srg_UUdag}
\begin{gathered}
     \delta \hat{U}(\lambda) \,\hat{n}^{\tau}(\bm{p}, \sigma) \, \delta \hat{U}^{\dagger}(\lambda) = \\
     \frac{1}{4} \sum_{\substack{\sigma_1  \sigma_2 \sigma_3 \\ \sigma_4 \sigma \sigma'}} \sum_{\substack{\tau_1  \tau_2 \tau_3 \\ \tau_4 \tau'}} \int \frac{dP^+ d\bm{P}^\perp}{(2\pi)^3} \frac{d\alpha d\bm{k}^\perp}{(2\pi)^3 \sqrt{\alpha (2 - \alpha)}} \frac{d\alpha'd\bm{k}'^\perp}{(2\pi)^3 \sqrt{\alpha' (2 - \alpha')}} \frac{1}{p^+ (P^+ - p^+ )} \left( \frac{P^+}{2}\right)^2 \\
     \times \Braket{\alpha, \bm{k}^\perp, \sigma_1, \tau_1, \sigma_2, \tau_2|\delta U(\lambda)|\frac{2 p^+}{P^+}, \bm{p}^\perp - \frac{p^+}{P^+}\bm{P}^\perp, \sigma, \tau, \sigma', \tau'} \\
     \times \Braket{\frac{2 p^+}{P^+}, \bm{p}^\perp - \frac{p^+}{P^+}\bm{P}^\perp, \sigma, \tau, \sigma', \tau'|\delta U^{\dagger}(\lambda)|\alpha', \bm{k}'^\perp, \sigma_3, \tau_3, \sigma_4, \tau_4} \\
     \times \hat{a}^{\dagger}(p_1) \hat{a}^{\dagger}(p_2) \hat{a} (p_4) \hat{a}(p_3) .
\end{gathered}
\end{equation}

Now we evaluate all four terms with respect to the nuclear ground state

\begin{equation}
    \ket{\Psi_A^{MF}} = \prod_{i<F} \hat{a}_i^\dagger \ket{\Omega},
\end{equation}

\noindent where $a_i^\dagger$ creates a nucleon in single-particle orbital indexed by $i$,
$\ket{\Omega}$ is the interacting vacuum state, and $F$ refers to the Fermi surface. We perform a
change of basis, re-writing plane-wave creation and annihilation operators in terms of
single-particle orbitals

\begin{equation}\label{change_of_basis}
    \hat{a}(\bm{p},\sigma,\tau) = \sum_{i} \psi_{i}^+(\bm{p},\sigma,\tau) \,\hat{a}_{i}.
\end{equation}
Where $\psi^+_{i}$ is the overlap between the plus-projected single particle light-front
wavefunctions of nucleus A with eigenvalues $i$ and the LF Dirac spinor.
\begin{equation}
    \psi_{i}^+(\bm{p},\sigma,\tau) = \frac{1}{\sqrt{2p^+}}
    \bar{u}(\bm{p},\sigma,\tau)\gamma^+\psi_i(\bm{p}) = \sqrt{\frac{2}{p^+}}u^\dagger(\bm{p},\sigma,\tau)\psi_i^+(\bm{p})
\end{equation}
Contractions of $a_i$ with respect to the nuclear ground state are given by,

\begin{equation}\label{nuclear_contraction}
    \wick{\c1 a^\dagger_i \c1 a_j} = \braket{\Psi_A^{MF}|\hat{a}^\dagger_i \hat{a}_j|\Psi_A^{MF}} = \delta_{ij}
\end{equation}

\noindent for $i,j<F$. Using Eqs. (\ref{change_of_basis}, \ref{nuclear_contraction}) we immediately
get the following:

\begin{equation}
    \braket{\Psi_A^{MF}|\mathbb{\hat{I}} \,\hat{n}^{\tau}(\bm{p}, \sigma) \, \mathbb{\hat{I}}|\Psi_A^{MF}} = \sum_{i<F} |\psi_i^+(\bm{p},\sigma,\tau)|^2.
\end{equation}

\noindent The second (interference) term gives:
\begin{equation}
\begin{gathered}
     \braket{\Psi_A^{MF}|\delta\hat{U}(\lambda) \,\hat{n}^{\tau}(\bm{p}, \sigma) \, \mathbb{\hat{I}}|\Psi_A^{MF}} = \\
     \,\frac{1}{2}\sum_{\substack{\sigma_1 \sigma_2 \\ \sigma'}} \sum_{\substack{\tau_1  \tau_2 \\ \tau'}} \sum_{ij<F}\int \frac{dP^+ d\bm{P}^\perp}{(2 \pi)^3} \frac{d\alpha d\bm{k}^\perp}{(2\pi)^3 \sqrt{\alpha (2 - \alpha)}} \frac{1}{ \sqrt{p^+ (P^+ - p^+)}} \frac{P^+}{2}\\
    \times \Braket{\alpha, \bm{k}^\perp, \sigma_1, \tau_1, \sigma_2, \tau_2|\delta U (\lambda)|\frac{2p^+}{P^+}, \bm{p}^\perp - \frac{p^+}{P^+}\bm{P}^\perp, \sigma, \tau, \sigma', \tau'} \\
    \times \, \psi_i^{+\dagger}(p_1) \psi_j^{+\dagger}(p_2) \left[\psi_j^+(\bm{p}_4,\sigma',\tau') \psi_i^+(\bm{p},\sigma,\tau) - \psi_i^+(\bm{p}_4,\sigma',\tau') \psi_j^+(\bm{p},\sigma,\tau) \right]. \\
\end{gathered}
\end{equation}

\noindent The third (conjugate interference) term gives:
\begin{equation}
\begin{gathered}
    \braket{\Psi_A^{MF}|\mathbb{\hat{I}} \,\hat{n}^{\tau}(\bm{p}, \sigma) \, \delta \hat{U}^{\dagger}(\lambda)|\Psi_A^{MF}} = \\
    \frac{1}{2}\sum_{\substack{\sigma_3 \sigma_4 \\ \sigma'}} \sum_{\substack{\tau_3  \tau_4 \\ \tau'}} \sum_{ij<F}\int \frac{dP^+ d\bm{P}^\perp}{(2 \pi)^3} \frac{d\alpha' d\bm{k}'^\perp}{(2\pi)^3 \sqrt{\alpha' (2 - \alpha')}} \frac{1}{ \sqrt{p^+ (P^+ - p^+)}}\frac{P^+}{2}\\
    \times \Braket{\frac{2p^+}{P^+}, \bm{p}^\perp - \frac{p^+}{P^+}\bm{P}^\perp, \sigma, \tau, \sigma', \tau' | \delta U^{\dagger}(\lambda) |\alpha', \bm{k}'^\perp, \sigma_3, \tau_3, \sigma_4, \tau_4} \\
    \times \, \psi_{i}^{+ \dagger}(\bm{p},\sigma,\tau) \psi_{j}^{+ \dagger}(\bm{p}_2,\sigma',\tau') \left[ \psi_{j}^+(p_4) \psi_{i}^+(p_3) - \psi_{i}^+(p_4) \psi_{j}^+(p_3) \right]. \\
\end{gathered}
\end{equation}

\noindent The fourth (quadratic) term gives:
\begin{equation}\label{srg_UUdag_mf}
\begin{gathered}
     \braket{\Psi_A^{MF}|\delta\hat{U}(\lambda) \,\hat{n}^{\tau}(\bm{p}, \sigma) \, \delta \hat{U}^{\dagger}(\lambda)|\Psi_A^{MF}} = \\
     \frac{1}{4} \sum_{\substack{\sigma_1  \sigma_2 \sigma_3 \\ \sigma_4 \sigma \sigma'}} \sum_{\substack{\tau_1  \tau_2 \tau_3 \\ \tau_4 \tau'}} \sum_{ij < F} \int \frac{dP^+ d\bm{P}^\perp}{(2\pi)^3} \frac{d\alpha d\bm{k}^\perp}{(2\pi)^3 \sqrt{\alpha (2 - \alpha)}} \frac{d\alpha'd\bm{k}'^\perp}{(2\pi)^3 \sqrt{\alpha' (2 - \alpha')}} \frac{1}{p^+ (P^+ - p^+ )} \left( \frac{P^+}{2}\right)^2 \\
     \times \Braket{\alpha, \bm{k}^\perp, \sigma_1, \tau_1, \sigma_2, \tau_2|\delta U(\lambda)|\frac{2 p^+}{P^+}, \bm{p}^\perp - \frac{p^+}{P^+}\bm{P}^\perp, \sigma, \tau, \sigma', \tau'} \\
     \times \Braket{\frac{2 p^+}{P^+}, \bm{p}^\perp - \frac{p^+}{P^+}\bm{P}^\perp, \sigma, \tau, \sigma', \tau'|\delta U^{\dagger}(\lambda)|\alpha', \bm{k}'^\perp, \sigma_3, \tau_3, \sigma_4, \tau_4} \\
     \times \psi_{i}^{+ \dagger}(p_1) \psi_{j}^{+ \dagger}(p_2)\bigg[ \psi_{j}^{+}(p_4) \psi_{i}^{+}(p_3) - \psi_{i}^{+}(p_4) \psi_{j}^{+}(p_3) \bigg].
\end{gathered}
\end{equation}

For calculation of inclusive electron-nucleus scattering nuclear current matrix elements, we will
encounter two terms that involve the $\delta\hat{U}(\lambda)$ transformation. We write down the
results here.

\begin{equation}
\begin{gathered}
    \braket{\Psi_{i}^{MF}|\hat{a}(\bm{q}_1,s_1,t_1) \delta \hat{U}^\dagger(\lambda)|\Psi_A^{MF}} = \braket{\Psi_{A}^{MF}|\hat{a}_i^\dagger \hat{a}(\bm{q}_1,s_1,t_1) \delta \hat{U}^\dagger(\lambda)|\Psi_A^{MF}} = \\
    \frac{1}{2}\sum_{\sigma_2 \sigma_3 \sigma_4} \sum_{\tau_2 \tau_3  \tau_4}\sum_{j<F}\int \frac{dP^+ d\bm{P}^\perp}{(2 \pi)^3} \frac{d\alpha' d\bm{k}'^\perp}{(2\pi)^3 \sqrt{\alpha' (2 - \alpha')}} \frac{1}{ \sqrt{q^+ (P^+ - q^+)}}\frac{P^+}{2}\\
    \times \Braket{\frac{2q_1^+}{P^+}, \bm{q}_1^\perp - \frac{q_1^+}{P^+}\bm{P}^\perp, s_1, t_1, \sigma_2, \tau_2 | \delta U^{\dagger}(\lambda) |\alpha', \bm{k}'^\perp, \sigma_3, \tau_3, \sigma_4, \tau_4} \\
    \times \, \psi_{j}^{+ \dagger}(p_2) \left[ \psi_{j}^+(p_4) \psi_{i}^+(p_3) - \psi_{i}^+(p_4) \psi_{j}^+(p_3) \right]. \\
\end{gathered}
\end{equation}

\begin{equation}
\begin{gathered}
    \braket{\Psi_{ij}^{MF}|\hat{a}(\bm{q}_2,s_2,t_2) \hat{a}(\bm{q}_1,s_1,t_1) \delta \hat{U}^\dagger(\lambda)|\Psi_A^{MF}} = \\
    \braket{\Psi_A^{MF}|\hat{a}_i^\dagger \hat{a}_j^\dagger \hat{a}(\bm{q}_2,s_2,t_2) \hat{a}(\bm{q}_1,s_1,t_1) \delta \hat{U}^\dagger(\lambda)|\Psi_A^{MF}} = \\
    \frac{1}{2}\sum_{\sigma_3 \sigma_4} \sum_{\tau_3  \tau_4} \int \frac{d\alpha' d\bm{k}'^\perp}{(2\pi)^3 \sqrt{\alpha' (2 - \alpha')}} \frac{1}{ \sqrt{q_1^+ q_2^+}}\frac{q_1^+ + q_2^+}{2}\\
    \times \Braket{\frac{2q_1^+}{q_1^+ + q_2^+}, \bm{q}_1^\perp - \frac{q_1^+}{q_1^+ + q_2^+}(\bm{q}_1^\perp + \bm{q}_2^\perp), s_1, t_1, s_2, t_2 | \delta U^{\dagger}(\lambda) |\alpha', \bm{k}'^\perp, \sigma_3, \tau_3, \sigma_4, \tau_4} \\
    \times \, \left[ \psi_{j}^+(p_4) \psi_{i}^+(p_3) - \psi_{i}^+(p_4) \psi_{j}^+(p_3) \right]. \\
\end{gathered}
\end{equation}



\end{document}